\documentclass[aps,prd,longbibliography,nofootinbib,floatfix]{revtex4-2}
\usepackage{array}[=2016-10-06]

\usepackage{graphicx}
\usepackage{amsmath,amssymb,bm,mathtools}
\usepackage{siunitx}
\DeclareSIUnit\year{yr}
\DeclareSIUnit\hour{h}
\DeclareSIUnit\bit{bit}
\DeclareSIUnit\byte{B}
\DeclareSIUnit\dollar{\$}
\usepackage{booktabs}
\usepackage{tabularx}
\newcolumntype{L}[1]{>{\raggedright\arraybackslash}p{#1}}
\newcolumntype{C}[1]{>{\centering\arraybackslash}p{#1}}
\newcolumntype{Y}{>{\raggedright\arraybackslash}X}
\usepackage{multirow}
\usepackage{microtype}
\usepackage{natbib}
\usepackage{xparse}
\usepackage{tikz}
\usepackage{pgfplots}
\usepgfplotslibrary{groupplots,fillbetween}
\usetikzlibrary{arrows.meta,positioning,fit,calc,shapes.geometric,backgrounds}
\pgfplotsset{
  compat=1.18,
  every axis/.append style={
    line width=0.75pt,
    tick style={line width=0.65pt},
    tick label style={font=\small},
    label style={font=\small},
    legend style={font=\small,fill=white,draw=black,line width=0.45pt},
    legend cell align=left,
    grid=major,
    grid style={gray!22},
    major grid style={gray!28}
  }
}
\usepackage{hyperref}
\hypersetup{
  colorlinks=true,
  allcolors=blue,
  pdftitle={Operations, Maintenance, and Industrial Scaling of Megawatt-Class Orbital Data Centers},
  pdfauthor={Slava G. Turyshev},
  pdfsubject={Lifecycle physics, robotic servicing, logistics, and human-support thresholds for orbital data centers}
}

\newcommand{\Eq}[1]{Eq.~(\ref{#1})}

\newcommand{\Tab}[1]{Table~\ref{#1}}

\newcommand{\LEO}{\ensuremath{\mathrm{LEO}}}

\newcommand{\Pit}{P_{\mathrm{IT}}}
\newcommand{\Ptot}{P_{\mathrm{tot}}}
\newcommand{\Pcl}{P_{\Sigma}}
\newcommand{\PMW}{P_{\mathrm{MW}}}
\newcommand{\pn}{p_n}
\newcommand{\pc}{p_c}
\newcommand{\alphaOH}{\alpha_{\mathrm{OH}}}
\newcommand{\mkw}{m_{\mathrm{kW}}}
\newcommand{\mkwm}{m_{\mathrm{kW,maint}}}
\newcommand{\Trad}{T_{\mathrm{rad}}}
\newcommand{\Arad}{A_{\mathrm{rad}}}
\newcommand{\Apv}{A_{\mathrm{PV}}}
\newcommand{\sigSB}{\sigma_{\mathrm{SB}}}
\newcommand{\etaPMAD}{\eta_{\mathrm{PMAD}}}
\newcommand{\etaView}{\eta_{\mathrm{view}}}
\newcommand{\epsRad}{\epsilon_{\mathrm{rad}}}
\newcommand{\fSun}{f_{\odot}}
\newcommand{\lambdaCat}{\lambda_{\mathrm{cat}}}
\newcommand{\lambdaDom}{\lambda_{\mathrm{dom}}}
\newcommand{\pex}{p_{\mathrm{exc}}}
\newcommand{\pU}{p_{U}}
\newcommand{\Mdot}{\dot M}
\newcommand{\CRF}{\mathrm{CRF}}
\newcommand{\MTBF}{\ensuremath{\mathrm{MTBF}}}
\newcommand{\MTTR}{\ensuremath{\mathrm{MTTR}}}

\begin{document}

\title{Operations, Maintenance, and Industrial Scaling of MW-Class Orbital Data Centers}
\author{Slava G. Turyshev}
\affiliation{
Jet Propulsion Laboratory, California Institute of Technology,\\
4800 Oak Grove Drive, Pasadena, CA 91109-0899, USA
}%
\date{\today}

\begin{abstract}
Megawatt-class orbital data centers require continuous maintenance, replacement, inventory, and service capacity in addition to spacecraft power and thermal systems.  This paper formulates an integrated analytical lifecycle framework for permanent and transient failures, modular orbital replacement units, robotic servicing, spare inventory, scheduled technology refresh, correlated faults, cybersecurity, and optional human support.  The model combines nonhomogeneous component hazards, capacity-weighted availability, multiclass robotic-service capacity, Poisson base-stock inventory, replacement-flow accounting, and human-support break-even relations. For a reference \SI{1}{MW} cluster with 10 active \SI{100}{kW} nodes, 1 reserve node, and $\sim$200 \SI{5}{kW} compute cartridges, low, nominal, and high deployed-mass allocations span $\sim$50--75\,kg/kW; these are system allocations rather than closed flight designs.  The reference assumptions yield 70.2 random or life-limited interventions and 323--349 planned refresh operations per MW-year, for a total of 393--419 standardized operations per MW-year.  Calendar-cohort accounting gives a first-generation logistics range of 5.3--9.0\,t/(MW year), with a nominal case of approximately \SI{6.6}{t/(MW.year)}, and 560--700 productive robot-hours per MW-year.  Planned refresh exceeds random replacement under the stated component populations, hazards, and 3--15-year intervals.  At approximately 400 standardized operations per MW-year, the post-internal-recovery exception probability should be $\lesssim10^{-3}$, with an objective near $10^{-4}$ at large scale; terminal non-recovery $\pU$ requires a smaller mission-level allocation.  The target catastrophic-loss hazard for a \SI{100}{kW} node is $0.01$--$0.03~\mathrm{yr^{-1}}$.  Large-domain events require joint specification of event rate, affected capacity fraction, duration, and recovery state. Parametric workload and cost cases place contingency visits at tens of megawatts, periodic campaigns at tens to hundreds of megawatts, and dedicated personnel at several hundred megawatts to gigawatts.  These are sensitivity results rather than forecasts.  The reference first deployment is uncrewed, autonomously fault-managed, robotically maintainable, supported by class-specific inventory based on a six-month replenishment horizon, and compatible with later human access without permanent habitation.
\end{abstract}

\maketitle
\tableofcontents

\section{Introduction and motivation}
\label{sec:intro}

Orbital data centers (ODCs) have been proposed as solar-powered compute constellations, communications-integrated edge platforms, and eventually as industrial-scale computing infrastructure.  A physically grounded assessment must begin with the fact that orbital solar flux does not remove the need to launch the photovoltaic (PV) conversion system, reject essentially all consumed electrical power as heat, maintain a sustained communications path, tolerate the space environment, and replace obsolete or failed hardware.  A recent system-level analysis quantified these coupled power, storage, thermal, communication, and mass constraints and showed that a representative \SI{1}{MW} high-sunlight low Earth orbit (LEO) design---where the megawatt denotes delivered information technology (IT) power---requires approximately $5.64\times10^3~\mathrm{m^2}$ of beginning-of-life (BOL) PV area, $2.50\times10^3~\mathrm{m^2}$ of radiator area, and a total deployed mass of roughly $34$--$59~\mathrm{kg/kW}$ before a dedicated servicing architecture is added \cite{Turyshev2026ODC}.  Unless stated otherwise, operating capability and retained subsystem capacity are evaluated at end of life (EOL).  Related studies have explored tethered structures, distributed training, communications-integrated operation, and workload placement \cite{Aguera2025Suncatcher,Bargatin2025Tether,Gutierrez2025ESPI,Chabra2026OrbitalBrain,Thummala2025Compute,Aili2025CarbonNeutral}.

The lifecycle problem is more demanding than a conventional spacecraft replacement calculation.  A data center contains large populations of semiconductors, memories, storage devices, optical transceivers, converters, pumps, valves, mechanisms, and connectors.  It also experiences deliberate hardware churn because compute performance per watt and storage density improve on timescales shorter than a typical spacecraft bus life.  A multi-megawatt ODC therefore behaves simultaneously as a spacecraft constellation, a repairable production system, an automated warehouse, a high-availability network, and an orbital logistics customer.  If maintenance is omitted from the initial architecture, the resulting design either requires frequent whole-node disposal or contains nominally replaceable elements that are inaccessible to robotic service.

The analysis evaluates whether a servicing system can execute hundreds of standardized operations per MW-year, recover from internal failures, keep unresolved work within the cluster availability allocation, maintain inventory under launch delay, and contain correlated software and power faults.  Human support is evaluated through separate workload and cost thresholds.  Humans can diagnose unmodeled failures and adapt tools to damaged interfaces, but a crewed facility requires pressurized volume, life support, radiation protection, emergency return, food, waste, exercise, medical systems, and transportation.  Human presence is justified only when the value of reduced recovery time and expanded intervention capability exceeds these fixed costs.

This paper develops the operations, maintenance, logistics, robotic-servicing, and human-in-the-loop architecture required to close that lifecycle problem.  The analysis is conditional: it does not assume that terrestrial-user general computing is economically competitive in orbit.  Rather, it asks what sustainment system is required if a MW-class or larger ODC is deployed for a workload whose location value or broader economics have already justified the spacecraft.  The main contributions are:

\begin{enumerate}
\item a repairable-system model linking component hazard, replacement mass, robotic task demand, spare stock, and delivered-capacity availability;
\item a maintainability architecture based on nested orbital replacement unit (ORU) hierarchies and standardized power, thermal, optical, and mechanical interfaces;
\item quantitative requirements on robotic task success, repair capacity, reserve nodes, inventory lead time, and common-mode fault containment;
\item comparative lifecycle models for autonomous-without-service, uncrewed robotic, occasionally crewed, campaign-crewed, and permanently crewed architectures;
\item scale-dependent mission concepts from a \SI{1}{MW} pathfinder to gigawatt-class orbital industry; and
\item technology thresholds that distinguish demonstrated capability from reasonable extrapolation and speculation.
\end{enumerate}

The paper is organized as follows.  Section~\ref{sec:system} defines the architecture and performance metrics.  Section~\ref{sec:physics} summarizes the physical closure inherited from power, thermal, environmental, and structural constraints.  Sections~\ref{sec:measurement}--\ref{sec:logistics} develop operational observables, maintainable hardware, robotics, reliability, and logistics.  Sections~\ref{sec:performance} and~\ref{sec:uncertainty} provide sensitivity and uncertainty analyses.  Human architectures and orbital geography are treated in Secs.~\ref{sec:human} and~\ref{sec:geography}.  Alternative concepts, representative missions, technology development, risks, and strategic implications follow in Secs.~\ref{sec:alternatives}--\ref{sec:strategy}.  Detailed derivations and parameter tables are collected in the appendices.

Table~\ref{tab:abbreviations} provides a compact reference for the cross-cutting abbreviations used throughout the analysis.  It is intentionally selective and does not replace definitions at first use.

\begin{table}[h!]
\caption{Major abbreviations used repeatedly in the manuscript.  Brief role descriptions are included only where they aid navigation.}
\label{tab:abbreviations}
\centering
\small
\setlength{\tabcolsep}{4.5pt}
\renewcommand{\arraystretch}{1.04}
\begin{tabularx}{\textwidth}{@{}L{0.13\textwidth} L{0.35\textwidth} L{0.11\textwidth} Y@{}}
\toprule
Abbreviation & Full term or role & Abbreviation & Full term or role \\
\midrule
ADCS & attitude determination and control system & BOL / EOL & beginning of life / end of life \\
ECC & error-correcting code & ECLSS & environmental control and life support system \\
GEO/\allowbreak LEO/\allowbreak MEO & geostationary, low, and medium Earth orbit & IT & information technology \\
MMOD & micrometeoroids and orbital debris & MTBF / MTTR & mean time between failures / mean time to repair or restore \\
ODC & orbital data center & ORU & orbital replacement unit \\
PMAD & power management and distribution & PV & photovoltaic generation \\
RF & radio frequency & SEE / TID & single-event effect / total ionizing dose \\
\bottomrule
\end{tabularx}
\end{table}

\section{System definition, operational architecture, and figures of merit}
\label{sec:system}

\subsection{Reference system boundary}

The system boundary includes every orbital element required to deliver a specified aggregate IT power $\Pcl$: compute and storage hardware, local and inter-node networking, power generation and conditioning, energy storage, thermal transport and radiators, flight avionics, an attitude determination and control system (ADCS), propulsion, optical and radio-frequency (RF) communications terminals, service robots, spare storage, cargo interfaces, reserve capacity, and end-of-life disposal capability.  Ground operations, ground communications infrastructure, launch, return, and crew transportation enter the lifecycle cost and logistics model but not the deployed mass $M_\Sigma$ unless the hardware remains in orbit.

Because lifecycle intensities and economic values are quoted per delivered megawatt while the physical power equations use watts, define the dimensionless normalization
\begin{equation}
\PMW\equiv\frac{\Pcl}{\SI{1}{MW}}.
\label{eq:PMWdefinition}
\end{equation}
Thus $\Pcl$ retains units of watts in power, area, mass, and capacity-fraction equations, whereas $\PMW$ is the numerical number of delivered megawatts and is used in all per-MW labor, logistics, and economic normalizations.  The distinction prevents an implicit factor of $10^6$ from entering intensive quantities.

Table~\ref{tab:nomenclature} summarizes the core notation used across the physical, reliability, logistics, and human-support models.

\begin{table}[tbp]
\caption{Core symbols used in the analysis.  Power and capacity quantities are EOL delivered values unless explicitly identified as BOL quantities.}
\label{tab:nomenclature}
\centering
\small
\setlength{\tabcolsep}{4.5pt}
\renewcommand{\arraystretch}{1.00}
\begin{tabularx}{\textwidth}{@{}L{0.13\textwidth} C{0.12\textwidth} Y@{}}
\toprule
Symbol & Units & Meaning \\
\midrule
\multicolumn{3}{@{}l}{\emph{Power and architecture}} \\
$\Pit$ & W & Delivered IT power of one node or payload interface \\
$\Pcl$ & W & Aggregate delivered IT power of the cluster \\
$\PMW$ & 1 & Dimensionless delivered-power normalization, $\Pcl/(\SI{1}{MW})$ \\
$\Ptot=\alphaOH\Pit$ & W & Total regulated electrical load including non-IT overhead \\
$\pn$, $\pc$ & W & Nominal IT power of an independently deployable node and a replaceable compute cartridge \\
$\mkw$ & kg kW$^{-1}$ & Deployed system mass per delivered IT kilowatt \\
\addlinespace[2pt]
\multicolumn{3}{@{}l}{\emph{Reliability and availability}} \\
$N_i$, $m_i$ & --, kg & Installed population and replacement mass of component class $i$ \\
$h_i(t)$ & yr$^{-1}$ & Age-dependent permanent-failure hazard of one item in class $i$ \\
$\Lambda_i$, $\Lambda$ & yr$^{-1}$ & Aggregate event rate of class $i$ and of the maintenance system \\
$\mathrm{MTBF}_{\rm eq,i}$ & yr & Equivalent exponential MTBF, $\lambda_i^{-1}$, used only as a summary metric \\
$T_{r,i}$ & yr & Planned refresh interval for component class $i$ \\
$A_{\rm cap}$ & -- & Time-averaged fraction of aggregate IT power delivered to users \\
$\lambdaCat$, $\lambdaDom$ & yr$^{-1}$ & Catastrophic node-loss rate and rate of a defined large-domain event; temporary events also require affected fraction and duration \\
$\pex$ & -- & Probability that a standardized operation enters external recovery after ordinary internal retry, tool change, and alternate-robot recovery \\
$\pU$ & -- & Probability that an operation remains terminally unresolved after all qualified recovery defenses \\
\addlinespace[2pt]
\multicolumn{3}{@{}l}{\emph{Logistics, labor, and economics}} \\
$\Mdot_{\rm up}$ & kg yr$^{-1}$ & Delivered replacement, refresh, consumable, and packaging mass flow \\
$L_i$, $S_i$ & yr, units & Resupply lead time and orbital base-stock level for spare class $i$ \\
$H_R$, $H_H$ & h yr$^{-1}$ & Total productive robotic and onsite human labor demand \\
$\ell_M$ & t/(MW yr) & Logistics intensity normalized by $\PMW$ \\
$\ell_R$ & h/(MW yr) & Productive robotic-work intensity normalized by $\PMW$ \\
$\ell_H$ & crew h/(MW yr) & Onsite human-work intensity normalized by $\PMW$ \\
$v$ & \$/(MW yr) & Economic value of one delivered MW-year of usable compute capacity \\
$C_H$, $C_{H,0}$ & \$/yr & Total and fixed annualized cost of a human-support architecture \\
$c_{H,1}$ & \$/(MW yr) & Variable human-support cost per delivered MW-year \\
\bottomrule
\end{tabularx}
\end{table}

The baseline is a distributed cluster of $N_n$ independently isolatable nodes,
\begin{equation}
N_n=\frac{\Pcl}{\pn},
\label{eq:Nnode}
\end{equation}
with $N_c=\Pcl/\pc$ active compute cartridges.  The reference numerical case uses
\begin{equation}
\Pcl=\SI{1}{MW},\qquad \pn=\SI{100}{kW},\qquad \pc=\SI{5}{kW},
\label{eq:referencepowers}
\end{equation}
so that ten active nodes contain two hundred active cartridges and $\PMW=1$.  One additional \SI{100}{kW} reserve node is carried in the first cluster.  At larger scale the reserve is pooled and expressed as a fraction $r_n$ of active capacity rather than as one full spare per MW.

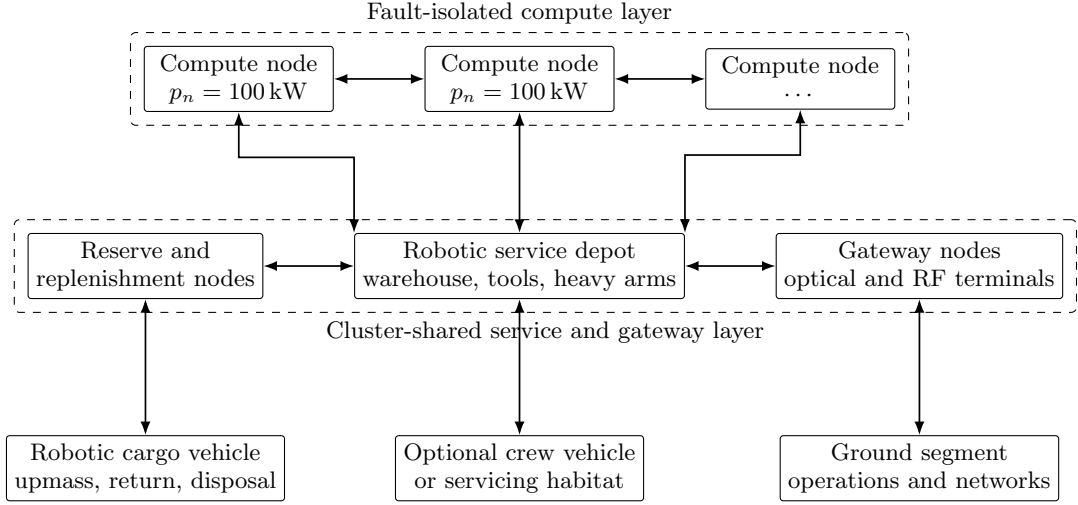
\begin{figure}[htbp]
\centering
\begin{tikzpicture}[
  node distance=7mm and 12mm,
  box/.style={draw,rounded corners=1.2pt,minimum height=8mm,align=center,font=\small,inner sep=3pt},
  layer/.style={draw,dashed,rounded corners=2pt,inner sep=5pt},
  arr/.style={-{Latex[length=2.0mm]},line width=0.65pt},
  bi/.style={{Latex[length=1.8mm]}-{Latex[length=1.8mm]},line width=0.65pt}
]
\node[box,minimum width=25mm] (c1) {Compute node\\$\pn=\SI{100}{kW}$};
\node[box,minimum width=25mm,right=of c1] (c2) {Compute node\\$\pn=\SI{100}{kW}$};
\node[box,minimum width=25mm,right=of c2] (c3) {Compute node\\$\cdots$};
\node[layer,fit=(c1)(c3),label={[font=\small]above:Fault-isolated compute layer}] (compute) {};

\node[box,minimum width=30mm,below=16mm of c2] (depot) {Robotic service depot\\warehouse, tools, heavy arms};
\node[box,minimum width=28mm,left=of depot] (reserve) {Reserve and\\replenishment nodes};
\node[box,minimum width=28mm,right=of depot] (gateway) {Gateway nodes\\optical and RF terminals};
\node[layer,fit=(reserve)(gateway),label={[font=\small]below:Cluster-shared service and gateway layer}] (shared) {};

\node[box,minimum width=30mm,below=18mm of reserve] (cargo) {Robotic cargo vehicle\\upmass, return, disposal};
\node[box,minimum width=30mm,below=18mm of depot] (crew) {Optional crew vehicle\\or servicing habitat};
\node[box,minimum width=30mm,below=18mm of gateway] (ground) {Ground segment\\operations and networks};

\draw[bi] (c1.south) -- ++(0,-6mm) -| (depot.north west);
\draw[bi] (c2.south) -- (depot.north);
\draw[bi] (c3.south) -- ++(0,-6mm) -| (depot.north east);
\draw[bi] (reserve) -- (depot);
\draw[bi] (depot) -- (gateway);
\draw[bi] (cargo) -- (reserve);
\draw[bi] (crew) -- (depot);
\draw[bi] (gateway) -- (ground);
\draw[bi] (c1) -- (c2);
\draw[bi] (c2) -- (c3);
\end{tikzpicture}
\caption{Reference maintainable orbital data center (ODC) architecture; RF denotes radio frequency.  Compute nodes are independent power, thermal, avionics, and fault-containment domains.  High-volume cartridge work is performed locally; rare spares, heavy manipulators, cargo transfer, and optional human access are pooled in a service layer.  The service depot improves maintainability but is not required for immediate survival of any node.}
\label{fig:architecture}
\end{figure}

Figure~\ref{fig:architecture} shows the operational decomposition.  Compute nodes remain independently survivable; a service and logistics layer pools rare spares and heavy robots; gateway nodes concentrate external communications; and an optional crew vehicle or servicing habitat attaches to the service layer rather than to each compute node.  This separation prevents the maintenance depot from becoming a single point of failure for power, thermal control, or flight safety.

\subsection{Operating states and service objective}

The principal service metric is delivered-capacity availability, not merely spacecraft survival.  For event class $e$ with occurrence rate $\Lambda_e$, affected IT power $p_e$, and mean unserved duration $\tau_e$, the first-order loss of delivered capacity is
\begin{equation}
1-A_{\rm cap}\simeq \frac{1}{\Pcl T_y}\sum_e \Lambda_e p_e\tau_e,
\qquad T_y=8760~\mathrm{h}.
\label{eq:Acap}
\end{equation}
This form counts a one-hour loss of \SI{5}{kW} and a one-hour loss of \SI{100}{kW} in proportion to the actual capacity removed.  It is therefore more informative for a distributed data center than binary spacecraft availability.

The cluster transitions among nominal, planned-maintenance, degraded, thermal-safe, power-safe, cyber-quarantine, collision-avoidance, survival, and retirement states.  A standard hierarchy is:
\begin{enumerate}
\item device-level correction, retry, scrubbing, or micro-reset;
\item cartridge-level workload migration, de-energization, and isolation;
\item node-level isolation and activation of hot or cold reserve capacity;
\item cluster-level load shedding, service prioritization, and gateway rerouting;
\item survival mode preserving avionics, communications, thermal circulation, and attitude control; and
\item retirement, sanitization, passivation, and transfer to a return or disposal vehicle.
\end{enumerate}
The ODC should tolerate loss of its high-rate space-to-ground path for days without losing thermal control, stored data, attitude, or the ability to receive emergency commands over a physically independent low-rate link.

\subsection{Task allocation}

The architecture allocates work according to repeatability, geometric uncertainty, response-time requirement, and consequence of failure.  Deterministic high-frequency tasks are assigned to automation; standardized mechanical work is assigned to robots; low-frequency unmodeled work with irregular geometry is assigned to external robotic or human intervention.  Table~\ref{tab:taskallocation} defines the baseline allocation.

\begin{table}[htbp]
\caption{Baseline allocation of operational and maintenance work.}
\label{tab:taskallocation}
\centering
\small
\renewcommand{\arraystretch}{1.00}
\begin{tabularx}{\textwidth}{@{}L{0.35\textwidth} L{0.29\textwidth} Y@{}}
\toprule
Task class & Primary execution & Required fallback \\
\midrule
Telemetry, trend analysis, workload migration, storage rebuild, network rerouting & Autonomous flight and cluster software & Ground engineering review and protected manual commands \\
Routine signed software and firmware rollout & Autonomous canary deployment and rollback & Ground-directed recovery using immutable safe image \\
Compute, storage, transceiver, and standardized converter replacement & Automated changer or local teleoperated robot & Visiting robotic servicer \\
Pump, valve, battery, power-distribution, and communications-terminal replacement & Teleoperated dexterous robot & Visiting heavy servicer or short-duration crew \\
External inspection and impact localization & Autonomous free-flyer and fixed cameras & Ground-directed free-flyer \\
Radiator or photovoltaic panel replacement & External arm or visiting servicer & Crew-supervised robotic operation \\
Stuck connector, deformed latch, unknown leak, damaged harness, or irregular structural repair & Multiple robots with specialized tools & Short-duration astronaut intervention \\
High-throughput inventory and cartridge flow at large scale & Automated warehouse & Resident technician supervision only for exceptions \\
First-of-a-kind upgrade integration & Ground-supervised robotics & Periodic or resident technicians \\
\bottomrule
\end{tabularx}
\end{table}

\subsection{Reference design point}

Table~\ref{tab:referencecase} collects the reference values used in subsequent numerical examples.  They are design allocations, not flight statistics.  The purpose of the paper is to identify the governing parameters and required validation tests; mission-specific values must be obtained from radiation testing, thermal-vacuum cycling, mechanism qualification, the U.S. National Aeronautics and Space Administration (NASA) Orbital Debris Engineering Model (ORDEM) and Meteoroid Engineering Model (MEM), and precursor flight data.

\begin{table}[htbp]
\caption{Reference \SI{1}{MW} maintainable orbital data center design point.  Mass entries are internally consistent planning allocations, not a closed spacecraft bill of materials.}
\label{tab:referencecase}
\centering
\small
\renewcommand{\arraystretch}{1.00}
\begin{tabularx}{\textwidth}{@{}L{0.22\textwidth} L{0.22\textwidth} Y@{}}
\toprule
Parameter & Reference value & Interpretation \\
\midrule
Orbit & high-sunlight \LEO, $\sim$550--700\,km & Low transport latency and short eclipse duration; orbit-specific debris and radiation analysis still required \\
Aggregate IT power $\Pcl$ & \SI{1}{MW} & Ten active \SI{100}{kW} nodes \\
Platform overhead $\alphaOH$ & 1.25 & \SI{1.25}{MW} total regulated load \\
Compute cartridge $\pc$ & \SI{5}{kW}, 20--35 kg & Two hundred active cartridges per MW \\
Reserve & one \SI{100}{kW} node; 5\% hot headroom & Immediate coverage of one node loss and local cartridge failures \\
Active hardware mass & 34--59 kg/kW & Physical and fixed-spacecraft range before servicing additions \cite{Turyshev2026ODC}; the low, nominal, and high cases below use 40, 49, and 59 t \\
Maintainability hardware & 3--7.5 kg/kW & Interfaces, segmentation, robots, tools, warehouse structure, and human-compatible worksite provisions; stocked ORUs excluded \\
First-line stocked ORUs & 2--2.5 kg/kW & Class-specific inventory allocation for the stated replenishment horizon; reserve node excluded \\
Reserve node & 4--6 t & One independent \SI{100}{kW} reserve compatible with the active-node mass cases \\
Total deployed planning cases & 49, 61.25, and 75 t & Low/nominal/high sums; reported as approximately 50--75 t/MW, not a closed flight mass \\
Resupply lead time & 180 days & Used for class-specific first-line spare planning, not a system-wide endurance guarantee \\
Design life & 10--15 yr & Bus and structural life; payload refreshed more rapidly \\
Compute refresh & 3--5 yr & Four-year reference \\
Target $A_{\rm cap}$ & 0.9995--0.9999 & Delivered-capacity objective with reserve, not single-node survival \\
\bottomrule
\end{tabularx}
\end{table}

\section{Physical principles and theoretical framework}
\label{sec:physics}

\subsection{Power, eclipse storage, and heat rejection}

The maintenance architecture cannot be separated from the physical size of the platform.  Let $\Pit$ denote delivered IT power at the payload interface and let
\begin{equation}
\Ptot=\alphaOH\Pit
\label{eq:Ptot}
\end{equation}
be the total regulated electrical load.  The overhead factor includes power conversion, thermal transport, avionics, attitude control, communications, robotics standby, and averaged propulsion loads.  In a sunlight fraction $\fSun$, photovoltaic output must support the instantaneous load and recharge energy used during eclipse.  With power management and distribution (PMAD) efficiency $\etaPMAD$, battery round-trip efficiency $\eta_b$, and EOL photovoltaic fraction $f_{\rm EOL,PV}$, the BOL photovoltaic power is
\begin{equation}
P_{\rm PV}^{\rm BOL}=\frac{\Ptot}{\etaPMAD f_{\rm EOL,PV}}
\left[1+\frac{1-\fSun}{\fSun\eta_b}\right].
\label{eq:PVpower}
\end{equation}
If $q_{\rm PV}=\eta_{\rm PV}\eta_{\rm geom}S_0$ is the effective areal electrical power density, then
\begin{equation}
\Apv^{\rm BOL}=\frac{P_{\rm PV}^{\rm BOL}}{q_{\rm PV}}.
\label{eq:PVarea}
\end{equation}
The corresponding battery mass for eclipse duration $t_{\rm ecl}$ is
\begin{equation}
M_b=\frac{\Ptot t_{\rm ecl}}
{e_b\rho_{\rm DoD}f_{\rm EOL,bat}},
\label{eq:batterymass}
\end{equation}
where $e_b$ is specific energy, $\rho_{\rm DoD}$ is allowable depth of discharge, and $f_{\rm EOL,bat}$ is retained usable capacity.  These relations reproduce the power-system structure developed for ODC feasibility studies \cite{Turyshev2026ODC,NASA2025SmallSpacecraft}.

A \SI{35}{min} eclipse at $\Ptot=\SI{1.25}{MW}$ requires \SI{0.729}{MWh} at the regulated bus.  An end-of-life usable system specific energy of \SIrange{75}{150}{Wh/kg} therefore implies approximately \SIrange{4.9}{9.7}{t} of storage per MW before redundancy.  In geostationary Earth orbit (GEO), a seasonal-eclipse maximum of \SI{69}{min} gives approximately \SI{1.44}{MWh} and \SIrange{9.6}{19.2}{t} per MW.  High-sunlight orbit selection is consequently not only a capital-mass choice; it removes a large population of life-limited, cycle-sensitive hardware from the maintenance flow.

Nearly all platform electrical power ultimately appears as low-grade heat.  With radiator emissivity $\epsRad$, deep-space view factor $\etaView$, operating temperature $\Trad$, and absorbed environmental flux $q_{\rm env}$, the required effective radiator area is
\begin{equation}
\Arad=\frac{\Ptot}
{\epsRad\etaView\sigSB\Trad^4-q_{\rm env}}.
\label{eq:radiatorarea}
\end{equation}
The logarithmic sensitivity to radiator temperature is approximately
\begin{equation}
\frac{\partial\ln\Arad}{\partial\ln\Trad}
=-\frac{4\epsRad\etaView\sigSB\Trad^4}
{\epsRad\etaView\sigSB\Trad^4-q_{\rm env}},
\label{eq:radtempsens}
\end{equation}
which is more negative than $-4$ when environmental absorption is non-negligible.  Raising radiator temperature therefore strongly reduces required radiator area and mass, but it reduces semiconductor timing margin and lifetime and increases leakage.  The allowable value is constrained by
\begin{equation}
T_j=\Trad+\Delta T_{\rm pkg}+\Delta T_{\rm interface}
+\Delta T_{\rm loop}+\Delta T_{\rm panel},
\label{eq:thermalbudget}
\end{equation}
not by radiator material alone \cite{Gilmore2002,Juhasz2002}.

Maintainability adds thermal penalties.  Dry-break couplings, valves, accumulators, redundant pumps, local isolation, and replaceable panels increase the effective radiator areal density and pumping load.  If the baseline mass-per-power is decomposed as
\begin{equation}
\mkw=m_{\mathrm{kW,PV}}+m_{\mathrm{kW,bat}}+m_{\mathrm{kW,rad}}
+m_{\mathrm{kW,fixed}}+\mkwm,
\label{eq:massdecomp}
\end{equation}
then $\mkwm$ is the added mass required for fault isolation and ORU-level replacement rather than whole-node replacement.

\subsection{Transient thermal protection during maintenance}

High-power cartridges have little time to survive a loss of coolant.  Let $C_{\rm th}=m_c c_{p,\rm eff}$ be the effective thermal capacitance of a cartridge after local spreading mass is included.  Immediately after loss of heat removal,
\begin{equation}
\frac{dT}{dt}\simeq \frac{\pc}{C_{\rm th}}.
\label{eq:heatup}
\end{equation}
For $\pc=\SI{5}{kW}$, $m_c=\SI{25}{kg}$, and $c_{p,\rm eff}=\SI{500}{J/(kg.K)}$, the initial temperature rise is approximately \SI{0.4}{K/s}.  A \SI{25}{K} margin is exhausted in about one minute.  Hardware protection must therefore isolate power on sub-second to few-second timescales; workload migration is useful for preserving service but is too slow to be the primary chip-protection mechanism.  A cartridge must be de-energized before any thermal interface is opened, and a node must contain enough bypass flow or thermal reserve to avoid destabilizing neighboring modules during a swap.

Electrical interfaces have a similar stored-energy constraint.  If a local direct-current (DC)-link capacitance $C_{\rm dc}$ is charged to voltage $V_{\rm bus}$, the stored energy is
\begin{equation}
E_{\rm dc}=\frac{1}{2}C_{\rm dc}V_{\rm bus}^2.
\label{eq:storedenergy}
\end{equation}
A representative $C_{\rm dc}=\SI{10}{mF}$ and $V_{\rm bus}=\SI{400}{V}$ contain \SI{800}{J}, sufficient to damage a blind-mate connector or robot end effector.  The replacement sequence must include positive isolation, controlled discharge, voltage verification, and a mechanical interlock that prevents connector motion until residual energy is below a specified safe threshold.

\subsection{Space-environment failure physics}

A terrestrial failure database does not directly transfer to orbit because the stress spectrum differs and because commercial processors are not generally designed for uncorrected radiation, vacuum-compatible materials, launch vibration, or repeated robotic connection cycles.  The principal environmental contributions are summarized below.

\paragraph{Radiation.}
The rate of a single-event effect (SEE) in device class $i$ is
\begin{equation}
R_{{\rm SEE},i}=N_i\int \Phi(E,\Omega)\,
\sigma_{{\rm SEE},i}(E,\Omega)\,dE\,d\Omega,
\label{eq:SEE}
\end{equation}
where $\Phi$ is the shielded particle spectrum and $\sigma_{\rm SEE}$ is the device or system cross section after architectural masking.  Correctable SEEs, link transients, and functional interrupts primarily reduce utilization; destructive latchup, cumulative total ionizing dose (TID), displacement damage, and latent degradation contribute to permanent hazard.  Mission specifications require orbit- and shielding-specific AE9/AP9-IRENE trapped-particle environment models, solar-particle models, and radiation-transport calculations rather than a universal dose number \cite{Ginet2013AE9AP9,Guild2021SpaceEnvironment}.  Recent irradiation studies of commercial edge-computing modules demonstrate useful tolerance but also reinforce the need for device-specific testing, watchdogs, current limiting, memory protection, and system-level fault containment \cite{Felix2024Jetson,Rodriguez2025Jetson}.

\paragraph{Micrometeoroids and orbital debris.}
Micrometeoroids and orbital debris (MMOD) produce an area-dependent impact hazard.  For exposed area $A$ and differential impact flux $F(d,v,\Omega)$, the expected number of mission-relevant impacts is
\begin{equation}
\mu_{\rm imp}=T\int_A\int F(d,v,\Omega)
P_{\rm fail}(d,v,\theta,\mathcal{S})\,dd\,dv\,d\Omega\,dA,
\label{eq:mmod}
\end{equation}
where $\mathcal{S}$ denotes shielding and component geometry.  The zero-event probability is $\exp(-\mu_{\rm imp})$.  ORDEM~3.2 and the corresponding meteoroid model should be used for design rather than a single area-normalized puncture rate \cite{Vavrin2023ORDEM32}.  Because the exposed radiator area scales with power, the reference design requires segmentation, local isolation, and replaceable panels.  A large unsegmented coolant loop converts a local puncture into a domain-level loss.

\paragraph{Atomic oxygen and contamination.}
In LEO, the erosion depth of an exposed polymer may be represented by
\begin{equation}
\Delta \ell_{\rm AO}=\frac{Y_{\rm AO}\mathcal{F}_{\rm AO}}{\rho_m},
\label{eq:AO}
\end{equation}
where $Y_{\rm AO}$ is erosion yield, $\mathcal{F}_{\rm AO}$ is integrated atomic-oxygen fluence, and $\rho_m$ is material density.  Coating defects, ram orientation, ultraviolet exposure, and outgassed contamination strongly affect durability.  Protective coatings and flight-exposure data are therefore required for blankets, cable insulation, flexible-array substrates, optical surfaces, and robot covers \cite{Banks2003AO,deGroh2025MISSE}.

\paragraph{Thermal and mechanical cycling.}
Repeated temperature excursions drive solder, interconnect, seal, and composite fatigue.  For a strain-controlled joint, a Coffin--Manson-type relation may be written
\begin{equation}
\frac{\Delta\epsilon_p}{2}=\epsilon_f'(2N_f)^c,
\label{eq:coffinmanson}
\end{equation}
with mission-specific coefficients.  Damage accumulated across nonuniform cycles can be approximated by Miner's rule,
\begin{equation}
D=\sum_j\frac{n_j}{N_{f,j}},
\qquad D\gtrsim1\ \text{at nominal fatigue failure},
\label{eq:miner}
\end{equation}
while recognizing that sequence effects and radiation-assisted degradation can invalidate a purely linear model \cite{Coffin1954,Manson1953,Miner1945}.  Launch vibration and shock can create latent connector, solder, and deployment-mechanism damage that appears during commissioning rather than at launch.

\paragraph{Mechanisms and connector cycles.}
A connector or latch qualified for $n$ robotic cycles should be treated as a life-limited item.  A Weibull cycle model gives
\begin{equation}
R_c(n)=\exp\Big[-\Big(\frac{n}{\eta_c}\Big)^{\beta_c}\Big],
\label{eq:connectorweibull}
\end{equation}
where $\eta_c$ and $\beta_c$ are obtained from thermal-vacuum, contamination, misalignment, and representative-load testing.  Mechanism history shows that lubrication, preload, material pairing, debris generation, and sensor redundancy are major reliability drivers \cite{Fusaro1999Mechanisms}.  The interface-cycle qualification life should exceed expected operational cycles by a margin that includes repeated failed insertion attempts and ground acceptance testing.

\subsection{Large flexible structures and servicing dynamics}

A MW-class node has deployed dimensions of tens of meters.  For a beam-like appendage with flexural rigidity $EI$, linear mass density $\mu_s$, and length $L$, the first bending frequency scales as
\begin{equation}
f_1\sim\frac{\beta_1^2}{2\pi L^2}\sqrt{\frac{EI}{\mu_s}}.
\label{eq:flexmode}
\end{equation}
As $L$ grows, the $L^{-2}$ dependence rapidly drives structural modes into the bandwidth of attitude control and robotic manipulation.  High-force maintenance should therefore be performed through braced fixtures or rail-mounted manipulators that close reaction loads locally.  Free-flying robots are well suited to inspection but poorly suited to applying connector insertion forces to an unrestrained flexible node.  Input shaping, joint-torque limits, mode-aware attitude control, and maintenance keep-out zones are required design constraints.

\subsection{Repairable-system and lifecycle model}

For component class $i$ with installed count $N_i$ and age-dependent permanent-failure hazard $h_i(t)$, the expected random replacement rate is
\begin{equation}
\Lambda_{i,\rm rnd}(t)=N_i h_i(t).
\label{eq:failurerate}
\end{equation}
A useful nonhomogeneous form separates commissioning, random, and wearout behavior:
\begin{equation}
h_i(t)=h_{0,i}e^{-t/\tau_{0,i}}+h_{r,i}
+h_{w,i}\left(\frac{t}{T_{w,i}}\right)^{k_i}.
\label{eq:bathtub}
\end{equation}
The first term represents infant mortality, the second a quasi-random plateau, and the third wearout.  A constant mean time between failures (MTBF) approximation is appropriate only where the middle term dominates.

Random replacement mass flow is
\begin{equation}
\Mdot_{\rm rnd}=\sum_i N_i h_i m_i,
\label{eq:Mdotrandom}
\end{equation}
while planned refresh contributes
\begin{equation}
\Mdot_{\rm ref}=\sum_i\frac{M_i}{T_{r,i}}.
\label{eq:Mdotrefresh}
\end{equation}
Catastrophic node replacement contributes
\begin{equation}
\Mdot_{\rm cat}=N_n\lambdaCat M_n,
\label{eq:Mdotcat}
\end{equation}
where $M_n$ is the delivered mass of one node.  Including consumables and packaging gives
\begin{equation}
\Mdot_{\rm up}=\Mdot_{\rm rnd}+\Mdot_{\rm ref}+\Mdot_{\rm cat}
+\Mdot_{\rm cons}+\Mdot_{\rm pack},
\label{eq:Mdotup}
\end{equation}
where a proportional packaging model, when appropriate, is $\Mdot_{\rm pack}=\eta_{\rm pack}(\Mdot_{\rm rnd}+\Mdot_{\rm ref}+\Mdot_{\rm cat}+\Mdot_{\rm cons})$.  Table~\ref{tab:massflow} instead carries packaging as an explicit additive allocation, so the two representations are alternatives and are not applied simultaneously.  Equations~\eqref{eq:Mdotrandom}--\eqref{eq:Mdotup} are the core mass-flow relations used throughout the paper.  The additive reference uses a calendar-cohort generation policy for technology-bearing compute, storage, and networking hardware: a corrective replacement normally receives a compatible same-generation spare and does not erase the predeclared cohort refresh date.  For age-limited ORUs whose corrective replacement resets component age, random and planned renewal compete; in that case \Eq{eq:Mdotrefresh} is a conservative standalone allocation rather than an independently additive realized rate, and the competing renewal rates in Appendix~\ref{app:failuremodel} must be used.

For a repairable item with exponential failure rate $\lambda_i$, let $\MTTR_i$ denote the mean time to repair or replace the item and define the effective restoration rate $\mu_i=1/\MTTR_i$.  The intrinsic steady-state availability is
\begin{equation}
a_i=\frac{\mu_i}{\lambda_i+\mu_i}
=\frac{\MTBF_i}{\MTBF_i+\MTTR_i}.
\label{eq:unitavailability}
\end{equation}
This expression is inadequate when spares are stocked out or service robots queue; in those cases $\MTTR_i$ includes fault detection, workload drain, robot wait, physical work, post-installation test, and any delay awaiting resupply.  The purpose of reserve capacity is to make user-visible interruption depend on fast isolation and switchover rather than on the full physical repair time.

\subsection{Correlated faults and common-mode exposure}

Independent-component reliability is not enough.  Software updates, key compromise, common power buses, shared thermal headers, formation-control errors, collision events, and inventory defects can remove multiple nodes simultaneously.  For permanent or effectively mission-ending domain losses, a capacity-weighted delivered-life hazard may be written
\begin{equation}
\lambda_{\rm eff}^{\rm perm}=\lambda_{\rm ind}
+\sum_d\phi_d\lambda_{d}^{\rm perm},
\label{eq:correlatedhazard}
\end{equation}
where $\phi_d$ is the fraction of cluster capacity permanently lost in domain event $d$.  Temporary domain events require duration explicitly; their first-order availability contribution is
\begin{equation}
1-A_{\rm dom}\simeq\frac{1}{T_y}\sum_d\lambda_d\,\mathbb E[F_d T_d],
\label{eq:domainavailabilitymain}
\end{equation}
with event fraction $F_d$ and duration $T_d$.  Mission-level loss probability further depends on the distribution and dependence of $F_dT_d$, not only its mean.  The architecture should target small domain fractions by separating power, coolant, safety avionics, software rollout cohorts, cryptographic keys, inventory lots, and orbital neighborhoods.  A single event-rate target such as $\lambdaDom\lesssim10^{-3}~\mathrm{yr^{-1}}$ is therefore incomplete unless the event definition, affected fraction, duration, and recovery state are stated.

\subsection{Technology obsolescence as a competing hazard}

For compute hardware, economic obsolescence is often faster than physical wearout.  If frontier performance per watt improves at effective rate $g_{\rm obs}$ and the installed payload suffers permanent hazard $\lambda$, frontier-relative delivered compute over horizon $T$ is
\begin{equation}
E_{\rm rel}=U_{\rm eff}\Pit
\int_0^T e^{-(\lambda+g_{\rm obs})t}\,dt
=U_{\rm eff}\Pit\frac{1-e^{-(\lambda+g_{\rm obs})T}}
{\lambda+g_{\rm obs}}.
\label{eq:obsolescence}
\end{equation}
The corresponding penalty is
\begin{equation}
\Pi_{\rm life+obs}=\frac{(\lambda+g_{\rm obs})T}
{1-e^{-(\lambda+g_{\rm obs})T}}.
\label{eq:obspenalty}
\end{equation}
This result motivates rolling cartridge refresh.  Replacing an entire \SI{5}{t}, \SI{100}{kW} node every four years produces \SI{12.5}{t/(MW.year)} of planned flow, whereas refreshing only a \SI{5}{t/MW} cartridge population on the same cadence produces \SI{1.25}{t/(MW.year)}.  Under these mass and cadence allocations, whole-node refresh raises planned flow by a factor of ten and exceeds the reference logistics range unless node mass per kilowatt is substantially reduced.

\section{Health monitoring, diagnostics, and operational observables}
\label{sec:measurement}

An ODC is maintained through measurements before it is maintained through tools.  The sensing architecture must support three distinct functions: immediate protection, diagnosis and isolation, and long-horizon remaining-useful-life estimation.  Immediate protection requires deterministic local logic with bounded latency.  Diagnosis may combine model-based residuals, cross-node comparisons, and ground review.  Prognostics may use statistical or machine-learning methods, but should not be permitted to inhibit hard safety limits without an independently validated path.

\subsection{State estimation and fault residuals}

Let the hidden health state of a node be $\bm{x}_k$, commands and workload variables be $\bm{u}_k$, and telemetry be $\bm{y}_k$.  A generic stochastic model is
\begin{align}
\bm{x}_{k+1}&=\bm{f}(\bm{x}_k,\bm{u}_k)+\bm{w}_k,\\
\bm{y}_k&=\bm{h}(\bm{x}_k,\bm{u}_k)+\bm{v}_k,
\label{eq:healthstate}
\end{align}
where $\bm{w}_k$ represents process uncertainty and unmodeled degradation and $\bm{v}_k$ represents sensor noise, quantization, timing error, and communication loss.  A residual
\begin{equation}
\bm{r}_k=\bm{y}_k-\bm{h}(\hat{\bm{x}}_k,\bm{u}_k)
\label{eq:residual}
\end{equation}
is evaluated against temporal, cross-sensor, and physical-consistency tests.  For a linearized system, local diagnosability requires that failure modes produce distinguishable signatures in the residual space; a simple observability test uses the rank of
\begin{equation}
\mathcal{O}=\begin{bmatrix}C&CA&CA^2&\cdots&CA^{n-1}\end{bmatrix}^{\!T}.
\label{eq:observability}
\end{equation}
Critical functions should not rely on a single sensor that is itself indistinguishable from the failure it monitors.  At least two physically diverse observables are desirable for coolant flow, bus isolation, latch state, stored-energy discharge, pressure-boundary integrity, and robot capture.

Spacecraft fault-protection practice already uses autonomous monitors, responses, safe modes, and model-based diagnosis \cite{Morgan2005FaultProtection,Rasmussen2008FaultProtection,Aaseng2023VSM}.  The ODC extension is one of scale: thousands of devices and hundreds of repetitive replacements create a statistical monitoring problem closer to industrial process control, while flight safety still requires spacecraft-grade determinism.

\subsection{Required telemetry}

The minimum measurement set should include:
\begin{itemize}
\item corrected and uncorrected memory errors, processor machine-check events, watchdog resets, current-limiting events, and checkpoint rollback counts;
\item optical power, bit-error rate, retry count, clock drift, switch queue occupancy, and link margin;
\item bus voltage, phase or current imbalance, converter efficiency, harmonic content, insulation resistance, and arc signatures;
\item battery temperature, impedance, cell divergence, coulombic efficiency, and estimated usable capacity;
\item coolant pressure, flow, temperature, quality or void fraction, pump current, valve state, accumulator position, and leak-detection residuals;
\item radiator infrared maps, panel strain, impact-acoustic or piezoelectric events, and fluid inventory;
\item photovoltaic current--voltage curves, string mismatch, deployment state, surface temperature, and insulation leakage;
\item attitude sensors, structural accelerometers, mechanism motor current, joint torque, backlash estimates, and flexible-mode response;
\item robot camera confidence, force--torque residuals, joint temperature, calibration drift, tool inventory, and insertion-force history; and
\item radiation dosimetry, charging, local contamination indicators, and space-weather context.
\end{itemize}

A representative \SI{100}{kW} node may contain $N_{s,n}=10^4$ telemetry channels when device counters and distributed thermal sensors are included.  The ten-node reference MW therefore contains $N_{s,\rm MW}=10^5$ channels.  At 1--10 samples s$^{-1}$ and 4--8 bytes per sample, uncompressed health data are approximately 35--690 GB per MW-day, or 35--690 TB/day at \SI{1}{GW}.  This volume is modest relative to internal network traffic but large enough to require local retention over a bounded window, with event-triggered excerpts, health-state summaries, and statistically sufficient features sent to Earth.

\subsection{False alarms, missed detections, and alert aggregation}

Scale makes naive thresholding untenable.  For the reference MW, $N_{s,\rm MW}=10^5$ channels tested once per second produce $8.64\times10^9$ tests per day.  To limit purely statistical false alerts to 0.1 per day, the mean per-test false-alarm probability must satisfy
\begin{equation}
p_{\rm FA}<\frac{0.1}{N_{s,\rm MW}(86400)}\simeq1.16\times10^{-11}.
\label{eq:falsealarm}
\end{equation}
At 10 Hz the bound is $1.16\times10^{-12}$.  For the symmetric Gaussian test in \Eq{eq:pfa}, these limits correspond to approximately 6.79 and 7.11 standard deviations, respectively; the one-node, 1-Hz value would be 6.44 standard deviations.  Such thresholds can make slowly developing faults difficult to detect.  The practical solution is hierarchical evidence accumulation: local hard limits for immediate safety, persistence and rate-of-change logic for component protection, multivariate residuals for diagnosis, and fleet-level statistical comparisons for prognostics.

For a scalar residual with healthy distribution $\mathcal{N}(0,\sigma_r^2)$ and fault mean shift $\mu_f$, a symmetric threshold $\gamma$ has
\begin{align}
p_{\rm FA}&=2Q\!\left(\frac{\gamma}{\sigma_r}\right),\label{eq:pfa}\\
p_{\rm D}&=Q\!\left(\frac{\gamma-\mu_f}{\sigma_r}\right)
+Q\!\left(\frac{\gamma+\mu_f}{\sigma_r}\right),
\label{eq:detectiontrade}
\end{align}
where $Q$ is the Gaussian tail function.  Sensor placement and model quality should be traded against this detection curve, not selected solely by channel count.

\subsection{Prognostics and replacement scheduling}

For slowly degrading items, maintenance should be scheduled from the distribution of remaining useful life rather than from a single point estimate.  If $T_f$ is the predicted failure time, the decision to replace at next service opportunity $t_s$ should depend on
\begin{equation}
P(T_f<t_s+L_{\rm contingency}\mid\mathcal{Y}_k),
\label{eq:RULdecision}
\end{equation}
where $\mathcal{Y}_k$ is the telemetry history and $L_{\rm contingency}$ is the time required to recover if the prediction is wrong.  Batteries, pumps, reaction wheels, optical terminals, converters, and connectors should each have explicit health indicators and confidence intervals.  For high-value but low-frequency failures, Bayesian updating across the entire fleet is particularly valuable: a single removed unit can change the posterior hazard assigned to every similar unit.

\subsection{Software maintenance, cybersecurity, and command authority}

The tenant workload network, maintenance network, and spacecraft safety-control plane should be separated by both architecture and authority.  Tenant software must have no path to propulsion, high-voltage switching, thermal isolation valves, robot motion, cryptographic root keys, or flight termination and disposal functions.  The safety plane should retain an immutable recovery image and a low-rate command path independent of the high-throughput gateway network.

A software or firmware rollout proceeds through ground qualification, cryptographic signing, deployment to a noncritical canary, a defined observation period, expansion to one fault domain at a time, and automatic rollback on power, thermal, error, or performance excursions.  Let a latent update defect have probability $p_u$ and let fraction $f_u$ of cluster capacity be exposed before detection.  The expected update-induced capacity exposure is proportional to
\begin{equation}
E[\phi_u]\simeq p_u f_u.
\label{eq:updateexposure}
\end{equation}
Canarying does not necessarily reduce $p_u$, but it bounds $f_u$ and therefore the correlated-loss term in \Eq{eq:correlatedhazard}.  Keys and update-signing authorities should be geographically and organizationally separated, with two-person or two-authority control for deorbit, broad power isolation, root-key replacement, and cluster-wide firmware changes.  Space-specific cybersecurity guidance is available from the U.S. National Institute of Standards and Technology and the Consultative Committee for Space Data Systems, but mission architecture must also account for the safety impacts of security controls themselves \cite{Scholl2023NIST8270,NIST2023HSN,CCSDS2022SDLS}.

\section{Space-maintainable compute and platform architecture}
\label{sec:maintainable}

\subsection{Limitations of terrestrial rack architectures}

The logical abstractions of a terrestrial data center---server, rack, row, fabric, failure domain, and rolling upgrade---remain useful.  The physical rack does not.  Terrestrial racks assume gravity, human hands, convective air cooling, frequent manual cable routing, benign radiation, and a building that supplies pressure containment, fire suppression, and access.  An orbital implementation should instead use enclosed but unpressurized service bays, conduction or liquid cooling, rigid backplanes, protected optical paths, blind-mate connectors, robot-readable geometry, and extraction envelopes that remain clear over the full mission.

Pressurizing the compute volume to reproduce terrestrial maintenance adds pressure vessels, micrometeoroid shielding, atmosphere management, fire and smoke control, airlocks, and crew exposure to high-voltage and coolant systems.  Convection inside a pressure vessel redistributes heat but does not eliminate the external radiator.  If humans are introduced, a separate habitat or crew vehicle connected to an unpressurized service worksite limits the human-rated system boundary.

\subsection{Nested orbital replacement unit hierarchy}

A maintainable ODC should use at least four replacement levels:
\begin{enumerate}
\item \textit{L1 cartridges:} compute, memory, storage, optical transceivers, local controllers, and sensors in the 1--35 kg class;
\item \textit{L2 subsystem ORUs:} converters, power-distribution units, batteries, pumps, valve manifolds, switches, and avionics in the 50--150 kg class;
\item \textit{L3 external ORUs:} radiator panels, photovoltaic power units, communications terminals, robot arms, and propulsion modules in the 100--500 kg class; and
\item \textit{L4 nodes:} independently deployable 50--250 kW spacecraft in the multi-tonne class, replaced only after catastrophic or economically irreparable failure.
\end{enumerate}
The hierarchy should be designed so that semiconductor and storage failures stop at L1, most power and thermal failures stop at L2, and impacts or mechanism damage stop at L3.  L4 replacement is a reserve and logistics function, not the routine refresh mechanism.

\subsection{Interface requirements}

A standard ORU interface should provide:
\begin{itemize}
\item capture-first, connector-second mechanical sequencing;
\item passive coarse alignment, fine kinematic registration, and machine-readable fiducials;
\item keyed, de-energized high-voltage blind-mate contacts with a verified discharge path;
\item redundant low-voltage control, identification, and health-monitor contacts;
\item protected optical interfaces with self-test and contamination covers;
\item double-shutoff, dripless thermal or fluid couplings where a dry conductive interface is insufficient;
\item a positive latch with both mechanical and telemetered state indication;
\item a secondary extraction or release feature usable after primary-latch failure;
\item standard grapple points and load paths that bypass sensitive electronics; and
\item serialized configuration data, cycle count, radiation history, and insertion-force history stored in nonvolatile memory.
\end{itemize}

The interface should be tested as a coupled system.  A connector that is individually qualified can still fail when harness stiffness, thermal distortion, robot calibration, contamination, and structural motion are combined.  Cooperative servicing aids and standard utility interfaces have long been recognized as enabling technologies for on-orbit servicing \cite{IDSS2022,ISO24330,NASA2026RRM}.  ODC hardware should extend this philosophy to high-cycle electrical, optical, and thermal connections.

\subsection{Thermal segmentation and panel optimization}

The reference node contains at least four independently isolatable thermal loops, each with multiple radiator panels.  Let a radiator of total area $A$ be divided into $n_p$ panels.  If a puncture or panel-level failure rate per area is $\phi_p$ and one failed panel removes fraction $1/n_p$ of heat-rejection capability until repaired, the capacity-loss contribution scales approximately as $\phi_p A\tau_p/n_p$.  If each panel adds valve, sensing, header, and structural mass $m_v$, a simplified segmentation objective is
\begin{equation}
J_p(n_p)=n_p m_v+C_A\frac{\phi_p A\tau_p}{n_p},
\label{eq:panelobjective}
\end{equation}
where $C_A$ converts unavailable capacity into an equivalent mass or cost.  The continuous optimum is
\begin{equation}
n_p^*=\Big(\frac{C_A\phi_p A\tau_p}{m_v}\Big)^{1/2}.
\label{eq:paneloptimum}
\end{equation}
The square-root scaling captures the governing trade: larger area, longer repair delay, and higher capacity value favor finer segmentation; heavy valves and headers favor fewer panels.  Detailed design must replace $C_A$ with actual reserve, workload, and thermal-transient constraints, but the result explains why the optimum is neither one monolithic panel nor thousands of tiny fluid sectors.

A reference \SI{100}{kW} node inherits approximately \SI{250}{m^2} of radiator area from the \SI{1}{MW} anchor.  Ten panels of \SIrange{20}{30}{m^2} provide a practical first division.  Losing one panel then removes roughly 10\% of node heat rejection, which can be covered by node-level throttling and cluster reserve while the panel is replaced.

\subsection{Node and cartridge granularity}

Small nodes reduce the fraction of cluster capacity lost in one event but duplicate avionics, propulsion, communication, and structural mass.  Large nodes reduce duplication but require larger reserve blocks and create more difficult servicing targets.  A compact design objective for node power is
\begin{equation}
J_n(\pn)=\frac{C_{\rm loc}M_{\rm loc}}{\pn}
+C_{\rm res}\frac{\pn}{\Pcl}
+C_{\rm dom}\phi(\pn),
\label{eq:nodeobjective}
\end{equation}
where $M_{\rm loc}$ is fixed local mass per node, the second term represents the cost of carrying reserve capacity at least as large as one node, and $\phi(\pn)$ captures any increase in correlated domain or handling complexity with node size.  Ignoring the last term gives the square-root scaling
\begin{equation}
\pn^*\propto\sqrt{M_{\rm loc}\Pcl}.
\label{eq:nodeoptimum}
\end{equation}
Thus the stationary node size under the simplified objective may increase with total cluster power while remaining much smaller than the cluster.  The \SI{100}{kW} reference is not universal; it is a reasonable mid-band value for a first MW because it limits one-node loss to 10\% while avoiding the fixed-mass duplication of hundreds of kilowatt-class spacecraft.

Cartridge power is governed by a different trade.  Smaller cartridges reduce the power removed by a single failure and simplify launch packaging, but increase connector count, robot operations, and interface mass.  Five to ten kilowatts per cartridge is the reference range because it corresponds to 100--200 active cartridges per MW, making weekly rolling refresh manageable at MW scale while keeping individual masses within the capabilities of compact automated changers.

\subsection{Maintainability mass penalty}

The preliminary maintainability-hardware allocation, excluding stocked ORUs and the reserve node, is
\begin{equation}
\mkwm\simeq \SIrange{3}{7.5}{kg/kW}.
\label{eq:maintmass}
\end{equation}
Table~\ref{tab:maintmass} separates this hardware from the \SIrange{2}{2.5}{kg/kW} first-line stock allocation.  The combined service-architecture increment is therefore approximately 5--10 kg/kW before reserve capacity.  These are architecture allocations, not measured mass properties.

\begin{table}[htbp]
\caption{Illustrative maintainability mass allocation.  L1 denotes the cartridge-level replacement tier.  Stocked ORUs and the reserve node are shown separately so that inventory is not hidden inside the hardware allowance.  Ranges are architecture-level design allocations, not measured flight values.}
\label{tab:maintmass}
\centering
\small
\renewcommand{\arraystretch}{1.02}
\begin{tabularx}{\textwidth}{@{}L{0.45\textwidth} C{0.07\textwidth} Y@{}}
\toprule
Provision & kg/kW & Main function \\
\midrule
Cartridge frames, backplanes, latches, and extraction clearances & 1--2 & Converts electronics into high-cycle cartridge-level replacement units \\
Thermal segmentation, valves, dry breaks, accumulators, and bypasses & 1--3 & Contains leaks and permits panel or pump replacement \\
Robots, tools, metrology, calibration targets, \& recovery fixtures & 0.5--1.5 & Provides local and external physical maintenance \\
Warehouse structure, access volume, cargo restraint, and quarantine fixtures & 0.3--0.5 & Houses and handles stocked and removed ORUs; stock mass itself is excluded \\
Human-compatible grapple, isolation, \& worksite provisions & 0.2--0.5 & Preserves future contingency servicing without a habitat \\
\midrule
Maintainability-hardware subtotal & 3--7.5 & Excludes stocked ORUs and reserve node \\
First-line stocked ORUs (separate) & 2--2.5 & Class-specific planning inventory for the stated replenishment horizon \\
Service-architecture subtotal & 5--10 & Excludes the independent reserve node \\
\bottomrule
\end{tabularx}
\end{table}

To avoid combining unrelated interval endpoints, the rounded 50--75 t/MW planning envelope uses three explicit sums.  The low, nominal, and high cases are, respectively,
\begin{equation}
\begin{split}
M_{\Sigma,\rm low}&=40+3+2+4=49~\mathrm{t},\\
M_{\Sigma,\rm nom}&=49+5+2.25+5=61.25~\mathrm{t},\\
M_{\Sigma,\rm high}&=59+7.5+2.5+6=75~\mathrm{t},
\end{split}
\label{eq:masscases}
\end{equation}
where each sum is active hardware + maintainability hardware + stocked ORUs + one reserve node.  These internally consistent cases explain the reported rounded range; they do not demonstrate integrated structural, thermal-hydraulic, propulsion, servicing-depot, docking, shielding, or qualification-margin closure.

\section{Robotic servicing architecture}
\label{sec:robotics}

Space robotics has demonstrated several constituent capabilities, but not the service rate and repeated-operation reliability required by an ODC.  Dextre performs external maintenance on the International Space Station (ISS), including battery and camera replacement, under ground teleoperation \cite{NASA2023Dextre}.  NASA's Robotic Refueling Mission used Dextre and specialized tools to cut blankets, manipulate caps and valves, inspect interfaces, and execute fluid-transfer precursor operations \cite{NASA2026RRM}.  Hubble servicing demonstrated the value of a spacecraft deliberately designed around replaceable instruments and astronaut-compatible interfaces \cite{NASA2026HubbleSM4}.  Commercial Mission Extension Vehicles have repeatedly docked to geostationary client satellites for life extension, and ADRAS-J demonstrated close inspection of a noncooperative object under modern safety constraints \cite{Northrop2021MEV2,Astroscale2026ADRASJ}.  These missions demonstrate teleoperation, rendezvous, docking, and modular servicing at limited operation counts.  They do not demonstrate autonomous high-cycle cartridge exchange, repair of a disabled maintenance robot, or hundreds of operations per year at the required availability.

\subsection{Robot classes}

A scalable ODC uses specialized machines rather than a single universal humanoid.

\paragraph{Automated cartridge machinery.}
The highest-volume tasks---compute, storage, and transceiver exchange---should be performed by a constrained mechanism analogous to an automated library or tool changer.  It should remove a cartridge, place it in a quarantine cassette, retrieve a replacement, mate it under force and alignment control, verify electrical isolation and optical cleanliness, perform pressure-decay or thermal-contact tests, and return the bay to service.  A specialized machine has fewer uncontrolled degrees of freedom, higher repeatability, and simpler verification than a general-purpose arm.

\paragraph{Fixed dexterous manipulators.}
Power, thermal, battery, switch, and avionics ORUs require six- or seven-degree-of-freedom arms with force--torque sensing, machine vision, exchangeable tools, and access to overlapping work envelopes.  At least two independent manipulators should be able to reach every critical L2 ORU or one should be able to reposition the other through a recovery fixture.

\paragraph{Rail or crawling transport robots.}
Mobile robots should move parts between the orbital warehouse and service cells, inspect behind equipment, carry tools, and recover disabled free-flyers.  Rails or hard translation paths are preferred for high-force tasks because they provide a reaction path and deterministic navigation.

\paragraph{External arms and visiting servicers.}
Large radiator panels, photovoltaic units, terminals, propulsion modules, and complete local robots require an external arm or visiting servicing vehicle.  The visiting vehicle also provides an independent capability to grapple a dead node, install a structural patch, transfer pallets, and move failed hardware to a return or disposal vehicle.

\paragraph{Free-flying inspection vehicles.}
Small free-flyers provide optical, infrared, and geometric inspection, leak localization, impact assessment, and docking surveillance.  They should not be the sole means of high-force repair.  Their propellant plume, collision risk, charging, and navigation uncertainties require keep-out zones and passive capture provisions.

\subsection{Robotic task flow}

A standard cartridge replacement sequence is:
\begin{enumerate}
\item confirm spare identity, configuration, radiation qualification, and software compatibility;
\item drain workload and verify state replication or checkpoint completion;
\item isolate electrical power and coolant, discharge stored energy, and confirm safe state with independent sensors;
\item capture the cartridge before unlatching utility interfaces;
\item extract under force and displacement limits while monitoring harness, connector, and structure residuals;
\item inspect the bay and removed unit, including optical faces and thermal seals;
\item insert and latch the replacement using coarse and fine registration;
\item perform electrical isolation, continuity, optical margin, leak or contact-pressure, and built-in self-tests;
\item install a trusted software image, admit the cartridge to a quarantine network, and execute a burn-in workload; and
\item return capacity gradually while updating inventory and component-history records.
\end{enumerate}
A failed step should leave the system in a mechanically captured and electrically safe condition.  The sequence must not depend on uninterrupted ground contact.

\subsection{Workload, queueing, and service capacity}

Let standardized maintenance tasks arrive at aggregate rate $\Lambda$ and require mean productive service time $\bar t_s$.  With $c$ equivalent robots, each providing $H_y$ productive hours per year, the utilization is
\begin{equation}
\rho_R=\frac{\Lambda\bar t_s}{cH_y}.
\label{eq:robotutil}
\end{equation}
A design that operates continuously near $\rho_R=1$ has unbounded sensitivity to burst arrivals, robot outages, and failed first attempts.  A planning limit of $\rho_R\lesssim0.6$--0.7 is appropriate before explicit queueing simulation.

For an $M/M/c$ approximation with arrival rate $\lambda_q$, individual service rate $\mu_q$, and offered load $a=\lambda_q/\mu_q$, the Erlang-C waiting probability is
\begin{equation}
P_W=\frac{\dfrac{a^c}{c!}\dfrac{c}{c-a}}
{\displaystyle\sum_{k=0}^{c-1}\frac{a^k}{k!}
+\dfrac{a^c}{c!}\dfrac{c}{c-a}},
\label{eq:erlangC}
\end{equation}
with mean queue wait
\begin{equation}
W_q=\frac{P_W}{c\mu_q-\lambda_q}.
\label{eq:queuewait}
\end{equation}
Real maintenance demand is not Poisson and service times are not exponential: technology-refresh campaigns are scheduled, impact events are bursty, and complex repairs have heavy-tailed duration.  Equations~\eqref{eq:erlangC}--\eqref{eq:queuewait} are therefore screening relations, followed by discrete-event simulation for mission design \cite{Gross2018Queueing}.

The reconciled reference workload developed in Sec.~\ref{sec:reliability} is 393--419 standardized operations and approximately 560--700 productive robot-hours per MW-year.  If $H_y\simeq3500$ productive hours per robot-year after calibration, maintenance, charging, and scheduling, one MW does not need a full robot by mean workload.  It nevertheless needs at least three or four deployed robotic capabilities to avoid a single-point maintenance failure.  At \SI{100}{MW}, the mean demand is 16--20 productive equivalents; imposing the stated $\rho_{\max}=0.6$--0.7 queueing margin requires approximately 23--33 capacity-sized equivalents before specialization and reserve.  At \SI{1}{GW}, the corresponding range is approximately 229--333 before specialized heavy servicers and recovery reserve.

\subsection{Task-success requirement}

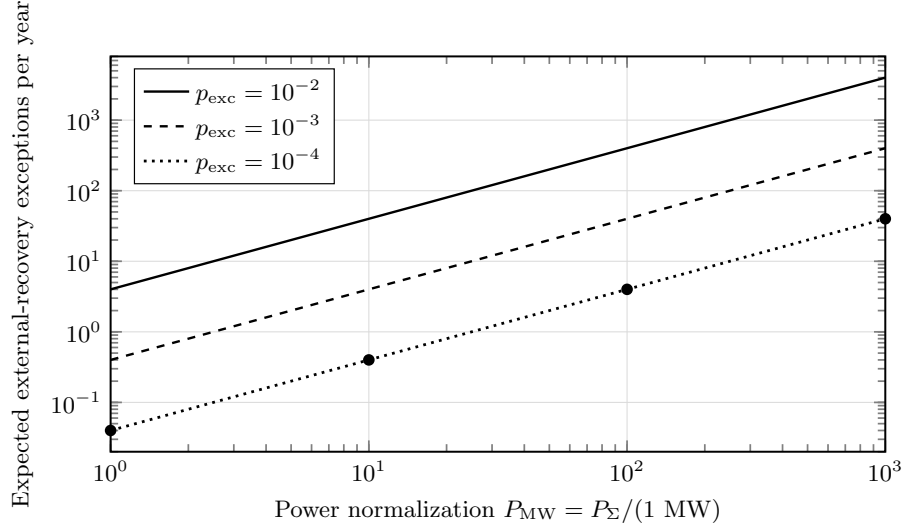
\begin{figure}[htbp]
\centering
\begin{tikzpicture}
\begin{axis}[
  width=0.66\textwidth,
  height=0.38\textwidth,
  xmode=log,ymode=log,
  xmin=1,xmax=1000,
  ymin=0.02,ymax=8000,
  xlabel={Power normalization $P_{\rm MW}=P_\Sigma/(1~\mathrm{MW})$},
  ylabel={Expected external-recovery exceptions per year},
  legend pos=north west,
  domain=1:1000,
  samples=100
]
\addplot[solid,line width=0.9pt] {400*x*1e-2};
\addlegendentry{$p_{\rm exc}=10^{-2}$}
\addplot[dashed,line width=0.9pt] {400*x*1e-3};
\addlegendentry{$p_{\rm exc}=10^{-3}$}
\addplot[dotted,line width=1.05pt] {400*x*1e-4};
\addlegendentry{$p_{\rm exc}=10^{-4}$}
\addplot[only marks,mark=*,mark size=1.8pt] coordinates {(1,0.04) (10,0.4) (100,4) (1000,40)};
\end{axis}
\end{tikzpicture}
\caption{Expected post-internal-recovery robotic exceptions for 400 standardized operations per MW-year; $\PMW=1000$ corresponds to a \SI{1}{GW} cluster.  Industrial scale converts a seemingly high per-task success probability into a substantial exception workload.  The plotted $p_{\rm exc}$ is not the terminal-unresolved probability $p_U$.}
\label{fig:exceptions}
\end{figure}

Three probabilities must not be conflated: $p_1$ is first-attempt failure; $\pex$ is the probability that an operation still requires external recovery after ordinary automatic retry, tool change, and alternate-robot recovery; and $\pU$ is the probability that the operation remains terminally unresolved after all qualified onboard, visiting-servicer, and human-contingency defenses.  Let $N_{\rm op}$ be standardized robotic operations per MW-year.  At dimensionless scale $\PMW$ defined by \Eq{eq:PMWdefinition}, the expected number of post-internal-recovery exceptions is
\begin{equation}
N_{\rm exc}=N_{\rm op}\PMW\pex.
\label{eq:exceptions}
\end{equation}
For $N_{\rm op}=400$, a 99\% completion rate leaves four exceptions per MW-year and four thousand per GW-year.  A 99.9\% rate leaves 400 per GW-year.  A 99.99\% rate leaves 40 per GW-year.  Figure~\ref{fig:exceptions} shows this scale law.

A first MW should require
\begin{equation}
\pex\lesssim10^{-3},\qquad \text{goal }\pex\lesssim10^{-4},
\label{eq:robotrequirement}
\end{equation}
for standardized operations, provided a separately sized external-recovery process clears that workload.  The terminal probability $\pU$ must instead be allocated from mission exposure.  Under the deliberately simple homogeneous-independent diagnostic,
\begin{equation}
P(N_U\ge1)=1-(1-\pU)^{N_{\rm life}},
\label{eq:terminalunresolved}
\end{equation}
and $N_{\rm life}=400\times12=4800$ operations gives $\pU\le1.07\times10^{-5}$ for a 5\% mission probability and $\pU\le2.09\times10^{-6}$ for 1\%.  These numbers are exposure diagnostics, not universal flight requirements; class dependence and common-cause recovery failure require an explicit mission model.  At hundred-megawatt scale, $\pex\sim10^{-4}$ becomes a more stringent external-exception objective unless a periodic campaign is planned to clear the resulting backlog.

\subsection{Maintenance-robot recovery architecture}

Maintenance infrastructure must be designed as a repairable system with explicit recovery states and replacement interfaces.  The recovery hierarchy is:
\begin{enumerate}
\item passive rails, capture sockets, winches, and tow points that work after local power loss;
\item standardized replaceable joint, camera, controller, cable, and end-effector modules;
\item overlapping fixed-arm work envelopes;
\item mobile robots that can deliver parts to fixed arms and fixed arms that can repair mobile robots;
\item at least one full cold-spare robot or equivalent modular spare per service cell;
\item a visiting servicer able to replace the complete local robotics package; and
\item human intervention only after these layers fail.
\end{enumerate}

If each of $n$ robots has independent availability $a_R$ and any $k$ are sufficient for critical recovery, fleet availability is
\begin{equation}
A_{k|n}=\sum_{j=k}^{n}\binom{n}{j}a_R^j(1-a_R)^{n-j}.
\label{eq:kofn}
\end{equation}
For four robots with $a_R=0.95$ and a requirement that any two remain available, $A_{2|4}=0.9995$.  This calculation is optimistic because shared software, tool storage, rails, and power can create correlated failures.  Critical recovery therefore needs design diversity and passive mechanical paths in addition to numerical redundancy.

\subsection{Autonomy versus teleoperation}

LEO round-trip propagation delay is only milliseconds, but end-to-end command, video encoding, ground routing, and operational authorization produce larger latencies.  Direct teleoperation remains practical for difficult tasks if contact is available.  GEO introduces $\sim$ \SI{0.24}{s} round-trip light time, and lunar distance approximately \SI{2.56}{s}; these delays favor supervised autonomy, guarded motion, and local contact detection rather than continuous joystick control.

High-volume replacement should be autonomous after a task plan is approved.  Novel mechanical work should be teleoperated with local collision avoidance and force limiting.  Fully autonomous structural repair, free-form cable routing, and recovery from severe deformation remain speculative because perception confidence, contact mechanics, and tool-selection uncertainty are difficult to certify.  The architecture should exploit autonomy where geometry is designed and repeatable rather than expecting general artificial intelligence to compensate for poor maintainability.

\section{Reliability, failure, and replacement requirements}
\label{sec:reliability}

\subsection{Preliminary component allocation}

Table~\ref{tab:componentallocation} gives a reference population and annual hazard allocation for one delivered MW.  The numbers are deliberately presented as design inputs.  The annual hazards include permanent failure or life-limited replacement that cannot be corrected by software; transient radiation events are handled separately in utilization.  The preferred reliability representation remains the age-dependent hazard $h_i(t)$ in \Eq{eq:bathtub}.  For intuitive comparison only, define an equivalent exponential mean time between failures,
\begin{equation}
\mathrm{MTBF}_{\mathrm{eq},i}\equiv\lambda_i^{-1},
\label{eq:mtbfeq}
\end{equation}
where $\lambda_i$ is the representative constant annual hazard used in the allocation.  Equation~\eqref{eq:mtbfeq} is not used to suppress infant mortality, wearout, common cause, or age-distribution effects; it is only a compact summary of the plateau-hazard assumption.  Every row requires qualification and precursor-flight validation under NASA reliability and maintainability practices \cite{NASA2017RM,Modarres2017,Rausand2004}.

\begin{table}[tbp]
\caption{Reliability allocation for random and life-limited orbital replacement units (ORUs) per delivered MW.  The equivalent exponential value $\mathrm{MTBF}_{\rm eq}=\lambda^{-1}$ is a summary only; the nonhomogeneous model in \Eq{eq:bathtub} remains the preferred representation.}
\label{tab:componentallocation}
\centering
\small
\renewcommand{\arraystretch}{1.05}
\setlength{\tabcolsep}{4pt}
\begin{tabularx}{\textwidth}{@{}Y C{0.09\textwidth} C{0.13\textwidth} C{0.15\textwidth} C{0.13\textwidth}@{}}
\toprule
Replacement class & Population $N_i$ & Item hazard $\lambda_i$ (yr$^{-1}$) & $\mathrm{MTBF}_{\rm eq}$ (yr) & Events $\Lambda_i$ (yr$^{-1}$) \\
\midrule
\SI{5}{kW} compute cartridge & 200 & 0.12 & 8.3 & 24.0 \\
Storage module & 700 & 0.03 & 33 & 21.0 \\
Optical/electrical transceiver & 400 & 0.025 & 40 & 10.0 \\
Fabric switch unit & 40 & 0.08 & 12.5 & 3.2 \\
Power converter & 160 & 0.015 & 67 & 2.4 \\
Power-distribution unit & 40 & 0.035 & 29 & 1.4 \\
Photovoltaic power/deployment module & 100 & 0.01 & 100 & 1.0 \\
Battery module, unscheduled & 20 & 0.04 & 25 & 0.8 \\
Pump & 40 & 0.04 & 25 & 1.6 \\
Valve or thermal controller & 300 & 0.006 & 167 & 1.8 \\
Radiator panel or puncture event & 100 & 0.001 & 1000 & 0.1 \\
Communications terminal & 8 & 0.05 & 20 & 0.4 \\
Attitude-control, avionics, or propulsion unit & 50 & 0.025 & 40 & 1.25 \\
Major robot unit or joint set & 12 & 0.10 & 10 & 1.2 \\
\midrule
Total & -- & -- & -- & 70.2 \\
\bottomrule
\end{tabularx}
\end{table}

Table~\ref{tab:componentservice} translates the same event allocation into productive robotic service time and annual replacement mass.

\begin{table}[tbp]
\caption{Servicing and replacement-mass consequences of the reliability allocation in Table~\ref{tab:componentallocation}.  Productive service time runs from local fault confirmation and safing through replacement verification; queueing, resupply delay, and unresolved recovery are excluded.}
\label{tab:componentservice}
\centering
\small
\renewcommand{\arraystretch}{1.05}
\setlength{\tabcolsep}{5pt}
\begin{tabularx}{\textwidth}{@{}Y C{0.18\textwidth} C{0.14\textwidth} C{0.18\textwidth}@{}}
\toprule
Replacement class & Service time (h/event) & Unit mass $m_i$ (kg) & Annual mass $\dot M_i$ (kg/yr) \\
\midrule
\SI{5}{kW} compute cartridge & 1.0--1.5 & 25 & 600 \\
Storage module & 0.5--1.0 & 1.5 & 31.5 \\
Optical/electrical transceiver & 0.5--1.0 & 0.8 & 8 \\
Fabric switch unit & 1--2 & 8 & 25.6 \\
Power converter & 2.5--5 & 20 & 48 \\
Power-distribution unit & 3--6 & 40 & 56 \\
Photovoltaic power/deployment module & 6--12 & 50 & 50 \\
Battery module, unscheduled & 2.5--5 & 35 & 28 \\
Pump & 4--8 & 30 & 48 \\
Valve or thermal controller & 2--5 & 10 & 18 \\
Radiator panel or puncture event & 8--20 & 150 & 15 \\
Communications terminal & 6--12 & 100 & 40 \\
Attitude-control, avionics, or propulsion unit & 3--8 & 40 & 50 \\
Major robot unit or joint set & 8--20 & 100 & 120 \\
\midrule
Total & -- & -- & 1138 \\
\bottomrule
\end{tabularx}
\end{table}

Under the declared reference populations and hazard allocations, Tables~\ref{tab:componentallocation} and~\ref{tab:componentservice} imply 70.2 random intervention opportunities and \SI{1.14}{t/(MW.year)} of direct ORU mass, respectively.  These are allocation consequences, not estimated flight rates.  Multiplying each event rate by its service-duration range gives the direct random-service demand
\begin{equation}
H_{R,\mathrm{rnd}}=\sum_i\Lambda_i t_{s,i}
\simeq\SIrange{87}{172}{robot.h/(MW.year)},
\label{eq:randomrobothours}
\end{equation}
before travel, queueing, failed attempts, or spare retrieval from another service zone.  The phase-level assumptions behind the service-duration ranges are reported in \Tab{tab:servicephases}.  Allowing for harnesses, seals, contamination, diagnostic removals, commissioning infant mortality, and uncertainty gives a random replacement planning range
\begin{equation}
\Mdot_{\rm rnd}\simeq \SIrange{1.3}{2.0}{t/(MW.year)}.
\label{eq:randomrange}
\end{equation}
The dominant random mass term in this allocation is the compute-cartridge population.  The ranking and magnitude remain conditional on the assumed populations, hazards, fault-containment boundary, and replacement masses; flight data may change them materially.

\subsection{Planned refresh flow}

Table~\ref{tab:refresh} gives representative refresh intervals.  The design should refresh only the technology-bearing subassembly while retaining the frame, cold plate, high-current bus structure, and robot-compatible chassis whenever possible.

\begin{table}[tbp]
\caption{Representative planned refresh cadence and standardized work per delivered MW.  Operation and mass-flow rows form a conservative additive planning envelope for the reference population.  Robot-hours are a lower-bound-compatible realization using the class-level minimum service times in Tables~\ref{tab:componentservice} and~\ref{tab:servicephases}; calibrated mission distributions will generally be broader.}
\label{tab:refresh}
\centering
\small
\setlength{\tabcolsep}{4.5pt}
\renewcommand{\arraystretch}{1.05}
\begin{tabularx}{\textwidth}{@{}Y C{0.12\textwidth} C{0.16\textwidth} C{0.17\textwidth} C{0.16\textwidth}@{}}
\toprule
Hardware class & Interval (yr) & Operations (yr$^{-1}$) & Mass flow (t yr$^{-1}$) & Robot work (h yr$^{-1}$) \\
\midrule
Accelerator/compute cartridges & 4 & 50 & 1.25 & 50 \\
Storage modules & 5 & 140 & 0.21 & 70 \\
Transceivers and switches & 6--7 & 70--75 & 0.10 & 55 \\
Converters and power distribution & 10 & 20 & 0.4--0.6 & 52 \\
Battery modules & 6 & 3--4 & 0.12 & 9 \\
Pumps, valves, and controllers & 10 & 30--35 & 0.3--0.6 & 76 \\
Communications, flight-control avionics, and robots & 8--10 & 5--10 & 0.3--0.6 & 30 \\
PV/radiator service kits and replaceable subassemblies & 10--15 & 5--15 & 0.2--0.5 & 80 \\
\midrule
Direct total & -- & 323--349 & 2.88--3.98 & $\simeq422$ \\
\bottomrule
\end{tabularx}
\end{table}

Compute and storage account for 190 planned operations per MW-year, while power, thermal, communications, and robotic hardware dominate the noncompute refresh mass.  The PV/radiator row refers to replaceable service kits and technology-bearing subassemblies, not complete panels.  The installed populations contain approximately \SI{5}{t} of PV modules and \SI{15}{t} of radiator panels; assigning a service-kit fraction $f_{\rm kit}=0.15$--0.25 over a 10--15-year cadence gives
\begin{equation}
\Mdot_{\rm panel,kit}=\frac{f_{\rm kit}(M_{\rm PV}+M_{\rm rad})}{T_{\rm panel}}
\simeq\SIrange{0.20}{0.50}{t/(MW.year)},
\label{eq:panelkitflow}
\end{equation}
which is the Table~\ref{tab:refresh} allocation.  Full replacement of all \SI{20}{t} of panels would instead require approximately 1.33--2.0 t/(MW yr) and is not credited here.  The operation total follows directly as $50+140+(70$--$75)+20+(3$--$4)+(30$--$35)+(5$--$10)+(5$--$15)=323$--349 operations per MW-year.  The corresponding mass rows sum to
\begin{equation}
\Mdot_{\rm ref}\simeq\SIrange{2.88}{3.98}{t/(MW.year)}.
\label{eq:refreshrange}
\end{equation}
The lower-bound-compatible robot-hour column gives $H_{R,\rm ref}\simeq422~\mathrm{h/(MW\,yr)}$.  Adding \Eq{eq:randomrobothours} and an auxiliary allowance $H_{R,\rm aux}\simeq50$--100 h/(MW yr) for inspection, calibration, inventory transfer, cargo handling, and post-maintenance trend review gives
\begin{equation}
H_R=H_{R,\rm rnd}+H_{R,\rm ref}+H_{R,\rm aux}
\simeq\SIrange{559}{694}{robot.h/(MW.year)},
\label{eq:totalrobothours}
\end{equation}
which is rounded to the 560--700 robot-h/(MW yr) planning range used elsewhere.  The central lifecycle result is
\begin{equation}
\Mdot_{\rm ref}\gtrsim\Mdot_{\rm rnd}
\label{eq:refreshdominates}
\end{equation}
for accelerator-heavy ODCs under the declared population, hazard allocation, refresh intervals, and additive planning policy.  This is not a universal ordering: an age-replacement policy, longer qualified intervals, or higher random hazard can reverse it.  Random failures shape availability and spare stock; planned refresh shapes the sustained industrial throughput.  At \SI{1}{MW}, one planned compute cartridge is replaced approximately each week.  At \SI{100}{MW}, about 100 planned compute-cartridge exchanges per week plus storage and networking modules require automated material handling.

\subsection{Catastrophic node loss}

For ten \SI{100}{kW} nodes of mass $M_n=\SIrange{4}{6}{t}$, expected whole-node replacement is
\begin{equation}
\Mdot_{\rm cat}=10\lambdaCat M_n.
\label{eq:catnumerical}
\end{equation}
At $M_n=\SI{5}{t}$, $\lambdaCat=0.01$, 0.03, and $0.10~\mathrm{yr^{-1}}$ imply 0.5, 1.5, and \SI{5}{t/(MW.year)}, respectively.  A long-lived architecture should therefore target
\begin{equation}
\lambdaCat\lesssim0.01\text{--}0.03~\mathrm{yr^{-1}}
\quad\text{per \SI{100}{kW} node},
\label{eq:cattarget}
\end{equation}
with
\begin{equation}
\lambdaDom\lesssim10^{-3}~\mathrm{yr^{-1}}
\label{eq:domtarget}
\end{equation}
for events capable of removing a large fraction of the cluster.  If $\lambdaCat$ remains near $0.1~\mathrm{yr^{-1}}$, whole-node replacement becomes comparable to all other logistics and the architecture is not mature enough for industrial scaling.

\subsection{Reserve capacity and replacement delay}

Suppose a node-loss rate of 0.2 events per MW-year removes 10\% of a \SI{1}{MW} cluster.  Without orbital reserve, an Earth replacement delay of 180 days contributes
\begin{equation}
1-A_{\rm cap}\simeq0.2(0.1)\frac{180}{365}
=9.9\times10^{-3}.
\label{eq:noreserveexample}
\end{equation}
If a cold reserve node is activated in one day,
\begin{equation}
1-A_{\rm cap}\simeq0.2(0.1)\frac{1}{365}
=5.5\times10^{-5}.
\label{eq:reserveexample}
\end{equation}
The reserve node changes the service interruption from a supply-chain timescale to a fault-management timescale.  Physical repair may still take weeks without violating the service-level objective.

A reserve policy can be formulated as a capacity newsvendor problem.  Let random simultaneous unavailable power be $X$, reserve power be $R$, annualized reserve cost per MW be $c_{\rm res}$, and shortage value be $v_s$.  Minimize
\begin{equation}
C(R)=c_{\rm res}R+v_s\,\mathbb{E}[(X-R)_+].
\label{eq:reserveopt}
\end{equation}
For a continuous distribution, the optimum satisfies
\begin{equation}
F_X(R^*)=1-\frac{c_{\rm res}}{v_s},
\label{eq:reservequantile}
\end{equation}
showing explicitly why higher compute value or stronger availability requirements justify more reserve.  A first MW carries one full node because the capacity distribution is coarse; at larger scale, 5--10\% pooled reserve is likely more efficient.

\subsection{Transient faults and delivered utilization}

Correctable radiation and software events reduce throughput rather than necessarily driving replacement.  If checkpoint interval is $\tau_{\rm cp}$, checkpoint time is $t_{\rm cp}$, correctable-event rate is $R_{\rm SEE}$, and recovery time is $t_r$, then for small losses
\begin{equation}
U_{\rm eff}\simeq U_0\left[1-\frac{t_{\rm cp}}{\tau_{\rm cp}}
-R_{\rm SEE}\left(t_r+\frac{\tau_{\rm cp}}{2}\right)\right].
\label{eq:Ueff}
\end{equation}
The checkpoint optimum for a workload with failure rate $R_f$ and checkpoint cost $t_{\rm cp}$ has the familiar square-root form $\tau_{\rm cp}\sim\sqrt{2t_{\rm cp}/R_f}$ under simplified assumptions.  In an ODC, checkpoint placement also couples to storage wear, network traffic, and cross-node correlated faults.  Local error-correcting codes (ECC), retry, and micro-restart should handle high-frequency recoverable events so that full workload rollback remains rare.

\section{Logistics, spare parts, and material disposition}
\label{sec:logistics}

Maintenance is operationally useful only if a replacement part, compatible tool, and functioning service path are available when needed.  The logistics architecture must therefore be designed jointly with reliability, not added after the component hazard model is complete.  Three time scales matter: the local response time to isolate a failure, the orbital repair delay after a spare is released, and the replenishment lead time from Earth.  Reserve compute decouples the first two, while orbital inventory decouples the second and third.

\subsection{Base-stock sizing under uncertain demand}

Let class $i$ experience an approximately Poisson replacement demand with annual rate $\Lambda_i$ and deterministic resupply lead time $L_i$.  Demand during lead time has mean and variance
\begin{equation}
\mu_{L,i}=\Lambda_i L_i,
\qquad
\sigma^2_{L,i}=\Lambda_i L_i.
\label{eq:leadpoisson}
\end{equation}
For moderate $\mu_{L,i}$, a normal-quantile approximation gives the initial base stock
\begin{equation}
S_i\simeq\left\lceil \Lambda_i L_i+z_q\sqrt{\Lambda_i L_i}\right\rceil,
\label{eq:basestock}
\end{equation}
where $z_q$ is the standard-normal quantile associated with a class-specific cycle service level $q$ over one lead-time interval.  It is not a system-wide probability of six-month endurance.  Exact Poisson quantiles should be used for rare, high-consequence items and compound-Poisson or negative-binomial models when common-cause events, uncertain lead time, synchronized renewal, manufacturing lots, or finite cargo opportunities create overdispersion.

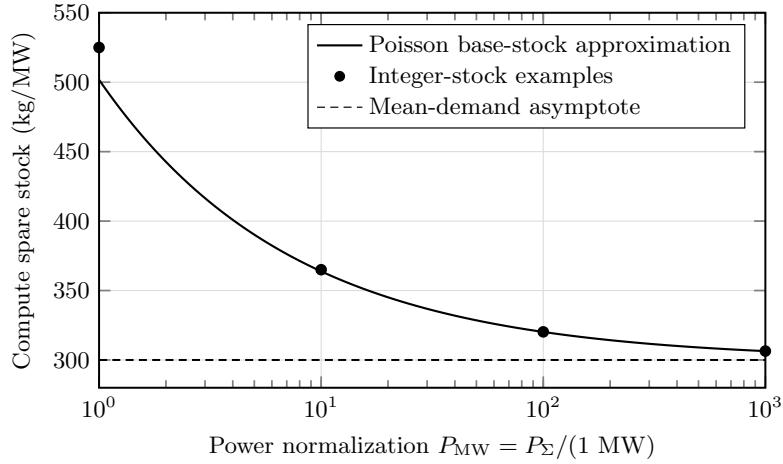
\begin{figure}[htbp]
\centering
\begin{tikzpicture}
\begin{axis}[
width=0.58\textwidth,
height=0.365\textwidth,
xmode=log,
xlabel={Power normalization $P_{\rm MW}=P_\Sigma/(1~\mathrm{MW})$},
ylabel={Compute spare stock (kg/MW)},
xmin=1,xmax=1000,
ymin=280,ymax=550,
legend pos=north east
]
\addplot[domain=1:1000,samples=180,thick]
{25*(24*0.5 + 2.33*sqrt(24*0.5/x))};
\addlegendentry{Poisson base-stock approximation}
\addplot[only marks,mark=*,mark size=1.8pt] coordinates {
(1,525) (10,365) (100,320.25) (1000,306.4)};
\addlegendentry{Integer-stock examples}
\addplot[densely dashed,domain=1:1000] {300};
\addlegendentry{Mean-demand asymptote}
\end{axis}
\end{tikzpicture}
\caption{Pooling of compute-cartridge spares for a six-month replenishment planning horizon, a replacement rate of 24 cartridges per MW-year, \SI{25}{kg} per cartridge, and an approximately 99\% class-specific cycle service level under stationary Poisson demand.  This is not a plant-level stockout guarantee.  The $P_{\rm MW}^{-1/2}$ safety-stock term is important for the first few megawatts but becomes a modest increment at industrial scale.}
\label{fig:sparepooling}
\end{figure}

For compute cartridges, let $\Lambda_{c,1}=24~\mathrm{yr^{-1}}$ denote the replacement rate of a one-MW reference population.  At scale $\PMW$, the aggregate rate is $\Lambda_c(\PMW)=\Lambda_{c,1}\PMW$.  With $L=0.5~\mathrm{yr}$ and $q\simeq0.99$ ($z_q=2.33$),
\begin{equation}
S_c(\PMW=1)=21
\label{eq:spareexample}
\end{equation}
cartridges, or approximately \SI{525}{kg} at \SI{25}{kg} per cartridge.  Pooling reduces safety stock per delivered MW because the mean term scales as $\PMW$ while the uncertainty term scales as $\sqrt{\PMW}$:
\begin{equation}
\frac{m_c S_c(\PMW)}{\PMW}
\simeq m_c\Lambda_{c,1}L+
\frac{m_c z_q\sqrt{\Lambda_{c,1}L}}{\sqrt{\PMW}}.
\label{eq:pooling}
\end{equation}
Because $\PMW$ is dimensionless, the left-hand side is reported numerically as kilograms of stock per delivered MW.  The example falls from approximately \SI{525}{kg/MW} at \SI{1}{MW} to \SI{365}{kg/MW} at \SI{10}{MW}, \SI{320}{kg/MW} at \SI{100}{MW}, and an asymptote near \SI{300}{kg/MW}.  Pooling is even more valuable for rare, expensive ORUs, provided a transport robot can move the spare to the failed node without creating excessive repair delay.  Figure~\ref{fig:sparepooling} makes the scale dependence explicit: most of the per-MW inventory benefit is obtained between the first few megawatts and the first industrial-scale cluster, after which the mean-demand term dominates.

The Poisson result is not a substitute for common-cause analysis.  A single radiation storm, production lot defect, coolant contamination event, or firmware-induced overstress can consume multiple identical spares.  For such cases, the inventory distribution is a mixture,
\begin{equation}
D_i=D_{i,\mathrm{ind}}+B_i K_i,
\label{eq:compounddemand}
\end{equation}
where $B_i$ is a Bernoulli indicator for a domain event and $K_i$ is the number of items consumed if it occurs.  A useful planning rule is to hold the statistical base stock of \Eq{eq:basestock} plus one complete recovery set for each specified domain-level failure that cannot be repaired by substitution.

\subsection{Inventory echelons and storage architecture}

A practical system uses three inventory echelons:
\begin{enumerate}
\item \emph{Line-side stock}: approximately 30 days of high-frequency L1 cartridges and end effectors colocated with each service cell, minimizing robot travel and local response time.
\item \emph{Central orbital warehouse}: six to twelve months of common parts, at least one of every critical rare ORU, robot joint modules, radiator repair hardware, fluid reserve, and quarantined failed units awaiting disposition.
\item \emph{Earth production and launch stock}: one to two replenishment cycles of qualified hardware, with launch-delay and production-yield margin.  This stock is not equivalent to ground production inventory because any design change must preserve orbital interface compatibility and flight qualification.
\end{enumerate}

For the reference MW, a first-line stocked-ORU allocation excluding the full reserve node is $\sim$ \SIrange{2}{2.5}{t}; this mass is carried explicitly and is not part of the maintainability-hardware subtotal in \Tab{tab:maintmass}.  At a packaged density of \SIrange{150}{250}{kg/m^3}, the static spare volume is \SIrange{8}{17}{m^3}; allowing aisles, robotic access, shock-isolating containers, failed-unit quarantine, tools, packaging, and one refresh campaign raises useful logistics volume to $\sim$ \SIrange{20}{40}{m^3}.  Inventory location is a reliability decision.  A centralized warehouse minimizes total stock but creates a transport dependency; fully distributed stock removes that dependency but duplicates rare parts.  The preferred architecture distributes high-frequency and survival-critical parts and pools low-frequency, high-mass hardware.

Each unit must retain a traceable digital thread: manufacturing lot, radiation lot acceptance, test history, launch vibration exposure, time in orbit, thermal cycles, connector cycles, firmware compatibility, and storage environment.  A spare whose configuration is unknown is not available capacity.  Inventory reconciliation therefore combines machine-readable identity, mass and silhouette checks, electrical built-in test, and location confirmation by independent cameras or fiducials.

\subsection{Annual mass flow and cargo cadence}

The annual delivered logistics mass is decomposed as
\begin{equation}
\Mdot_{\rm up}=
\Mdot_{\rm rnd}+
\Mdot_{\rm ref}+
\Mdot_{\rm cat}+
\Mdot_{\rm cons}+
\Mdot_{\rm pack},
\label{eq:totalupmass}
\end{equation}
where the terms denote random ORUs, planned refresh, catastrophic-node replacement, consumables and propellant, and delivery/return packaging.  The reference ranges are summarized in \Tab{tab:massflow}.

\begin{table}[htbp]
\caption{First-generation annual logistics allocation per delivered MW.  The component rows enter the additive balance in \Eq{eq:totalupmass}.  Their ranges are epistemic design bounds rather than probability quantiles; the direct bound sum and a reproducible nominal realization are reported separately.}
\label{tab:massflow}
\centering
\small
\renewcommand{\arraystretch}{1.08}
\begin{tabularx}{\textwidth}{@{}L{0.28\textwidth} C{0.16\textwidth} Y@{}}
\toprule
Flow class & Mass [t/(MW yr)] & Dominant drivers \\
\midrule
Random cartridge-through-panel replacements & 1.3--2.0 & Compute cartridges, power/thermal modules, robot parts, infant mortality \\
Whole-node loss & 0.5--1.5 & $\lambdaCat=0.01$--$0.03~\mathrm{yr^{-1}}$ for a \SI{5}{t}, \SI{100}{kW} node \\
Planned technology refresh & 2.88--3.98 & Direct sum of the refresh rows in \Tab{tab:refresh} \\
Consumables and service materials & 0.1--0.5 & Propellant, coolant reserve, filters, lubricants, patch materials, tools \\
Packaging \& delivery overhead & 0.5--1.0 & Launch restraint, shielding, cargo adapter, return/disposal container \\
\midrule
Direct additive envelope & 5.28--8.98 & Arithmetic lower- and upper-bound sums; rounded to 5.3--9.0 \\
Nominal reference realization & 6.6 & $1.50+0.75+3.38+0.25+0.70$ t/(MW yr) \\
First-generation scale-up objective & $\leq8$ & Requires measured performance below the conservative upper bound \\
Mature industrial objective & 3--5 & Requires chassis reuse, lower node-loss hazard, and longer qualified refresh intervals \\
\bottomrule
\end{tabularx}
\end{table}

Equation~\eqref{eq:totalupmass} is additive, so the direct component bounds give 5.28--8.98 t/(MW yr), rounded to 5.3--9.0.  The individual intervals are design ranges rather than independent statistical confidence intervals; the nominal 6.6 t/(MW yr) case in \Tab{tab:massflow} provides a consistent reference point.  The corresponding annual flows are \SIrange{5.3}{9.0}{t} at \SI{1}{MW}, \SIrange{53}{90}{t} at \SI{10}{MW}, \SIrange{530}{900}{t} at \SI{100}{MW}, and \SIrange{5.3}{9.0}{kt} at \SI{1}{GW}.  A \SI{1}{GW} system therefore requires continuous cargo delivery of approximately \SIrange{14.5}{24.7}{t/day}, together with automated storage and material handling.  For a cargo vehicle with useful delivered payload $M_v$ and manifest utilization $\eta_v$, the nominal flight rate is
\begin{equation}
N_{v,\mathrm{yr}}=\frac{\Mdot_{\rm up}}{\eta_v M_v}.
\label{eq:cargocadence}
\end{equation}
Cadence must be sized not only to mean mass but also to launch slips and common-cause demand.  At least two transportation providers or independently recoverable launch opportunities are desirable once a single missed mission can exhaust critical stock.

\subsection{Propellant and fluid reserve}

For a platform of mass $M$ undergoing small annual velocity increment $\Delta v$, electric-propulsion propellant is approximately
\begin{equation}
M_{\rm prop}\simeq M\frac{\Delta v}{I_{\rm sp}g_0},
\label{eq:propellant}
\end{equation}
valid for $\Delta v\ll I_{\rm sp}g_0$.  A \SI{50}{t} per-MW platform, $I_{\rm sp}=\SI{1500}{s}$, and $\Delta v=\SI{10}{m/s/yr}$ require roughly \SI{34}{kg/(MW.yr)}; \SI{50}{m/s/yr} requires approximately \SI{170}{kg/(MW.yr)}.  These values are smaller than hardware refresh but are sensitive to altitude, drag, conjunction policy, formation geometry, and logistics-vehicle rendezvous.  Propellant interfaces should therefore be standardized and robotically serviceable even if the initial mission carries lifetime propellant.

Coolant is nominally closed-loop.  Routine consumption should be negligible, but the warehouse should retain enough compatible fluid and accumulator volume to refill at least one isolated loop after a puncture or maintenance spill.  The ability to recover and purify fluid becomes attractive only when the cumulative delivered-fluid cost exceeds the mass, power, contamination risk, and maintenance requirements of the recovery plant.

\subsection{Return, refurbishment, cannibalization, and disposal}

For failed unit $i$, return to Earth is favored when
\begin{equation}
C_{\rm ret,i}+C_{\rm diag,i}+C_{\rm requal,i}
< C_{\rm new,i}+C_{\rm disp,i}+V_{\rm info,i},
\label{eq:returncriterion}
\end{equation}
where $V_{\rm info,i}$ is written on the right as the replacement cost of losing root-cause information; equivalently it may be subtracted from the return side.  Early in a program, $V_{\rm info}$ can dominate because a returned failure may reveal a fleet-wide latent defect.  Once a failure mode is statistically understood, only a sample of repetitive failures needs return.

On-orbit refurbishment is favored when the high-value chassis, cold plate, optical bench, motor, or shielding can be reused and a small board, seal, bearing, or connector causes most failures.  Let $f_r$ be recovered value fraction, $m_r$ the refurbishment-kit mass, $H_r$ the robotic work, and $c_{R,h}$ the fully allocated cost per productive robot-hour.  Let $C_{\rm kit}$ include the manufacture, qualification, and delivery of the kit mass $m_r$.  Refurbishment dominates replacement if
\begin{equation}
C_{\rm kit}+c_{R,h} H_r+C_{\rm test}
< f_r(C_{\rm build}+C_{\rm launch})+\Delta C_{\rm risk},
\label{eq:refurbcriterion}
\end{equation}
where $\Delta C_{\rm risk}$ is the reliability premium for a refurbished item.  This relation favors reusable cold plates, enclosures, robot chassis, and power magnetics before it favors semiconductor-level repair.

Cannibalization should begin at modest scale because it requires little process equipment: failed units become sources of known-good motors, controllers, sensors, connectors, optical modules, and mechanical subassemblies.  Bulk-material recycling requires much larger throughput.  Mixed electronics are difficult to separate and requalify; by contrast, aluminum structure, copper conductors, standardized radiator sheet, and pressure-vessel material may justify orbital recycling when scrap flow reaches hundreds of tonnes per year and an orbital manufacturing base already exists.  Recycling should not be credited as a first-generation logistics reduction unless the process has an explicit mass, power, yield, contamination, and qualification budget.

Low-value hardware in LEO should be cryptographically sanitized, batteries discharged and isolated, pressure vessels and propellant systems passivated, enclosed so that no loose fragments are released, and transferred to a controlled reentry or compliant disposal vehicle.  The preferred disposal unit is a container, not an independent failed ORU: deliberate release of small hardware would convert maintenance waste into an orbital-debris source.  GEO and cislunar disposal require orbit-specific graveyard or transfer strategies, with long-lived custody and tracking included in operating cost.

\section{Quantitative performance and sensitivity analysis}
\label{sec:performance}

The preceding relations can be organized into a small set of intensive figures of merit that remain meaningful from one MW to one GW.  All per-MW quantities are normalized by the dimensionless $\PMW$, not by $\Pcl$ in watts:
\begin{align}
\ell_M &\equiv \frac{\Mdot_{\rm up}}{\PMW}
&&[\mathrm{t/(MW\,yr)}],
\label{eq:ellM}\\
\ell_R &\equiv \frac{H_R}{\PMW}
&&[\mathrm{robot\,h/(MW\,yr)}],
\label{eq:ellR}\\
\ell_H &\equiv \frac{H_H}{\PMW}
&&[\mathrm{crew\,h/(MW\,yr)}],
\label{eq:ellH}\\
\chi_R &\equiv \frac{n_R h_y}{H_R}
&&\text{robotic capacity margin},
\label{eq:chiR}\\
\rho_S &\equiv \frac{S}{\Lambda L}
&&\text{inventory coverage ratio},
\label{eq:rhoS}\\
\mathcal{H} &\equiv \frac{\PMW b_H}{C_H}
&&\text{human-presence value ratio}.
\label{eq:humanratio}
\end{align}
Here $n_R$ is the number of productive robot-equivalent servers, $h_y$ their productive hours per year, and $b_H$ is the annual human-attributable benefit per delivered MW, with units \$/[MW yr].  A design should have $\chi_R>1$ under the 95th-percentile maintenance workload, $\rho_S>1$ for common parts with additional safety stock, and permanent human presence becomes economically admissible only when $\mathcal H>1$.

\subsection{Node granularity as a coupled optimum}

Node power controls both duplicated spacecraft mass and the service impact of one node failure.  A compact objective per delivered kilowatt is
\begin{equation}
J(p_n)=m_0+\frac{1000M_{\rm loc}}{p_n}
+w_A\frac{p_n}{\Pcl}+w_L\lambda_n(p_n)\tau_n(p_n),
\label{eq:granularityobjective}
\end{equation}
where $M_{\rm loc}$ is fixed local spacecraft mass, $w_A$ converts single-node lost fraction into an equivalent design penalty, and the last term represents lifecycle cost and availability effects.  If $\lambda_n\tau_n$ is weakly dependent on node power and the dominant trade is $a/p_n+b p_n$, the stationary point is
\begin{equation}
p_n^*=\sqrt{\frac{a}{b}}
=\sqrt{\frac{1000M_{\rm loc}\Pcl}{w_A}}.
\label{eq:granularityoptimum}
\end{equation}
The equation is not a universal optimum because $w_A$, robot reach, launch packaging, radiator segmentation, and node hazard are mission-specific.  It does show why neither very small nor monolithic nodes are automatically favored.  For the first MW, \SI{100}{kW} is a reasonable fault-domain anchor: one node removes 10\% nameplate before reserve, while fixed bus duplication remains potentially tolerable.  At 100 MW or more, larger physical service cells may contain many independently isolated 100-kW electrical and thermal domains.

\subsection{Availability budget}

A representative delivered-capacity budget is given in \Tab{tab:availabilitybudget}.  The budget is intentionally expressed as fractional lost capacity rather than binary outage.  The target is an architectural allocation, not a measured forecast.

\begin{table}[htbp]
\caption{Illustrative annual delivered-capacity unavailability allocation for a mature cluster.}
\label{tab:availabilitybudget}
\centering
\small
\renewcommand{\arraystretch}{1.08}
\begin{tabularx}{\textwidth}{@{}L{0.29\textwidth} C{0.18\textwidth} Y@{}}
\toprule
Contributor & Allocation to $1-A_{\rm cap}$ & Design implication \\
\midrule
Cartridge migration \& local restart & $2\times10^{-5}$ & Hot headroom and sub-minute isolation \\
Planned hardware replacement & $2\times10^{-5}$ & Drain work before de-energization; service one domain at a time \\
Independent node loss with reserve & $6\times10^{-5}$ & Reserve activation within approximately one day \\
Network/gateway degradation & $5\times10^{-5}$ & Dual fabrics and geographically diverse gateways \\
Power and thermal derating & $5\times10^{-5}$ & Segmented loops, N+2 conversion, graceful throttling \\
Cyber/software rollback events & $5\times10^{-5}$ & Canary updates and bounded blast radius \\
Conjunction \& formation maneuvers & $2\times10^{-5}$ & Checkpointing and maneuver-aware scheduling \\
Correlated residual and margin & $2.3\times10^{-4}$ & Diversity, independent command, domain separation \\
\midrule
Total illustrative allocation & $5.0\times10^{-4}$ & $A_{\rm cap}=99.950\%$ \\
Stretch objective & $10^{-4}$ & Requires much smaller common-mode and logistics tails \\
\bottomrule
\end{tabularx}
\end{table}

Table~\ref{tab:availabilitybudget} shows that four-nines delivered capacity cannot be obtained by shortening routine cartridge replacement from four hours to one hour if correlated software, power, or thermal events contribute $10^{-3}$ of lost capacity.  The primary control variables are reserve activation time, fault-domain size, command diversity, and upper quantiles of repair delay.

\subsection{Scale laws for logistics and robots}

\begin{figure}[htbp]
\centering
\begin{tikzpicture}
\begin{groupplot}[
 group style={group size=2 by 1,horizontal sep=1.7cm},
 width=0.42\textwidth,height=0.32\textwidth,
 xmode=log,xmin=1,xmax=1000,
 xlabel={Power normalization $P_{\rm MW}$}
]
\nextgroupplot[ylabel={Annual logistics mass (t/yr)},ymode=log,ymin=3,ymax=1e4]
\addplot[thick,domain=1:1000,samples=150] {5.3*x};
\addplot[thick,dashed,domain=1:1000,samples=150] {9.0*x};
\addplot[draw=none,name path=A,domain=1:1000] {5.3*x};
\addplot[draw=none,name path=B,domain=1:1000] {9.0*x};
\addplot[gray!25] fill between[of=A and B];
\addlegendentry{5.3--9.0 t/(MW yr)}
\nextgroupplot[ylabel={Productive robot equivalents},ymode=log,ymin=0.1,ymax=300]
\addplot[thick,domain=1:1000,samples=150] {560*x/3500};
\addplot[thick,dashed,domain=1:1000,samples=150] {700*x/3500};
\addplot[draw=none,name path=C,domain=1:1000] {560*x/3500};
\addplot[draw=none,name path=D,domain=1:1000] {700*x/3500};
\addplot[gray!25] fill between[of=C and D];
\addlegendentry{560--700 h/(MW yr)}
\end{groupplot}
\end{tikzpicture}
\caption{Linear first-order scale laws for annual logistics and mean productive robot-equivalent demand.  The plotted robot curves divide annual work by 3500 productive hours per robot-year; capacity-sized deployed fleets must additionally include the $\rho_{\max}=0.6$--0.7 utilization margin, task compatibility, geography, and recovery reserve.}
\label{fig:scalelaws}
\end{figure}
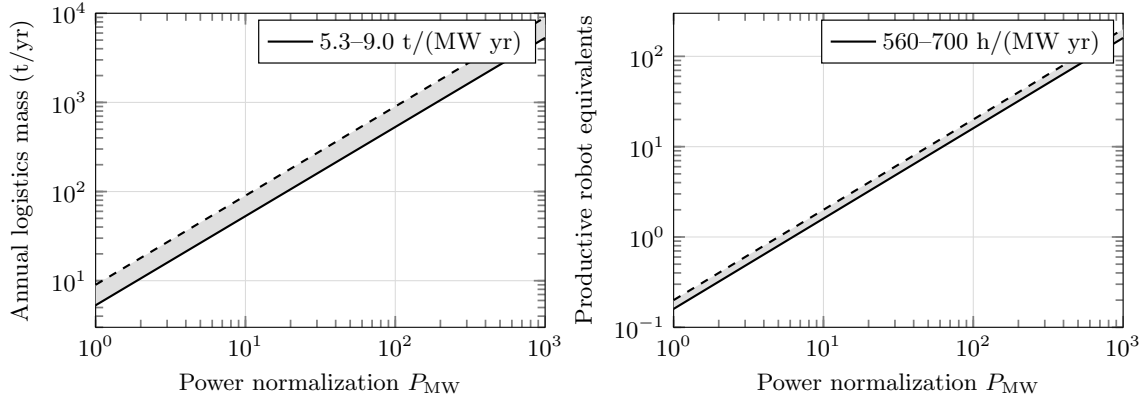

If standardized operations scale approximately linearly with power, the expected robotic workload is
\begin{equation}
H_R(\PMW)=\ell_R\PMW,
\qquad
\ell_R\simeq\SIrange{560}{700}{h/(MW.yr)}.
\label{eq:robotlinearscale}
\end{equation}
At $h_y\simeq\SI{3500}{productive~h/yr}$, the mean workload is 0.16--0.20 robot equivalents per MW.  Redundancy and specialized tools dominate the first MW, so three or four deployed robots are needed despite sub-unit average workload.  At scale, a useful planning relation is
\begin{equation}
N_R\simeq \max\Big[N_{R,\min},
\frac{\ell_R\PMW}{\rho_{\max}\eta_R h_y}\Big]+N_{R,\rm reserve},
\label{eq:robotfleet}
\end{equation}
where $\eta_R$ is the fraction of work that the selected robot class can perform and $\rho_{\max}=0.6$--0.7 is the planning utilization from Sec.~\ref{sec:robotics}.  A \SI{100}{MW} cluster has 16--20 mean productive equivalents but requires approximately 23--33 capacity-sized equivalents before specialization and reserve; a GW system has 160--200 mean equivalents and requires approximately 229--333 on the same utilization basis.  Figure~\ref{fig:scalelaws} shows the mean productive-equivalent demand; deployable fleet sizing lies above that curve because of utilization margin, task compatibility, geography, and recovery reserve.  Mean annual flow becomes nearly linear in delivered power, whereas minimum robot diversity and recovery redundancy keep the first few megawatts above the asymptotic curve.

\subsection{Sensitivity ranking}

For a response $Y$ and parameter $x$, define the local normalized sensitivity
\begin{equation}
S_x^Y\equiv\frac{\partial\ln Y}{\partial\ln x}.
\label{eq:elasticity}
\end{equation}
The architecture-level sensitivities in \Tab{tab:sensitivities} identify variables worth technology investment.

\begin{table}[htbp]
\caption{Leading local sensitivities around the reference architecture.  Photovoltaic and orbital data center terms are written in full where space permits.}
\label{tab:sensitivities}
\centering
\small
\renewcommand{\arraystretch}{1.00}
\begin{tabularx}{\textwidth}{@{}L{0.19\textwidth} L{0.17\textwidth} L{0.13\textwidth} Y@{}}
\toprule
Response & Parameter & Elasticity or dependence & Interpretation \\
\midrule
Photovoltaic mass & deployed specific power $SP_{\rm PV}$ & $-1$ & Doubling array specific power halves the PV mass term \\
Radiator area & $T_{\rm rad}$ & $-4q_{\rm emit}/q_{\rm net}=-5.20$ & Reference-point elasticity for $q_{\rm emit}=650.95$ and $q_{\rm net}=500.95~\mathrm{W/m^2}$; constrained by junction temperature and transport drop \\
Radiator area & absorbed environ- mental flux $q_{\rm env}$ & $q_{\rm env}/q_{\rm net}=+0.30$ & Reference-point logarithmic elasticity; becomes singular as absorbed flux approaches emitted flux \\
Random replacement mass & component hazard $\lambda_i$ & $+1$ & Qualification and derating act linearly on routine logistics \\
Base stock safety term & scale $\PMW$ & $-1/2$ per MW & Pooling benefit saturates at large scale \\
Exceptions & $p_{\rm exc}$ & $+1$ & A decade in task reliability is a decade in human/complex-repair demand \\
Whole-node logistics & $\lambdaCat$ & $+1$ & Must be below a few percent per node-year \\
Crew threshold & $C_{H,0}$ & $+1$ & Shared habitat or transport directly lowers the required orbital data center scale \\
Crew threshold & $b_H-c_{H,1}$ & $-1$ & Higher capacity value \& larger human benefit lower threshold \\
Queue delay & utilization $\rho_R$ & nonlinear & Diverges as robotic service capacity approaches demand \\
\bottomrule
\end{tabularx}
\end{table}

The sensitivity analysis identifies robot exception rate, catastrophic-node hazard, radiator temperature capability, and chassis reuse as the principal lifecycle parameters.  Routine manipulation time has secondary importance once service-capacity margin is adequate.  Permanent-crew economics depend primarily on fixed human-support cost and the expected value of recovery from low-probability, high-consequence events rather than standard cartridge-exchange time.

\section{Uncertainty, systematic error, and validation}
\label{sec:uncertainty}

The numerical results above are not point predictions.  They are architecture-level estimates conditioned on component populations, hazard rates, service times, resupply lead times, and value assumptions that are poorly known before a precursor fleet operates.  A mission-specific study must distinguish three classes of uncertainty:
\begin{enumerate}
\item \emph{aleatory variability}: irreducible randomness in radiation events, impacts, launch slips, individual component life, and repair duration;
\item \emph{epistemic uncertainty}: incomplete knowledge of hazard distributions, robot performance, thermal-cycle damage, and software common-mode behavior; and
\item \emph{systematic model error}: omitted coupling, nonstationarity, selection bias in returned hardware, and incorrect transfer of terrestrial or small-spacecraft statistics to MW-class systems.
\end{enumerate}
The third class is the most dangerous because it does not shrink by averaging more identical units if the entire model is biased.

\subsection{Telemetry noise and calibration drift}

For measurement vector $\bm y$, the covariance used in diagnosis should be decomposed as
\begin{equation}
\bm C_y=\bm C_{\rm sensor}+\bm C_{\rm quant}+\bm C_{\rm timing}
+\bm C_{\rm model}+\bm C_{\rm environment},
\label{eq:telemetrycov}
\end{equation}
where the last two terms are often correlated across channels.  Temperature gradients bias flow and pressure inference; structural motion biases robot vision and force estimation; radiation can create simultaneous sensor and controller upset; and changing workload alters electrical and thermal baselines.  A fixed threshold calibrated on the ground is therefore insufficient.  Each critical measurement chain should include on-orbit calibration stimuli, reference standards, or cross-comparison against a physically diverse sensor.

Sensor drift is especially consequential for prognostics.  If a health indicator $z(t)=d(t)+b(t)+n(t)$ contains true degradation $d$, bias $b$, and zero-mean noise $n$, a replacement rule based on $z$ cannot distinguish degradation from bias without an independent reference.  For batteries, converters, pumps, and thermal loops, deliberate low-amplitude system identification during scheduled quiet windows can estimate impedance, efficiency, pressure-flow curves, or thermal conductance and separate slowly varying bias from actual wear.

\subsection{Hazard-rate uncertainty and sparse failures}

For component class $i$ with $k_i$ observed failures in accumulated exposure $E_i$ item-years, a Gamma prior $\lambda_i\sim\mathrm{Gamma}(a_i,b_i)$ with rate parameter $b_i$ and a Poisson likelihood produce
\begin{equation}
\lambda_i\mid k_i,E_i
\sim\mathrm{Gamma}(a_i+k_i,b_i+E_i).
\label{eq:bayeshazard}
\end{equation}
The posterior width remains large when failures are rare.  Zero observed failures does not establish a zero hazard; it establishes an upper bound determined by exposure.  With a noninformative exponential approximation, observing no events in $E$ item-years gives a 95\% upper bound near $3/E$.  Demonstrating a catastrophic-node hazard below $0.01~\mathrm{yr^{-1}}$ at 95\% confidence therefore requires of order 300 node-years with no event, or a validated decomposition into lower-level hazards and strong prior evidence.  A pathfinder with ten nodes for one year cannot empirically establish an industrial catastrophic-loss rate.

Age aggregation is another systematic.  The fleet event rate is
\begin{equation}
\Lambda_i(t)=\int_0^\infty n_i(a,t)h_i(a)\,da,
\label{eq:agepopulation}
\end{equation}
where $n_i(a,t)$ is the age distribution.  Treating $h_i$ as constant masks commissioning infant mortality and synchronized wearout after a batch refresh.  Rolling replacement should intentionally desynchronize age and manufacturing lots so that a wearout knee or latent lot defect does not become a cluster-wide maintenance surge.

\subsection{Repair-time distributions and queueing bias}

Mean task time is a poor predictor of service capacity if the duration distribution is heavy-tailed.  Let service time $T_s$ have squared coefficient of variation $c_s^2$.  For a single-server approximation, the Pollaczek--Khinchine relation is
\begin{equation}
W_q=\frac{\Lambda\mathbb E[T_s^2]}{2(1-\rho)}
=\frac{\rho\,\mathbb E[T_s](1+c_s^2)}{2(1-\rho)},
\label{eq:PK}
\end{equation}
with $\rho=\Lambda\mathbb E[T_s]$.  A small number of jammed latches, ambiguous leaks, or failed robots can therefore dominate delay even when the median cartridge exchange is fast.  Mission models should retain task classes and empirical duration distributions; replacing them with one exponential mean will generally underestimate both queue delay and spare occupancy.

A further bias arises because successful routine tasks are numerous and well measured, whereas rare terminally unresolved tasks may be censored by node disposal or crew rescue.  The operational database should record every retry, alternate-tool use, intervention, and abandoned procedure.  ``Task success'' must be defined before flight and must include restoration, leak test, built-in test, and safe packaging of the removed unit, not merely mechanical insertion.

\subsection{Correlated and adversarial events}

The independence assumption is optimistic for at least six mechanisms:
\begin{enumerate}
\item a production-lot defect affects many nominally redundant ORUs;
\item one firmware image or root key spans multiple nodes;
\item a radiation storm increases simultaneous upset and latchup risk;
\item a debris-generating event changes the local impact environment;
\item a shared logistics or ground-network outage delays multiple repairs; and
\item an adversary intentionally targets common credentials, update infrastructure, or gateway services.
\end{enumerate}
A useful stress model is a marked domain-event process.  Events occur with rate $\lambda_d$ and remove random capacity fraction $F_d$ for duration $T_d$.  The capacity loss is
\begin{equation}
1-A_d=\frac{\lambda_d}{T_y}\mathbb E[F_d T_d].
\label{eq:domainavailability}
\end{equation}
Upper quantiles of $F_dT_d$, rather than only its mean, should be tested against service-level and financial limits.  Architecture reviews should include specified bounding cases: loss of all nodes sharing one firmware release, all thermal loops sharing one fluid lot, one gateway region, one robot software stack, or one launch provider.

\subsection{Cost and value uncertainty}

Human thresholds and return economics depend on uncertain future launch, build, compute-value, insurance, and service prices.  The correct output is therefore a boundary surface, not a single date or power level.  For a response $Y(\bm\theta)$ with parameter covariance $\bm C_\theta$, local uncertainty propagation is
\begin{equation}
\sigma_Y^2\simeq
\nabla_{\theta}Y^{T}\bm C_\theta\nabla_{\theta}Y,
\label{eq:uncertaintyprop}
\end{equation}
while nonlinear thresholds should be evaluated by Monte Carlo or global sensitivity methods.  Correlated cost assumptions must be preserved: cheaper launch may coincide with higher launch frequency and lower inventory, while a shared depot may lower crew cost and also improve robotic recovery.  Table~\ref{tab:uncertaintybudget} ranks the quantities that most strongly limit decision confidence.  Catastrophic-node hazard and the distinct $p_{\rm exc}$ and $p_U$ rates require the longest statistical exposure, while integrated manifests are needed to prevent systematic undercounting of refresh and packaging mass.

\begin{table}[htbp]
\caption{Principal uncertainty and systematic-error budget for an orbital data center architecture study.}
\label{tab:uncertaintybudget}
\centering
\small
\renewcommand{\arraystretch}{1.08}
\begin{tabularx}{\textwidth}{@{}L{0.20\textwidth} C{0.11\textwidth} L{0.29\textwidth} Y@{}}
\toprule
Quantity & Precursor uncertainty & Dominant systematic & Required validation \\
\midrule
Permanent cartridge hazard & factor 2--4 & terrestrial-to-space transfer, infant mortality & radiation, thermal-vacuum, vibration, and multi-year fleet exposure \\
Catastrophic node hazard & order of magnitude & sparse statistics and unmodeled common cause & hundreds of node-years or validated fault-tree allocation \\
Robot task time & factor 1.5--3 & omission of retries and recovery & hundreds to thousands of end-to-end cycles in representative dynamics \\
Robot exception and terminal-unresolved rates & one to two decades & recovery-boundary definition, common cause, and censoring bias & adversarial fault insertion, end-to-end trials, and independent scoring \\
Radiator puncture/leak rate & factor 2--5 & area, orientation, shielding, and consequence threshold & Orbital-debris and meteoroid-model analysis plus panel impact testing and flight coupons \\
Spare demand & 30--100\% & overdispersion and synchronized lots & fleet Bayesian update and common-cause scenarios \\
Annual logistics mass & 30--60\% & packaging and refresh-scope omission & integrated manifests for a full refresh campaign \\
Crew annualized cost & factor 2--3 & shared infrastructure allocation & transportation/habitat reference architecture \\
Availability value $v$ & factor 3--10 & workload and market heterogeneity & contract-specific service and outage valuation \\
\bottomrule
\end{tabularx}
\end{table}

\subsection{Validation hierarchy}

The model should be retired progressively by evidence:
\begin{enumerate}
\item component radiation, thermal-vacuum, contamination, life-cycle, and connector testing;
\item coupled interface rigs with flight-like robot sensing, structural compliance, and worst-case thermal distortion;
\item hardware-in-the-loop autonomy and fault-management testing with injected sensor, software, power, and network faults;
\item a free-flying servicing demonstration with at least $10^2$ complete ORU cycles and deliberate off-nominal cases;
\item a 50--100 kW precursor node with replaceable compute, power, and thermal modules;
\item a one-MW pathfinder accumulating multiple node-years and at least one rolling refresh campaign; and
\item a 10-MW preindustrial cluster that validates pooled inventory, multi-robot scheduling, node recovery, and cargo operations.
\end{enumerate}
No single test closes the lifecycle case.  Statistical repetition is essential for high-cycle interfaces, while destructive and adversarial tests are essential for rare common-mode events.  With zero observed terminal-unresolved outcomes in $n$ representative independent trials, the exact one-sided 95\% upper bound is $1-0.05^{1/n}\simeq3/n$.  Thus approximately 2995 zero-outcome trials are required to place a marginal bound below $10^{-3}$ and 29,956 below $10^{-4}$; 100 free-flying cycles bound only about 2.95\%.  The proposed $10^2$ free-flight and $10^4$ ground-cycle exposures are development gates and mechanism-finding tests, not by themselves class-specific or common-mode qualification of $\pex$ or $\pU$.

\section{Human-in-the-loop architectures and break-even conditions}
\label{sec:human}

Routine ODC maintenance does not require human execution; its repetitive task volume is assigned to automated and robotic systems even when a habitat exists.  The potential value of humans lies in the residual task distribution: unmodeled failures, deformed interfaces, ambiguous fluid leaks, damaged robots, first-of-a-kind upgrades, and emergencies in which adaptive perception and tool use can restore a large asset faster than remote redesign and launch.  Five architectures are therefore distinguished: autonomous operation without physical servicing, uncrewed robotic servicing, robotic operation with occasional crew visits, campaign-based crew servicing, and semi-permanent or permanent onsite technicians.

\subsection{Architecture comparison}

Table~\ref{tab:humanarchitectures} compares the five service regimes on a common basis.  The comparison assumes robotic execution of routine work in every scalable architecture.  The differentiating variables are the recovery time for low-frequency nonstandard failures and the fixed infrastructure required to provide that response.

\begin{table}[h!]
\caption{Comparison of autonomous, robotic, and human-support architectures.  Routine high-volume work remains robotic in every scalable regime; the distinction is the recovery architecture for low-frequency nonstandard failures.}
\label{tab:humanarchitectures}
\centering
\small
\renewcommand{\arraystretch}{1.07}
\setlength{\tabcolsep}{4.5pt}
\begin{tabularx}{\textwidth}{@{}L{0.17\textwidth} L{0.30\textwidth} L{0.26\textwidth} Y@{}}
\toprule
Regime & Service model and response & Added infrastructure & Adoption criterion \\
\midrule
A: autonomous, no external service & software recovery, sparing, and load migration; failed hardware is retired until whole-node replacement & onboard redundancy and safing only & acceptable only when physical refresh and latent-failure flow remain small \\
B: uncrewed robotic baseline & local machinery and visiting servicers handle routine work in hours to days; difficult exceptions may persist for weeks or months & service depot, redundant robots, cargo interfaces, and orbital spares & robotic work and logistics intensity, exception probability, fleet availability, and resupply delay \\
C: occasional crew visit & same robotic baseline; crew launched for one high-value contingency, typically with weeks of mobilization & \SIrange{1}{5}{t} of docking, isolation, access, and worksite provisions; crew vehicle supplies habitation & contingency-visit criterion in \Eq{eq:visitcriterion} \\
D: campaign-based crew service & robotic normal operation; technicians clear backlogs and execute upgrades during 30--90 day campaigns & approximately \SIrange{20}{50}{t} service module or habitat plus crew vehicle & annualized campaign benefit exceeds transport, habitation, and campaign cost \\
E: semi-permanent or permanent technicians & robotic routine flow with continuous human oversight; accessible exceptions recovered in hours to days & approximately \SIrange{60}{130}{t}, \SIrange{10}{30}{t/yr} logistics, and \SIrange{80}{200}{kW} support power & workload and economic thresholds are both satisfied \\
\bottomrule
\end{tabularx}
\end{table}

A contingency event illustrates the availability leverage.  Suppose one event per year removes fraction $f=0.1$ of capacity.  The capacity availability attributable to that event is
\begin{equation}
A=1-f\frac{\tau}{T_y}.
\label{eq:humanrepairavailability}
\end{equation}
For repair delays of 90, 30, 7, and 1 days, $A$ is 97.53\%, 99.18\%, 99.81\%, and 99.973\%, respectively.  This does not imply that a resident crew produces four-nines cluster availability: it shows that human value is concentrated in rare events with large affected fraction and long autonomous recovery delay.  Fault containment and reserve remain necessary because even resident technicians cannot instantaneously repair a cluster-wide power, thermal, or software failure.

\subsection{Onsite labor model}

Let $N_t$ be standardized maintenance operations per MW-year, $\pex$ the probability that a procedure remains unresolved after automatic retry and alternate-robot recovery, $h_e$ human labor per unresolved event, and $h_c$ planned human campaign labor per MW-year.  With the normalization in \Eq{eq:PMWdefinition}, the onsite technical intensity is
\begin{equation}
\ell_H\equiv\frac{H_H}{\PMW}=N_t\pex h_e+h_c.
\label{eq:humanlabor}
\end{equation}
Using $N_t=350$--500, $\pex=10^{-3}$--$5\times10^{-3}$, $h_e=8$--20 h, and $h_c=20$--40 h gives the following low, nominal, and high cases
\begin{equation}
\begin{aligned}
\ell_H^{\rm low}&=350(10^{-3})(8)+20=22.8,\\
\ell_H^{\rm nom}&=400(10^{-3})(12)+30=34.8,\\
\ell_H^{\rm high}&=500(5\times10^{-3})(20)+40=90,
\end{aligned}
\qquad \left[\frac{\mathrm{crew\,h}}{\mathrm{MW\,yr}}\right].
\label{eq:humanlaborrange}
\end{equation}
The resulting range is 23--90 crew-h/(MW yr), with a nominal planning value near 35.  A bounding case with $\pex=10^{-2}$, $N_t=500$, $h_e=20$ h, and $h_c=40$ h reaches 140 crew-h/(MW yr), showing that robot exception probability rather than routine exchange time controls human intervention workload.

A four-to-six-person technical crew may provide approximately $6\times10^3$--$10^4$ productive technical hours per year after sleep, exercise, medical, training, housekeeping, administration, emergency drills, and non-maintenance duties.  The workload-only threshold for continuous occupation is
\begin{equation}
P_{\mathrm{MW},H,\mathrm{work}}^*\simeq
\frac{H_{\rm crew,prod}}{\ell_H}
\simeq67\text{--}440,
\qquad
P_{\mathrm{MW},H,\mathrm{work}}^{*,\rm nom}\simeq230,
\label{eq:humanworkthreshold}
\end{equation}
so that $P_{H,\rm work}^*=(\SI{1}{MW})P_{\mathrm{MW},H,\rm work}^*$.  Below this scale, the maintenance workload does not fill a resident crew's productive capacity unless the crew also supports other orbital assets or is retained principally for response latency.

\subsection{Economic threshold}

For unstructured event class $j$ with system rate $\Lambda_j$, affected capacity fraction $f_j$, autonomous or remote repair time $t_{A,j}$, and onsite-human repair time $t_{H,j}$, the human-attributable availability gain is
\begin{equation}
\Delta A=\frac{1}{T_y}\sum_j\Lambda_j f_j
\left(t_{A,j}-t_{H,j}\right).
\label{eq:humanavailabilitygain}
\end{equation}
Let $c_{R,h}$ be the fully allocated cost per robot-hour and $c_M$ the delivered-and-disposed logistics cost per tonne.  The annual robotic and logistics costs are
\begin{equation}
C_{\rm robot}=\PMW c_{R,h}\ell_R,
\qquad
C_{\rm logistics}=\PMW c_M\ell_M.
\label{eq:robotlogcost}
\end{equation}
If local humans change the corresponding intensities by $\Delta\ell_R$ and $\Delta\ell_M$, and provide other avoided upgrade or low-probability-event cost $s_0$, define the net annual benefit per delivered MW,
\begin{equation}
b_H\equiv v\Delta A+c_{R,h}\Delta\ell_R+c_M\Delta\ell_M+s_0,
\qquad
[b_H]=\mathrm{\$/(MW\,yr)}.
\label{eq:humanbenefitintensity}
\end{equation}
The human-support cost may contain fixed and size-dependent terms,
\begin{equation}
C_H(\PMW)=C_{H,0}+c_{H,1}\PMW.
\label{eq:humancostscale}
\end{equation}
Permanent presence breaks even when
\begin{equation}
\PMW b_H\geq C_{H,0}+c_{H,1}\PMW,
\label{eq:humanbreakeven}
\end{equation}
with economic threshold
\begin{equation}
P_{\mathrm{MW},H,\mathrm{econ}}^*=\frac{C_{H,0}}{b_H-c_{H,1}},
\qquad b_H>c_{H,1}.
\label{eq:humanthreshold}
\end{equation}
The applicable transition is
\begin{equation}
P_{\mathrm{MW},H}^*=\max\!\left(P_{\mathrm{MW},H,\mathrm{work}}^*,P_{\mathrm{MW},H,\mathrm{econ}}^*\right),
\qquad P_H^*=(\SI{1}{MW})P_{\mathrm{MW},H}^*,
\label{eq:combinedhumanthreshold}
\end{equation}
because a resident crew must be justified by both useful workload and economic value.  Total power is therefore a scale proxy, not the fundamental cause of the transition: the controlling quantities are $\ell_R$, $\ell_M$, $\pex$, the repair-time reduction in \Eq{eq:humanavailabilitygain}, $v$, transportation and logistics cost, and the fixed human-support cost.

\begin{figure}[htbp]
\centering
\begin{tikzpicture}
\begin{axis}[
width=0.60\textwidth,height=0.365\textwidth,
xlabel={Human-attributable availability gain $\Delta A$ (percentage points)},
ylabel={Permanent-crew break-even power (MW)},
xmin=0.2,xmax=2.2,ymin=100,ymax=2500,
legend pos=north east
]
\addplot[thick,domain=0.2:2.2,samples=180]
{500/(100*(x/100)+0.5)};
\addlegendentry{$C_{H,0}=0.5$ B\$/yr}
\addplot[thick,dashed,domain=0.2:2.2,samples=180]
{800/(100*(x/100)+0.5)};
\addlegendentry{$C_{H,0}=0.8$ B\$/yr}
\addplot[thick,dotted,domain=0.2:2.2,samples=180]
{1500/(100*(x/100)+0.5)};
\addlegendentry{$C_{H,0}=1.5$ B\$/yr}
\end{axis}
\end{tikzpicture}
\caption{Illustrative permanent-crew threshold from \Eq{eq:humanthreshold}, using $v=100$ million \$/MW-yr, $s=0.5$ million \$/MW-yr, and $c_{H,1}=0$.  The result is governed by uncertain fixed crew cost, repair-latency reduction, and the fraction of availability that humans actually recover; it is not a universal power threshold.}
\label{fig:humanthreshold}
\end{figure}
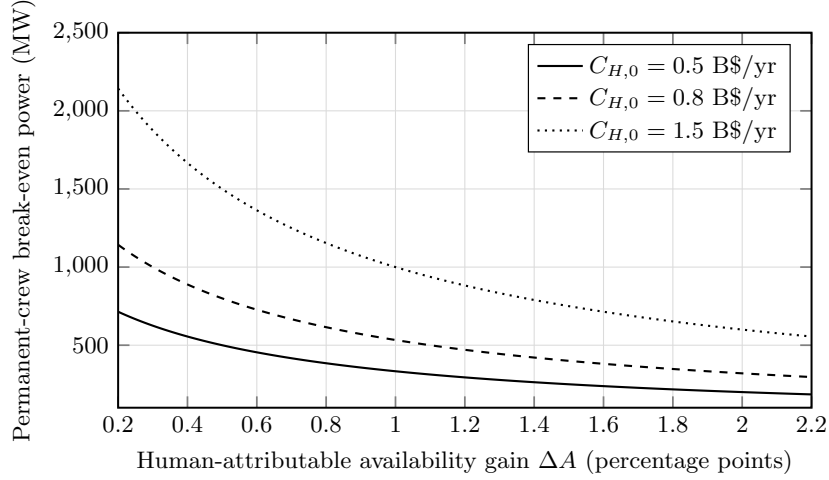

For the simplified fixed-cost case $c_{H,1}=0$, write $b_H=v\Delta A+s$ where $s$ aggregates the last three terms of \Eq{eq:humanbenefitintensity}.  Illustrative ranges $C_{H,0}=0.5$--1.5~billion \$/yr, $v=25$--150~million \$/MW-yr, $\Delta A=0.005$--0.02, and $s=0.2$--1.0~million \$/MW-yr produce a broad \SIrange{0.1}{5}{GW} envelope.  A central case $C_{H,0}=0.8$~billion \$/yr, $v=100$~million \$/MW-yr, $\Delta A=0.01$, and $s=0.5$~million \$/MW-yr gives
\begin{equation}
P_{\mathrm{MW},H,\mathrm{econ}}^*\simeq533,
\qquad
P_{H,\mathrm{econ}}^*\simeq\SI{533}{MW}.
\label{eq:humancentral}
\end{equation}
This is an economic sensitivity example, not a market forecast or an independently established transition scale.  Figure~\ref{fig:humanthreshold} shows the governing sensitivity: fixed crew-support cost and the human-attributable increment in delivered availability move the break-even scale through the several-hundred-megawatt to gigawatt regime, even before location-dependent transport and radiation costs are added.

For campaign-based servicing with campaign frequency $f_{\rm camp}$, mission cost $C_{\rm camp}$, and campaign logistics cost $C_{\rm log,camp}$, the annualized criterion is
\begin{equation}
f_{\rm camp}(C_{\rm camp}+C_{\rm log,camp})
<\PMW b_{\rm camp},
\label{eq:campaigncriterion}
\end{equation}
where $b_{\rm camp}$ is the per-MW annual value of cleared exceptions, accelerated upgrades, and restored availability.  A campaign avoids the continuous fixed cost $C_{H,0}$ but retains mobilization delay.

A short-duration contingency mission has a still lower threshold.  If a visit costing $C_v$ plus dedicated logistics $C_{\rm log,v}$ restores power $P_r$ by time $\Delta t$ earlier than robotic or replacement recovery, avoids replacement cost $C_{\rm repl,avoided}$, and saves asset value $V_a$, define $P_{r,\rm MW}=P_r/(\SI{1}{MW})$.  The mission is justified when
\begin{equation}
C_v+C_{\rm log,v}
<P_{r,\rm MW}v\Delta t+V_a+C_{\rm repl,avoided}.
\label{eq:visitcriterion}
\end{equation}
For $C_v+C_{\rm log,v}=\$200$ million, $v=\$100$ million/MW-yr, $\Delta t=0.25~\mathrm{yr}$, and $V_a=C_{\rm repl,avoided}=0$, the availability-only threshold is $P_{r,\rm MW}=8$, or $P_r=\SI{8}{MW}$.  Occasional crew intervention can therefore satisfy the cost criterion at tens of megawatts even when permanent presence does not.

\subsection{Human maintenance task allocation}

Resident or visiting humans should concentrate on tasks with high geometric or diagnostic entropy:
\begin{itemize}
\item diagnose failures that do not match known telemetry signatures;
\item recover a stuck robot, tangled restraint, or blocked translation path;
\item repair deformed latches, damaged connectors, irregular harnesses, or failed release mechanisms;
\item locate and contain ambiguous fluid leaks or contamination;
\item install structural patches and nonstandard MMOD protection;
\item integrate first-of-a-kind upgrades and modify equipment not designed for the new configuration;
\item conduct forensic inspection of fleet-significant failures;
\item manage high-risk battery, pressure, propellant, and return-cargo operations; and
\item intervene after common-mode software, power, or maintenance-infrastructure failures.
\end{itemize}
They should not manually replace routine compute cartridges, move hundreds of storage units, perform normal inventory counts, or conduct repetitive exterior surveys.  Even a permanently crewed ODC remains robotically maintained.

The reference campaign crew consists of four to six cross-trained specialists: robotics/mechanisms, power/avionics, thermal/fluids, compute/networking, extravehicular activity operations and safety leadership, and logistics/medical cross-coverage.  At GW scale, \Eq{eq:humanlaborrange} implies \SIrange{23000}{90000}{crew.h/yr}, with a nominal value near \SI{35000}{crew.h/yr}.  At 1500--1700 productive technical hours per person-year, this corresponds to approximately 14--60 technical personnel across shifts, nominally about 20--25, if the onsite fraction remains unchanged.  This crew should live in a separate servicing depot or habitat, not inside the compute structure, so that high voltage, laser terminals, propulsion, coolant, large mechanisms, and debris-sensitive arrays can be isolated from the habitable volume.

\subsection{Habitation mass, power, and safety requirements}

Human presence adds systems that do not contribute IT power: pressure vessels, environmental control and life support system (ECLSS) hardware, water recovery, food storage, waste management, exercise, hygiene, fire protection, medical capability, radiation shelter, extravehicular activity or suitport capability, crew interfaces, and continuous emergency-return capability.  NASA human-system standards apply to any vehicle, habitat, payload, and equipment with which crew interact and therefore materially change design assurance and operations \cite{NASA3001V1,NASA3001V2}.  ISS-derived systems have demonstrated approximately 98\% water recovery, but food, packaging, filters, clothing, medical items, hardware spares, and waste remain significant recurring logistics \cite{NASA2023WaterRecovery,Leach2021ISSLogistics,Wieland2005ECLSS}.

For a four-person crew, NASA's packaged-food reference allocation of approximately $1.82~\mathrm{kg/(person\,day)}$ implies about \SI{2.66}{t/yr} before non-food logistics \cite{Perchonok2009FoodMass}.  Water closure reduces but does not eliminate make-up water and ECLSS maintenance.  Historical ISS manifests also show that science/outfitting, ECLSS hardware, operational supplies, clothing, and hygiene are material fractions of delivered mass; reducing food alone does not close human logistics \cite{Leach2021ISSLogistics}.

A \SI{100}{t} crew facility amortizes to \SI{100}{kg/kW} at \SI{1}{MW}, \SI{10}{kg/kW} at \SI{10}{MW}, \SI{1}{kg/kW} at \SI{100}{MW}, and \SI{0.1}{kg/kW} at \SI{1}{GW}.  The crew-facility mass contribution falls to approximately 1 kg/kW at 100 MW, but recurring cost, human risk, and transportation remain dominant until several hundred megawatts or until the infrastructure is shared.  This scaling supports crew-compatible interfaces at the first MW without permanent habitation.

The first ODC should therefore include docking or berthing targets, grapple fixtures, clear approach corridors, hard power/propulsion/laser/fluid isolation, crew-readable labels, handrails or restraint points at selected worksites, and load paths for an attached vehicle.  It should not include a permanent atmosphere, sleeping quarters, food systems, exercise equipment, medical facilities, or full human rating of every compute subsystem.

\section{Operational geography}
\label{sec:geography}

Orbit changes both the physical plant and the maintenance architecture.  The relevant variables are sunlight and eclipse, charged-particle environment, debris and meteoroid flux, atomic oxygen and drag, station-keeping, communications latency and geometry, cargo travel time, crew abort capability, and the energy required to deliver or dispose of hardware.  A human-support architecture that satisfies the LEO cost threshold may fail the corresponding GEO or cislunar threshold even when local personnel reduce repair time.

\subsection{Latency and control regime}

The geometric one-way propagation delay is
\begin{equation}
t_\ell=\frac{d}{c}.
\label{eq:lighttime}
\end{equation}
A LEO vehicle directly overhead has a geometric one-way delay of a few milliseconds.  GEO altitude gives approximately \SI{119}{ms} one way and at least \SI{238}{ms} round trip before terrestrial routing and processing.  Mean Earth--Moon distance gives approximately \SI{1.28}{s} one way and \SI{2.56}{s} round trip.  Continuous force-reflecting teleoperation becomes progressively less stable and less efficient as delay approaches the human sensorimotor time scale; supervisory autonomy, guarded motion, local collision avoidance, and task-level commands become mandatory beyond LEO \cite{Sheridan1992Teleoperation}.

Latency does not imply that humans must be local.  A well-designed robot can execute a standardized task autonomously at lunar distance.  Latency increases the value of local humans specifically for unstructured diagnosis and improvisation, but transportation, radiation, and emergency-return penalties increase faster than that value unless a broader local industrial settlement already exists.

\subsection{Eclipse and energy-storage requirements}

For eclipse duration $t_{\rm ecl}$, the regulated bus energy is $E_{\rm ecl}=\Ptot t_{\rm ecl}$.  The reference \SI{1}{MW} IT load with $\alphaOH=1.25$ and a \SI{35}{min} eclipse requires approximately \SI{0.73}{MWh} delivered from storage.  At an end-of-life usable system specific energy of \SIrange{75}{150}{Wh/kg}, storage mass is approximately \SIrange{4.9}{9.7}{t/MW} before redundancy.  Replacing that battery inventory every six years adds roughly \SIrange{0.8}{1.6}{t/(MW.yr)} and tens of scheduled operations.  A high-sunlight dawn--dusk LEO architecture is therefore favorable not only for initial mass but also for maintenance throughput.

GEO has high sunlight duty over the year but seasonal eclipses approaching roughly an hour.  Storage is used intensively during eclipse seasons rather than every orbit, changing both battery power and cycle-life design.  Cislunar periodic orbits and Lagrange-region trajectories require trajectory-specific shadow and thermal analyses; ``continuous sunlight'' should never be assumed from location name alone.

\subsection{Regime comparison}

Table~\ref{tab:orbitcomparison} summarizes the operational consequence of location.  In this comparison, medium Earth orbit (MEO) denotes regimes above conventional LEO and below GEO.  High-sunlight LEO minimizes transport latency and provides the shortest contingency transit and abort times among the listed regimes; higher Earth orbits and cislunar regimes increasingly favor supervisory autonomy because radiation, abort, replenishment, and disposal penalties rise faster than the value of low-latency local intervention for a standalone ODC.

\begin{table}[tbp]
\caption{Operational geography and reference maintenance doctrine.  LEO, MEO, and GEO denote low, medium, and geostationary Earth orbit.  Radiation and debris levels remain mission-specific.}
\label{tab:orbitcomparison}
\centering
\small
\setlength{\tabcolsep}{4pt}
\renewcommand{\arraystretch}{1.07}
\begin{tabularx}{\textwidth}{@{}L{0.16\textwidth} L{0.24\textwidth} L{0.28\textwidth} Y@{}}
\toprule
Regime & Operational advantages & Sustainment penalties & Reference doctrine and human access \\
\midrule
High-sunlight LEO & lowest latency; shortest cargo and crew transit; possible eclipse suppression; frequent launch opportunities & debris and conjunctions, atomic oxygen, drag, large exposed area, and flexible-structure maneuvers & autonomous/robotic baseline; shortest abort distance; lowest contingency- and campaign-access thresholds \\
Ordinary eclipsing LEO & accessible and low latency & multi-tonne batteries, $10^4$--$10^5$ cycles, thermal cycling, drag, and debris & robotic service with larger battery-refresh flow; generally inferior for a continuous IT load \\
Higher LEO or MEO & lower drag in selected shells; wider coverage & stronger trapped radiation in many regimes, greater delivery energy, and slower rescue & autonomy dominant; crew threshold rises by several times and demonstrated service paths are sparse \\
GEO & fixed ground geometry, low drag, and communications colocation & high radiation, seasonal eclipse, long disposal horizon, and $\sim$  \SI{0.24}{s} geometric round trip & robotic servicing and life extension; crew only for exceptionally valuable shared infrastructure \\
Cislunar or Lagrange-region & low atmospheric drag; proximity to possible lunar industry; selected favorable illumination geometries & solar-particle and galactic-cosmic-ray exposure, multi-day logistics, $\sim$ \SI{2.6}{s} Earth--Moon round trip, and sparse rescue & strong autonomy and standardized logistics; local humans only within a broader cislunar program \\
Deep-space or heliocentric & potentially favorable sunlight and thermal-sink geometry at selected locations & long communication and replenishment delay, radiation, and no practical rapid rescue & essentially autonomous; design for graceful depletion and local manufacture \\
\bottomrule
\end{tabularx}
\end{table}

The LEO debris environment must be evaluated with directional and size-resolved engineering models such as ORDEM rather than a single flux number \cite{Vavrin2023ORDEM32}.  Large-area ODCs may also alter conjunction operations because their ballistic coefficient, flexible appendages, and pointing constraints differ from compact satellites.  A maintenance maneuver must be compatible with thermal pointing and robot safing; collision avoidance is therefore part of the compute scheduler and not only a flight-dynamics function.

Higher altitude does not monotonically improve life.  Reduced drag can be offset by trapped radiation, slower disposal, higher transport cost, and reduced crew access.  GEO adds the possibility of commercially valuable colocation with communications infrastructure, which may justify a communications-integrated compute payload and shared servicer even when a standalone terrestrial-compute case would not close.

\subsection{Location-dependent human threshold}

Let the annualized human-support cost at location $r$ be
\begin{equation}
C_H(r,\PMW)=C_{H,0}(r)+c_{H,1}(r)\PMW,
\label{eq:locationcrewc}
\end{equation}
where $C_{H,0}(r)$ includes the fixed habitat, transportation, radiation-protection, abort, and operations costs.  The economic threshold becomes
\begin{equation}
P_{\mathrm{MW},H}^*(r)=
\frac{C_{H,0}(r)}{b_H(r)-c_{H,1}(r)},
\qquad b_H(r)>c_{H,1}(r).
\label{eq:locationthreshold}
\end{equation}
Distance can increase the repair-time benefit in $b_H$ because local humans avoid communication and mobilization delay, but it increases transportation, radiation protection, abort, and logistics cost and may reduce $v$ for Earth-serving workloads.  For the stated standalone sensitivity cases, the net effect is an upward scenario boundary: approximately 0.3--1 GW in LEO can become multiple GW in GEO or cislunar space.  The exception is a shared habitat and transport network whose fixed cost is already paid by communications, science, power, or resource-extraction activities.

\section{Alternative concepts and comparative trade studies}
\label{sec:alternatives}

Alternative ODC concepts should be compared with the same metrics: delivered IT power, mass per delivered kW, radiator area and temperature, capacity availability, external communication intensity, annual logistics mass, robot-hours, correlated-fault domain, disposal mass and cost, and lifecycle cost.  Any subsystem, operation, or cost excluded from the stated system boundary must be allocated explicitly.

\subsection{Whole-node replacement versus orbital replacement units}

The simplest lifecycle architecture discards an entire node after failure or obsolescence.  It minimizes interfaces and robot complexity but couples electronics refresh to structure, photovoltaic arrays, radiators, propulsion, and avionics.  If ten \SI{5}{t} nodes per MW are replaced on a four-year compute cycle,
\begin{equation}
\Mdot_{\rm node,ref}=\frac{10(\SI{5}{t})}{4~\mathrm{yr}}
=\SI{12.5}{t/(MW.yr)}.
\label{eq:wholenoderefresh}
\end{equation}
Refreshing only \SI{5}{t/MW} of compute cartridges every four years requires \SI{1.25}{t/(MW.yr)}.  The order-of-magnitude difference dominates the approximately 5--10 kg/kW service-architecture increment, including first-line stock but excluding the reserve node, over a multi-year life.  Whole-node replacement is appropriate for early demonstrations and catastrophic loss; it exceeds the reference logistics limits as a primary refresh strategy beyond a few megawatts.

\subsection{Free-flying constellation versus structural spine or tether}

A free-flying constellation provides strong physical fault isolation, independent launch and disposal, and flexible growth.  It pays repeated avionics, propulsion, metrology, and inter-satellite-link mass and requires frequent formation and conjunction management.  A structural spine or tether can share power, thermal transport, communications, and service translation and can provide passive gravity-gradient or sun-pointing behavior in selected orbits.  Tether-based multi-megawatt concepts have been proposed as a way to organize long distributed arrays \cite{Bargatin2025Tether}.  Their principal risks are flexible dynamics, deployment, collision-induced oscillation, tether severance, electrical grounding and charging, and a potentially large correlated fault domain.

The preferred industrial architecture may be hybrid: independently survivable 50--250 kW nodes temporarily berth to a passive logistics spine for servicing and high-capacity networking, but retain enough propulsion, power-safe capability, and thermal control to separate after a depot failure.  A rigid or tethered structure should not become the only path for coolant, power, or command unless its failure probability is demonstrably below the cluster domain-hazard budget.

\subsection{Photovoltaic, nuclear, and beamed-power supply}

Photovoltaics are technologically mature and scale directly with delivered electric power, but require thousands of square meters, pointing, deployment, degradation margin, and large structural inertia.  Fission power removes sunlight and eclipse dependence but introduces reactor, conversion machinery, shielding, high-temperature materials, launch safety, and a second thermal-rejection problem.

For electrical output $P_e$ and conversion efficiency $\eta_e$, reactor thermal power is
\begin{equation}
P_{\rm rx}=\frac{P_e}{\eta_e},
\qquad
Q_{\rm conv}=P_e\left(\frac{1}{\eta_e}-1\right).
\label{eq:nuclearheat}
\end{equation}
For $P_e=\SI{1.25}{MW}$ and $\eta_e=0.25$, $P_{\rm rx}=\SI{5}{MW}$ and conversion waste is \SI{3.75}{MW}, in addition to the \SI{1.25}{MW} ultimately dissipated by the ODC.  Rejecting all \SI{5}{MW} near \SI{350}{K} would require approximately four times the baseline radiator area.  A viable nuclear architecture must reject conversion heat at much higher temperature, for example several hundred square meters near \SI{700}{K}, while retaining the low-temperature IT radiator.  Fission becomes technically competitive only if high-temperature conversion, radiator systems, shielding geometry, and long-life rotating or static conversion close simultaneously.  It does not remove the heat-rejection requirement.

Beamed power moves generation mass to another platform or ground site and replaces PV area with receiver area, pointing, conversion loss, and an external service dependency.  It may be attractive inside a mature orbital power market, but a standalone ODC cannot credit the source platform as massless or free.  Beam interruption also becomes a correlated power hazard unless local ride-through and multiple sources are provided.

\subsection{Single-phase, two-phase, and passive thermal transport}

Passive heat pipes and loop heat pipes minimize moving parts and are attractive between chips, cold plates, and a replaceable cartridge boundary.  Long-distance MW transport generally requires many parallel paths or pumped loops.  Single-phase loops are comparatively diagnosable and controllable but incur pump power and larger mass flow,
\begin{equation}
\dot m=\frac{Q}{c_p\Delta T}.
\label{eq:singlephaseflow}
\end{equation}
Two-phase systems exploit latent heat and reduce mass flow, but introduce flow-regime instability, inventory control, noncondensable gas sensitivity, freeze/thaw concerns, and more difficult leak recovery.  The technically conservative first MW uses passive local spreading plus multiple isolated single-phase or well-characterized two-phase loops, rather than one novel cluster-wide thermal network.

\subsection{Specialized machinery versus general-purpose humanoid robots}

A general-purpose dexterous robot is valuable for exceptions, tool recovery, and equipment not designed for service.  It is inefficient for thousands of identical cartridge operations.  Specialized magazines, rails, automatic latches, and insertion machines offer higher speed, lower sensing complexity, smaller work envelopes, and easier statistical qualification.  The preferred fleet is heterogeneous: high-cycle dedicated machinery for L1 work, fixed dexterous arms for L2/L3 tasks, mobile inspection and transport robots, and a small number of general-purpose systems for exceptions.  Humanoid morphology is not an objective unless legacy human interfaces must be serviced.

\subsection{Pressurized compute versus unpressurized robotic service}

A pressurized compute hall permits conventional human access and some use of terrestrial hardware, but pressure containment, atmosphere, fire, contamination, and airlock systems add mass and risk while the external radiator remains.  Pressure also makes every penetration and service door a crew-safety interface.  The unpressurized architecture permits higher modularity, direct external rejection, and lower human-rating scope.  A separate pressurized service vehicle can bring humans, tools, and temporary environmental protection to selected work sites.  This split architecture is strongly preferred until human occupancy is continuous and the value of manual access is demonstrated.

\subsection{Common-metric trade summary}

Table~\ref{tab:concepttrade} reduces the alternatives to common lifecycle metrics.  The quantitative comparison favors cartridge-level ORU refresh, fault-isolated free-flying nodes, photovoltaic power in high-sunlight LEO, and specialized service machinery for the first industrial generation; whole-node refresh and a pressurized compute hall are not used as baselines because their recurring mass or habitation requirements exceed the benefit of reduced interface complexity.

\begin{table}[htbp]
\caption{Comparative technical judgment for major orbital data center alternatives.  Replacement levels L1--L3 denote cartridge, subsystem, and external-hardware tiers.}
\label{tab:concepttrade}
\centering
\small
\renewcommand{\arraystretch}{1.10}
\begin{tabularx}{\textwidth}{@{}L{0.20\textwidth} L{0.22\textwidth} L{0.26\textwidth} Y@{}}
\toprule
Concept & Main advantage & Principal limit & Technical judgment \\
\midrule
Whole-node disposal & minimum service interfaces & 10--15 t/(MW yr) refresh flow and large outage block & acceptable pathfinder strategy; not industrial baseline \\
Cartridge-level rolling refresh & order-of-magnitude lower refresh mass; small fault domains & high-cycle connector and robot qualification & essential at \SI{10}{MW} and above \\
Free-flying nodes & isolation, modular launch, independent disposal & duplicated bus mass and formation operations & preferred first architecture \\
Tether/spine & shared structure, service translation, possible passive geometry & flexible dynamics and correlated structural domain & promising hybrid after full-scale dynamics and severance tests \\
Photovoltaic power & mature conversion, modular arrays & large deployed area, degradation, eclipse sensitivity & baseline in high-sunlight low Earth orbit \\
Fission electric & continuous power independent of sunlight & conversion efficiency, two-temperature radiators, shielding, safety & potentially competitive at high scale; requires separate high-temperature heat rejection \\
Beamed power & shifts generation and enables shared utility & external dependency, pointing, receiver and source cost & long-term ecosystem option \\
Specialized service machinery & high throughput and qualifiable cycles & narrow task set & primary cartridge-level maintenance approach \\
General-purpose dexterous robot & handles exceptions and novel geometry & lower throughput and harder validation & reserve and subsystem/external-unit role \\
Pressurized compute hall & direct human access & large mass, fire/atmosphere risk, no radiator elimination & generally rejected \\
Separate crew depot & bounds human-rating scope and hazards & docking and translation complexity & preferred future human architecture \\
\bottomrule
\end{tabularx}
\end{table}

\section{Representative mission scenarios and industrial evolution}
\label{sec:missions}

Applying the same modular architecture at one, ten, one hundred, and one thousand MW produces different first-order operating requirements.  The linear mass, flow, and workload scalings are not forecasts because cargo cadence, depot geography, plane changes, gateway infrastructure, conjunction-management demand, local refurbishment, and common modes introduce nonlinearities.  The values below preserve \SI{100}{kW} fault domains for availability accounting, but an industrial implementation may package several domains on one larger spacecraft or service frame.  Fault-domain count should not be confused with launch-unit count.

\subsection{Scenario A: one-MW high-sunlight LEO pathfinder}

The first operational system consists of ten active and one reserve \SI{100}{kW} nodes, two or more gateway functions, a small service/warehouse module, three or four mutually recoverable robots, and class-specific common spares sized to the six-month replenishment planning horizon.  Total deployed mass including maintainability and reserve is $\sim$\SIrange{50}{75}{t}.  The \SIrange{5.3}{9.0}{t/yr} logistics envelope corresponds to roughly one to two annual cargo flights only for a vehicle delivering about 5--10 t of useful manifested payload; actual cadence follows \Eq{eq:cargocadence}.  The cluster is uncrewed; a visiting vehicle is not part of nominal operations.

The primary objectives are lifecycle validation and service characterization through completion of at least one rolling compute refresh, accumulating thousands of robot operations and connector cycles across the fleet, replacing pumps and converters, isolating a radiator-panel fault, demonstrating full-node reserve activation, and returning representative failures for forensic analysis.  Commissioning should intentionally exercise alternate command paths, loss of gateway, robot recovery, inventory discrepancy, and cyber rollback before high-value workloads are admitted.

A suitable operational sequence is:
\begin{enumerate}
\item deploy and commission nodes independently, initially at reduced thermal load;
\item establish optical mesh and out-of-band command paths;
\item commission the service module, warehouse, and robot metrology;
\item activate compute in 10--20\% increments while calibrating thermal and structural models;
\item hold one node unloaded as cold reserve until fleet hazard is characterized;
\item perform quarterly planned maintenance windows and continuous cartridge rolling replacement;
\item execute an integrated refresh campaign in mission years three to five; and
\item retire nodes or panels through controlled containerized disposal, not free release.
\end{enumerate}

\subsection{Scenario B: ten-MW preindustrial cluster}

A ten-MW cluster contains roughly one hundred active 100-kW domains and 5--10\% pooled reserve.  Deployed mass is approximately \SIrange{0.5}{0.75}{kt}, and annual logistics are \SIrange{53}{90}{t}.  Mean robotic work corresponds to only two productive robot equivalents, but five to seven deployed systems are required for geographic coverage, specialization, and mutual recovery.  A central warehouse of tens of tonnes and a dedicated visiting robotic servicer are required by the reference logistics and recovery architecture.

At this scale, the contingency-visit criterion can be satisfied if several megawatts remain unavailable for months.  It is not enough work to justify permanent technicians.  The principal stage gate is whether the service architecture can achieve $\pex\lesssim10^{-3}$, whole-node hazard below a few percent per year, and annual logistics below approximately \SI{8}{t/(MW.yr)} while preserving high utilization.

A communications-integrated GEO variant can also be evaluated at this scale if compute shares a relay architecture and processes network traffic, routing, caching, security, or satellite data.  The maintenance doctrine remains robotic because crew access is difficult.  A GEO servicer can carry rare ORUs and mission-extension hardware, but compute modules must tolerate longer replenishment and stronger radiation.

\subsection{Scenario C: one-hundred-MW LEO industrial facility}

At \SI{100}{MW}, approximately one thousand 100-kW fault domains deliver the load, although many may be grouped into larger serviced platforms.  Deployed mass is of order \SIrange{5}{7.5}{kt}, annual logistics are \SIrange{0.53}{0.90}{kt}, and a first-order capacity-sized fleet is approximately 25--35 robots before mission-specific geographic and task-class refinement.  Cargo arrives weekly or biweekly in reusable standardized containers.  Automated warehouse management, in-orbit acceptance testing, cartridge refurbishment, robot overhaul, and radiator/solar-panel repair become core infrastructure.

Periodic four-to-six-person technical campaigns of 30--90 days may satisfy the campaign criterion for backlog clearance, major upgrades, forensic work, and repair of the service infrastructure.  A permanently occupied habitat is not yet automatically justified: \Eq{eq:humanthreshold} depends on compute value and shared infrastructure.  The reference intermediate architecture is a small uncrewed servicing depot with intermittent crew occupancy.

The cluster should be divided into several independently maneuverable and cyber-isolated industrial zones.  A single logistics depot, software authority, thermal-fluid lot, or gateway region must not cover the full 100 MW.  At least two warehouses and service cells should be able to support survival-critical work if one zone is unavailable.

\subsection{Scenario D: gigawatt orbital-compute infrastructure}

A one-GW system contains ten thousand 100-kW fault domains, perhaps organized into 1--10 MW platforms and several geographically separated service complexes.  A linear mass extrapolation gives \SIrange{50}{75}{kt} deployed and \SIrange{5.3}{9.0}{kt/yr} logistics.  This corresponds to \SIrange{14.5}{24.7}{t/day} of replacement, refresh, packaging, and consumables.  Mean work is 160--200 productive robot equivalents, while the stated utilization margin alone raises capacity-sized demand to approximately 229--333 before task specialization and recovery reserve.  Continuous cargo flow, multiple tugs, orbital traffic management, local repair shops, standardized utility interfaces, and extensive inventory pooling are unavoidable.

In the stated high-value, shared-infrastructure sensitivity case, a resident technical workforce can become economically admissible in LEO at this scale.  The human role remains exception handling and industrial integration.  Chassis reuse, material recycling, local manufacture of simple structural and thermal parts, and automated requalification may reduce Earth upmass, but semiconductor fabrication remains an extremely ambitious separate industry.  Geographic separation is essential: a GW should not be treated as one fault domain, one orbit plane, one credential hierarchy, or one debris-risk concentration.

Table~\ref{tab:scenarios} collects the resulting scale progression.  The one-MW mission is primarily a lifecycle-validation experiment, ten megawatts establishes pooled logistics and a dedicated servicer, one hundred megawatts requires orbital industrial operations and can justify periodic crew campaigns, and a gigawatt requires port-like continuous material flow and multiple independent depots.

\begin{table}[htbp]
\caption{Representative scale progression for a high-sunlight low-Earth-orbit orbital data center.}
\label{tab:scenarios}
\centering
\small
\renewcommand{\arraystretch}{1.10}
\begin{tabularx}{\textwidth}{@{}L{0.075\textwidth} C{0.10\textwidth} C{0.115\textwidth} C{0.09\textwidth} C{0.15\textwidth} Y@{}}
\toprule
Delivered IT power & Deployed mass & Annual logistics & Deployed robots & Human doctrine & Required new capability \\
\midrule
\SI{1}{MW} & 50--75 t & 5.3--9.0 t/yr & 3--4 & none nominal & standardized replacement units, class-specific 6-month stock planning, reserve node, and one full refresh \\
\SI{10}{MW} & 0.5--0.75 kt & 53--90 t/yr & 5--7 & contingency visit possible & central warehouse, dedicated servicer, pooled node reserve \\
\SI{100}{MW} & 5--7.5 kt & 0.53--0.90 kt/yr & 25--35 & campaign criterion may be satisfied & automated warehouse, repair shops, multiple service zones, regular cargo \\
\SI{1}{GW} & 50--75 kt & 5.3--9.0 kt/yr & $\sim$250--350 & workforce criterion may be satisfied & continuous cargo operations, tugs, recycling, local manufacture, multiple depots and habitats \\
\bottomrule
\end{tabularx}
\end{table}

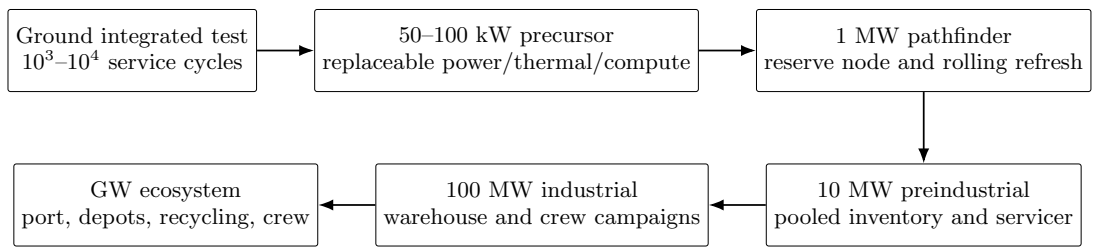
\begin{figure}[htbp]
\centering
\resizebox{0.80\textwidth}{!}{%
\begin{tikzpicture}[
phase/.style={draw,rounded corners=1pt,minimum width=2.7cm,minimum height=1.15cm,align=center,font=\small},
arr/.style={-{Latex[length=2.1mm]},line width=0.7pt},
node distance=8mm
]
\node[phase] (g) {Ground integrated test\\$10^3$--$10^4$ service cycles};
\node[phase,right=of g] (p) {50--100 kW precursor\\replaceable power/thermal/compute};
\node[phase,right=of p] (m1) {1 MW pathfinder\\reserve node and rolling refresh};
\node[phase,below=10mm of m1] (m10) {10 MW preindustrial\\pooled inventory and servicer};
\node[phase,left=of m10] (m100) {100 MW industrial\\warehouse and crew campaigns};
\node[phase,left=of m100] (gw) {GW ecosystem\\port, depots, recycling, crew};
\draw[arr] (g) -- (p);
\draw[arr] (p) -- (m1);
\draw[arr] (m1) -- (m10);
\draw[arr] (m10) -- (m100);
\draw[arr] (m100) -- (gw);
\end{tikzpicture}%
}
\caption{Capability-gated development sequence.  Each transition is justified by measured lifecycle performance, not only by successful power-up.  The displayed cycle counts are development exposures, not automatic qualification of $p_{\rm exc}$ or $p_U$; failure-rate, logistics, recovery, and common-mode evidence must be met before scaling the next order of magnitude.}
\label{fig:roadmap}
\end{figure}

Figure~\ref{fig:roadmap} expresses this progression as capability gates rather than calendar milestones.  The next power decade is entered only after measured failure rate, robot exception rate, logistics intensity, inventory performance, and common-mode containment close at the preceding scale.

\subsection{Cislunar shared-infrastructure scenario}

Under the present cost model, a cislunar ODC has a lower allocated fixed cost when integrated with a larger logistics and habitation system than when operated as a standalone Earth-cloud substitute.  Potential workloads include lunar navigation and communications, processing of surface and orbital sensor data, autonomous planning, telescope data reduction, and control of resource-extraction or manufacturing systems.  These workloads have location value and may tolerate lower space-to-Earth communication intensity.  The ODC would use supervisory autonomy, long-duration spares, robust radiation protection, and local tugs.  Resident technicians satisfy the cost criterion only if habitats, transport, medical capability, and emergency response are already supported by other activities; assigning their full fixed cost to the ODC raises the threshold above the reference early-deployment scales.

\section{Technology requirements and development pathways}
\label{sec:technology}

Technology readiness is system-specific.  A robotic arm, connector, radiator, or optical terminal may be flight proven individually while the integrated high-cycle service process remains immature.  The relevant question is whether the full chain---diagnosis, safing, access, manipulation, mating, leak and electrical test, software acceptance, inventory update, and failed-unit disposition---has been demonstrated at the required repetition and environmental scale.

\subsection{Demonstrated, extrapolated, and speculative capability}

Table~\ref{tab:readiness} separates flight heritage of constituent technologies from readiness of the integrated ODC service process.  Ground-teleoperated manipulation and rendezvous are comparatively mature, whereas high-cycle blind-mate thermal interfaces, autonomous recovery, robot cross-repair, and a complete maintainable MW cluster remain low-readiness integrated capabilities.

\begin{table}[htbp]
\caption{Readiness classification for major orbital data center lifecycle capabilities.  Approximate readiness levels are indicative and depend on the exact implementation.}
\label{tab:readiness}
\centering
\small
\renewcommand{\arraystretch}{1.10}
\begin{tabularx}{\textwidth}{@{}L{0.24\textwidth} C{0.10\textwidth} L{0.27\textwidth} Y@{}}
\toprule
Capability & Approx.\ readiness & Evidence boundary & ODC-specific gap \\
\midrule
Ground-teleoperated external manipulation and replacement-unit handling & 7--9 & International Space Station robotics, Dextre, Hubble servicing, and refueling demonstrations & autonomous repetition, uncrewed recovery, higher throughput \\
Rendezvous, inspection, docking, and life extension & 7--9 & cooperative and noncooperative inspection/docking missions & cargo transfer and service of large flexible ODC nodes \\
High-rate optical links and inter-satellite networking & 7--9 for constituent links & flight demonstrations and commercial optical inter-satellite links & aggregate sustained throughput, gateway scaling, serviceable terminals \\
Radiation-tolerant commercial compute with error-correcting codes and watchdogs & 4--7 & component and module tests; limited flight use & MW population, high temperature, long lifetime, lot control \\
Replaceable 5--10 kW compute cartridge & 3--5 & terrestrial liquid-cooled modules and spacecraft replacement-unit heritage & vacuum-qualified blind-mate power/optical/thermal interface \\
Modular high-power radiator with robotic isolation and replacement & 3--5 & spacecraft radiators and servicing components exist separately & hundreds of square meters per node, puncture recovery, repeated fluid mating \\
Autonomous high-cycle cartridge machinery & 3--4 & industrial automation and laboratory servicing rigs & microgravity dynamics, contamination, class-specific exception and terminal-recovery qualification \\
Robot repairing another service robot & 2--4 & modular terrestrial robots and limited space demonstrations & passive recovery, overlapping reach, flight-qualified joint exchange \\
One-MW uncrewed maintainable cluster & 2--4 & subsystem analyses and smaller spacecraft heritage & integrated power, thermal, structure, communications, logistics, and operations \\
Orbital repair shop and chassis refurbishment & 2--4 & ISS maintenance and ground industrial practice & automated inspection, requalification, contamination control \\
Bulk orbital recycling and local ODC-part manufacture & 1--3 & materials-processing experiments and terrestrial analogs & industrial throughput, quality assurance, feedstock variability \\
Permanent ODC technical workforce & 2--4 as an integrated concept & human spaceflight and industrial maintenance exist separately & economic justification, hazard separation, dedicated transport and return \\
\bottomrule
\end{tabularx}
\end{table}

The most mature demonstrations establish the feasibility of individual servicing tasks.  They do not establish the statistical reliability, cost, or throughput required by an ODC.  Conversely, the most speculative items---orbital recycling, robot self-repair at fleet scale, and resident industrial crews---are not required for the first MW.

\subsection{Performance thresholds}

Table~\ref{tab:requirements} is the principal stage-gate specification derived from the preceding physics, reliability, logistics, and robotics analysis.  The decisive thresholds are the catastrophic and defined large-domain event allocations, the post-internal-recovery exception probability, a separately mission-allocated terminal-unresolved probability, service-capacity margin, class-specific inventory coverage, and bounded logistics intensity; meeting only the nominal exchange-time targets is insufficient.

\begin{table}[htbp]
\caption{Quantitative stage-gate requirements for a first-generation maintainable orbital data center.  L1 and L2 denote cartridge- and subsystem-level replacement tiers.}
\label{tab:requirements}
\centering
\small
\renewcommand{\arraystretch}{0.98}
\begin{tabularx}{\textwidth}{@{}L{0.33\textwidth} L{0.26\textwidth} Y@{}}
\toprule
Metric & Requirement or objective & Rationale \\
\midrule
Catastrophic \SI{100}{kW} node hazard & $\leq0.01$--$0.03~\mathrm{yr^{-1}}$ & keeps whole-node upmass below other lifecycle flows \\
Permanent or long-duration large-domain event rate & $\lesssim10^{-3}~\mathrm{yr^{-1}}$, with $F_d$ and $T_d$ specified & prevents an undefined event-rate target from hiding affected fraction or duration \\
Post-internal-recovery robotic exception rate $p_{\rm exc}$ & $\leq10^{-3}$; goal $10^{-4}$ & bounds external-recovery workload from 400 operations/MW-year \\
Terminal-unresolved operation probability $p_U$ & mission allocated; homogeneous 1-MW, 12-year diagnostic $\leq1.07\times10^{-5}$ for 5\% any-event risk & prevents post-recovery failures from being conflated with recoverable exceptions \\
L1 service migration interruption & $<15$ min & keeps user impact small before physical replacement \\
L1 physical exchange and test & $<4$ h nominal & supports hundreds of operations with modest robot capacity \\
L2 power/thermal ORU replacement & $<24$ h nominal & permits one-domain service without long reserve occupancy \\
Full reserve-node activation & $<24$ h; stretch $<1$ h & decouples service availability from Earth resupply \\
Common-spare coverage & six-month class-specific planning horizon at 99\% cycle service level, plus explicit common-cause sets & tolerates nominal lead time without claiming a system-wide stockout guarantee \\
Robot service capacity margin & $\chi_R\geq1.5$ at 95th-percentile demand & avoids queue divergence and permits robot maintenance \\
Critical-work robot fleet availability & $\geq99.95\%$ & prevents the service system from becoming a dominant outage source \\
L1 connector qualification & $\geq20$ full mate cycles at worst-case alignment/temperature & covers ground test, retries, and mission exchanges \\
Service-machine motion life & $\geq10^5$ representative cycles & supports multi-MW shared machinery with margin \\
Coolant-interface leakage after mate & $\lesssim10^{-8}~\mathrm{kg/s}$ aggregate per isolated loop, or a loop-specific demonstrated equivalent & limits multi-year fluid loss and contamination \\
Maintainability hardware increment & $\leq7.5~\mathrm{kg/kW}$, excluding stock and reserve & keeps the hardware allowance consistent with the explicit mass allocation \\
First-line stocked-ORU allocation & $\leq2.5~\mathrm{kg/kW}$ for the stated planning horizon & prevents inventory mass from being hidden inside maintainability hardware \\
First-generation logistics intensity & $\leq8~\mathrm{t/(MW\,yr)}$ & bounds cargo cadence and cost \\
Mature logistics objective & $3$--$5~\mathrm{t/(MW\,yr)}$ & requires chassis reuse and low catastrophic hazard \\
Fleet false alerts & $<0.1$ actionable false alert/day per service zone & prevents operator saturation \\
Update exposure before rollback & $\leq$ one 5--10\% fault domain & bounds cyber/software common-mode loss \\
\bottomrule
\end{tabularx}
\end{table}

These thresholds are coupled.  For example, a four-hour exchange requirement is meaningful only if the exception metric includes ordinary alternate-robot recovery, the terminal-unresolved state is separately allocated, and reserve capacity masks the four hours.  A low leak rate is meaningful only if valves isolate the failed loop and sensors can distinguish interface leakage from inventory-estimation bias.

\subsection{Development program}

A technically efficient program begins with the interfaces and operations architecture rather than a monolithic MW demonstration:
\begin{enumerate}
\item \textit{Integrated ground service cell.}  Operate flight-like cartridges, pumps, valves, optical connectors, robot arms, magazines, and warehouse software in thermal-vacuum and representative structural dynamics.  Accumulate at least $10^4$ standard cycles and hundreds of deliberately off-nominal cases.
\item \textit{Environmental and radiation qualification.}  Test complete powered compute cartridges, not only individual devices, across TID, heavy-ion/proton SEE, thermal cycles, vacuum, contamination, and launch vibration.  Establish lot acceptance and post-irradiation performance-per-watt degradation.
\item \textit{Free-flying robotic precursor.}  Demonstrate inspection, capture, ORU exchange, fluid or dry thermal mating, alternate-tool recovery, failed-robot retrieval, and cargo transfer without crew proximity; treat the exposure as mechanism validation rather than statistical qualification of $p_U$.
\item \textit{50--100 kW maintainable node.}  Fly independently isolatable power, thermal, compute, and avionics ORUs with a replaceable radiator sector and at least one full technology upgrade.
\item \textit{One-MW pathfinder.}  Validate reserve activation, pooled logistics, multi-node software rollout, high-rate internal networking, one full rolling refresh, and controlled disposal.
\item \textit{Ten-MW preindustrial demonstration.}  Validate dedicated cargo/servicer operations, pooled inventory economics, multi-robot queueing, zone-level cyber isolation, and a contingency human-compatible worksite.
\end{enumerate}

The stage gates are data-driven.  A successful launch and year of nominal computation do not justify scaling if robot exceptions, logistics intensity, or correlated faults remain above the requirements in \Tab{tab:requirements}.

\subsection{Enabling research priorities}

Priority research topics are:
\begin{itemize}
\item high-temperature, radiation-tolerant computing and memory that preserve performance per watt while allowing hotter radiators;
\item high-cycle, self-aligning, debris-tolerant electrical, optical, and thermal interfaces with built-in verification;
\item modular radiator sectors with autonomous leak localization, isolation, and robotic replacement;
\item task-level autonomy with formal safety envelopes, uncertainty-aware perception, and recovery from partial insertion or latch failure;
\item robot modularity, passive recovery fixtures, and cross-repair between mechanically diverse service systems;
\item fleet Bayesian reliability estimation, digital twins, and configuration-controlled prognostics;
\item low-mass standardized cargo, return, and disposal containers compatible with multiple vehicles;
\item secure update, key hierarchy, and independent safety-control architectures that bound correlated cyber loss; and
\item automated acceptance testing and requalification of refurbished orbital hardware.
\end{itemize}

\section{Risks, limitations, and feasibility constraints}
\label{sec:risks}

The analysis supports a conditional maintainability architecture but does not establish overall commercial viability.  The underlying ODC must still close power, thermal, communication, lifetime, launch, and build-cost constraints.  Maintenance can increase delivered life but does not reduce the lower bounds set by radiator area, space-to-ground throughput, and mass per delivered power.

\subsection{Fundamental and near-fundamental constraints}

Several limits are not likely to disappear through operations innovation:
\begin{enumerate}
\item essentially all IT and platform electrical power becomes heat that must be radiated;
\item radiator heat flux scales as $T^4$ and is bounded by semiconductor junction temperature, thermal resistance, and environment;
\item eclipse energy is proportional to power and eclipse duration;
\item propagation delay is set by distance;
\item large exposed area creates impact and flexible-structure consequences that scale with area and span; and
\item Earth-serving workloads must cross a finite sustained communications interface.
\end{enumerate}
These constraints can be managed by orbit, workload, temperature capability, and architecture, but not eliminated.

\subsection{Engineering and program risks}

Table~\ref{tab:riskmatrix} translates the feasibility limits into program controls.  The dominant near-term risks are thermal common cause, unresolved robotic operations, correlated maintenance-software faults, and premature scale-up before low-probability lifecycle outcomes have been quantified; each requires architectural containment rather than margin alone.

\begin{table}[htbp]
\caption{Principal risks and the design response required before industrial scaling.}
\label{tab:riskmatrix}
\centering
\small
\renewcommand{\arraystretch}{0.98}
\begin{tabularx}{\textwidth}{@{}L{0.22\textwidth} L{0.22\textwidth} L{0.17\textwidth} Y@{}}
\toprule
Risk & Consequence & Pre-mitigation likelihood & Required response \\
\midrule
Radiator or thermal-loop common failure & rapid multi-node derating or loss & medium & multiple isolated loops, local thermal inertia, replaceable panels, independent fluid lots \\
Deployment or flexible-structure instability & inability to reach full power; collision or pointing loss & medium & incremental deployment, ground dynamic test, active damping, independent nodes \\
Robot task exceptions above $10^{-3}$ & growing backlog and frequent crew/replacement demand & high for first generation & dedicated machinery, interface redesign, alternate tools, statistical qualification \\
Robot fleet common software fault & maintenance paralysis & medium & diverse controllers, immutable recovery, passive capture, visiting servicer \\
Compute radiation or high-temperature degradation & low utilization and high cartridge flow & medium--high & powered module tests, shielding and derating, error-correcting codes, lot control, and graceful throttling \\
Catastrophic node hazard above 0.03/yr & whole-node logistics dominates & medium & fault containment, propulsion/power redundancy, precursor node-years \\
Launch or cargo interruption & stockout and extended reserve occupancy & medium & six-to-twelve-month stock, multiple providers, pooled rare spares \\
Cyber compromise or update common mode & correlated loss or unsafe actuation & medium & isolated safety plane, canaries, two-authority command, diverse images and keys \\
Debris-environment deterioration & rising impact and maneuver rate & uncertain & multiple orbital zones, replaceable shielding, current environment models, disposal discipline \\
Business case eroded by terrestrial improvement & stranded orbital infrastructure & high uncertainty & prioritize space-native and communications-integrated workloads and modular redeployment \\
Human accident or medical event & loss of crew and facility & low frequency, extreme consequence & separate habitat, safe-return redundancy, minimize required exposure \\
\bottomrule
\end{tabularx}
\end{table}

The highest programmatic risk is premature scale.  A one-MW pathfinder can survive a high logistics intensity or frequent engineering attention; a 100-MW facility cannot.  Scaling before interfaces and robot exception rates are measured converts each unresolved detail into hundreds of annual events.

\subsection{Model limitations}

The present model has five principal limitations.  First, the component allocation is not a flight bill of materials and is not derived from a specific accelerator architecture.  Second, hazard rates are design assumptions rather than on-orbit measurements.  Third, robot task times and exception probabilities require integrated testing in representative dynamics.  Fourth, transportation and human costs are parametric and may change by orders of magnitude.  Fifth, market value $v$ depends strongly on workload, utilization, data movement, and contract structure.

The analysis also assumes that failures are detected before they propagate, that removed units can be safely stored, and that regulatory and traffic-management approval exists for frequent rendezvous, disposal, and large flexible structures.  The 50--75 t/MW planning cases do not close primary structure and deployment, complete thermal-hydraulic networks and degraded-loop states, service-depot and robot-body mass, attitude-resolved drag and station keeping, rendezvous and docking interfaces, propulsion and installed-plant disposal, shielding, harness and power distribution, flexible-body control, MMOD containment, or growth and qualification margin.  Mission-specific studies must also include debris casualty risk, spectrum and optical coordination, launch and reentry safety, export and data-security constraints, and liability allocation.

\subsection{Feasibility judgment}

The present first-order power, storage, radiator, service-flow, and logistics balances are mutually consistent; however, an integrated spacecraft closure has not been demonstrated.  The 50--75 t/MW values are planning allocations rather than closed flight masses.  The least mature elements are the integrated high-cycle service interfaces, autonomous recovery, radiator servicing, structural and thermal-hydraulic closure, propulsion and disposal, and a validated cost and logistics model for hardware refresh.  A 10-MW system is a conditional extrapolation only after those elements are demonstrated.  A 100-MW facility requires orbital industrial operations rather than conventional mission operations.  A GW facility is a scenario concept that depends on a broad transportation, servicing, manufacturing, traffic-management, and possibly human-support ecosystem that requires independent validation beyond a successful MW pathfinder.

\section{Broader scientific opportunities and extensions}
\label{sec:broader}

The lifecycle architecture has value beyond terrestrial-user computing.  Space-native and communications-integrated workloads can justify smaller systems earlier because orbital location changes the value equation rather than merely substituting for a terrestrial building.

\subsection{Space-native scientific processing}

Earth-observation, planetary, heliophysics, and astronomy missions can process data near the sensor to perform calibration, event detection, adaptive compression, region-of-interest selection, transient follow-up, and autonomous retasking.  The relevant gain is reduction of external communication intensity and response latency.  A maintainable compute cluster could host multiple instruments and upgrade algorithms or accelerators without replacing the sensor spacecraft.  This configuration is applicable to data-rich constellations whose sensors outlive their original processors.

\subsection{Distributed observatories and sensor fusion}

An ODC can serve as a local correlation and control layer for distributed apertures, synthetic-aperture radar, radio interferometry, optical interferometry, and multi-satellite environmental monitoring.  Internal optical links carry high-rate intermediate products while only reduced science products cross the Earth interface.  The same fault-domain and timing architecture developed for compute maintenance supports resilient distributed instruments.  Precision timing and phase coherence may, however, impose stricter pointing, clock, and structural requirements than ordinary batch compute.

\subsection{Autonomous exploration and cislunar operations}

Cislunar navigation, surface operations, resource prospecting, robotics, and traffic coordination benefit from local compute and storage.  A maintainable ODC could provide regional digital twins, hazard maps, route planning, communications routing, and control of industrial equipment.  These applications tolerate or benefit from supervisory autonomy and may share tugs, depots, power, and crew.  Their business case should be evaluated as enabling infrastructure for local activity, not as a direct competitor to terrestrial cloud computing.

\subsection{Servicing infrastructure as a shared platform}

The service system can also support other orbital assets.  Standard cargo containers, robotic tools, warehouses, inspection vehicles, fluid interfaces, and crew worksites can support communications satellites, observatories, power platforms, tugs, and debris-removal missions.  Sharing lowers the fixed cost of rare spares and human presence and moves the crew threshold downward.  Shared servicing infrastructure reduces the fixed cost allocated to an individual ODC and lowers the workforce break-even threshold.

\section{Strategic implications and priorities}
\label{sec:strategy}

The strategic implication is that maintainability is not an optional lifecycle enhancement; it changes the feasible architecture.  A design that cannot replace compute without discarding radiators and power systems is unlikely to survive the technology-refresh economics of tens of megawatts.  A design whose service robot cannot be passively recovered is not autonomous.  A design whose update authority or coolant manifold spans the full cluster does not possess the fault isolation implied by its node count.

Five priorities follow.

\paragraph{Interface standardization before power scaling.}
A common mechanical, electrical, optical, thermal, software, and data-record interface creates competition among hardware suppliers and servicers and permits rolling upgrades.  Standardization should include failure recovery and verification, not only nominal mating geometry.

\paragraph{Characterization of low-probability outcomes.}
Programs should publish task-attempt distributions, retries, jams, leak tests, robot recovery, stockouts, and return diagnostics.  Industrial viability depends on unresolved-operation rates, recovery distributions, and common-cause events rather than single successful demonstrations.

\paragraph{Contingency human access without permanent habitation.}
Early nodes should carry grapple, isolation, docking, and worksite provisions.  This preserves a high-value contingency option at small mass cost.  Permanent habitation should be deferred until the workload and economic conditions in \Eq{eq:combinedhumanthreshold} are demonstrated.

\paragraph{Initial workload selection.}
Space-native preprocessing and edge computing embedded in communications systems reduce the external-data and shared-infrastructure penalties that dominate early ODC economics.  They also provide smaller, revenue- or mission-bearing platforms on which servicing and refresh technology can mature.

\paragraph{Quantitative stage gates.}
A transition from 1 to 10 MW should require measured logistics below \SI{8}{t/(MW.yr)}, post-internal-recovery exceptions below $10^{-3}$, an explicit terminal-unresolved allocation, node hazard below a few percent per year, successful full refresh, and bounded common-mode events with stated fraction and duration.  The 10-to-100 MW transition should require automated warehouse and cargo operations, multiple independent service zones, and evidence that the supply chain can sustain hundreds of tonnes per year.  Crew decisions should be based on measured residual workload, repair-time reduction, and capacity value.

\section{Conclusions}
\label{sec:conclusion}

A large orbital data center is a repairable industrial system in which power, thermal control, computing, robotics, inventory, transportation, and command authority are inseparable.  For a reference high-sunlight \SI{1}{MW} cluster, the minimum physical requirements include thousands of square meters of photovoltaic and radiator area and tens of tonnes of hardware.  The planning mass allocation assigns 3--7.5 kg/kW to maintainability hardware, 2--2.5 kg/kW to first-line stocked ORUs, and 4--6 t to one reserve node.  Internally consistent low, nominal, and high cases are 49, 61.25, and 75 t/MW, reported as $\sim$50--75 t/MW; these are system allocations, not closed flight-spacecraft masses.

Under the declared populations and hazards, the reference allocation implies approximately 70 random or life-limited ORU intervention opportunities per MW-year and 1.3--2.0 t/(MW yr) of direct random replacement.  The conservative additive refresh allocation is larger: 323--349 standardized operations and 2.88--3.98 t/(MW yr).  Adding whole-node loss, consumables, and packaging gives a first-generation planning envelope of 5.3--9.0 t/(MW yr), with a nominal reference realization near 6.6 t/(MW yr) and a mature objective of 3--5 t/(MW yr).  Under the stated 3--15-year cadence, cartridge-level rolling refresh is strongly favored; whole-node replacement as the normal four-year compute-upgrade method increases planned mass flow by approximately an order of magnitude.

Robotic servicing is the operational baseline.  Routine telemetry, fault isolation, workload migration, safe-mode response, and software rollback should be autonomous.  Standard cartridge, power, thermal, radiator, terminal, and cargo work should be performed by dedicated machines and teleoperated or supervised robots.  Visiting robotic vehicles should carry heavy tools, recover disabled nodes, and replace local robots.  Human intervention is allocated to low-frequency mechanical, fluid, structural, and diagnostic failures and to first-of-a-kind integration.

The quantitative robotic requirement is stringent.  At about 400 standardized operations per MW-year, a post-internal-recovery exception probability of 1\% creates four external-recovery cases per MW-year and thousands at gigawatt scale.  The exception probability $p_{\rm exc}$ should be below $10^{-3}$ and preferably $10^{-4}$, while terminal non-recovery $p_U$ requires a separate mission-level allocation of order $10^{-5}$ or lower for a multiyear first-MW exposure under the simple independent diagnostic.  Critical robot fleets should provide overlapping reach, passive recovery, modular joints, diverse controllers, and $\sim$ 99.95\% functional availability.  The catastrophic hazard of a 100-kW node should be held below 0.01--0.03 per year.  A large-domain event-rate allocation near $10^{-3}$ per year is meaningful only when its affected fraction, duration, and recovery state are also specified.

The first MW should be uncrewed, autonomous in flight safety, robotically maintainable, and supplied with class-specific spares sized to the stated six-month replenishment planning horizon.  It should carry one reserve 100-kW node, independent data and safety-control networks, sectionalized power, at least four isolatable thermal loops per node, replaceable radiator panels, standardized cargo containers, and interfaces for a future crew vehicle.  It should not carry a permanent atmosphere or habitat.

Human presence is a scale- and value-dependent option.  In the illustrative cases, the LEO contingency-visit criterion can be satisfied in the tens-of-megawatts regime when several megawatts of high-value capacity would otherwise remain unavailable for months; the campaign criterion can be satisfied in the tens-to-hundreds-of-megawatts regime; and the dedicated-workforce threshold occurs from several hundred megawatts to gigawatts, with a broad 0.1--5 GW sensitivity range.  These are scenario boundaries, not forecasts or independently established transitions.  Shared orbital habitats and transport lower the threshold; GEO and cislunar distance, radiation, and abort requirements raise it.

The required developments are high-temperature radiation-tolerant compute; standardized high-cycle electrical, optical, mechanical, and thermal interfaces; segmented replaceable radiators; statistically qualified autonomous service machinery; robots that can recover other robots; secure bounded software rollout; and a configuration-controlled logistics data system.  The staged development sequence proceeds through ground service cells, free-flying servicing demonstrations, a maintainable 50--100 kW node, a one-MW full-refresh pathfinder, and a ten-MW preindustrial cluster before hundred-megawatt construction.

The analysis therefore treats maintainability as a primary architecture constraint from the initial design.  The first ODC should not be crewed, but it should be designed so that crew can intervene.  At tens to hundreds of megawatts, the primary operational requirements are rolling refresh, robot exception control, common-mode containment, and annual material flow of hundreds of tonnes.  At gigawatt scale, the reference model requires shared cargo vehicles, tugs, depots, warehouses, repair facilities, recycling systems, and technical personnel.  Economic justification remains conditional on the stated workload and cost assumptions, while the engineering requirements are specified explicitly.

\section*{Acknowledgments}
The work described here was carried out at the Jet Propulsion Laboratory, California Institute of Technology, Pasadena, California, under a contract with the National Aeronautics and Space Administration.
\textcopyright\ 2026. California Institute of Technology. Government sponsorship acknowledged.

\appendix

\section{Capacity availability and reserve derivations}
\label{app:availability}

\subsection{Alternating-renewal availability}

For a repairable item with statistically independent up times $X$ and down times $Y$, the long-run time availability follows from the renewal-reward theorem:
\begin{equation}
A=\frac{\mathbb E[X]}{\mathbb E[X]+\mathbb E[Y]}.
\label{eq:renewalavailability}
\end{equation}
For exponential failure and repair with rates $\lambda$ and $\mu$, respectively,
\begin{equation}
A=\frac{\mu}{\lambda+\mu}
=\frac{\MTBF}{\MTBF+\MTTR}.
\label{eq:mtbfavailability}
\end{equation}
This binary measure is appropriate for one item but not for a capacity system.  Let state $s$ deliver power $P_s$.  The capacity reward is $r_s=P_s/P_\Sigma$, and stationary probabilities $\pi_s$ give
\begin{equation}
A_{\rm cap}=\sum_s\pi_s r_s.
\label{eq:markovreward}
\end{equation}
Equation~\eqref{eq:Acap} is the rare-event, low-overlap approximation to \Eq{eq:markovreward}.

\subsection{Independent nodes with reserve}

For $N$ identical active-capable nodes, each unavailable with probability $q$, and a service requirement of at least $K$ nodes, independent-node service availability is
\begin{equation}
A_{\rm svc}=\sum_{j=K}^{N}\binom{N}{j}(1-q)^j q^{N-j}.
\label{eq:binomialservice}
\end{equation}
A cluster with ten required nodes and one reserve has $N=11$, $K=10$.  If $q=0.01$, then
\begin{equation}
A_{\rm svc}=(0.99)^{11}+11(0.99)^{10}(0.01)
\simeq0.9948.
\label{eq:reservebinomialexample}
\end{equation}
This number is lower than a modern service target because $q=0.01$ is a full-node unavailability, not a catastrophic-loss probability masked by rapid reserve activation.  Reducing repair time from months to days lowers $q=\lambda\tau$ proportionally in the rare-failure limit.

If reserve activation is not instantaneous, capacity reward distinguishes the short transition interval from the longer physical repair interval.  For node failures at rate $\Lambda_n$, affected fraction $f_n$, activation time $\tau_a$, and probability $p_s$ that reserve stock is available, the leading loss is
\begin{equation}
1-A_{\rm cap,n}\simeq
\Lambda_n f_n\left[p_s\tau_a+(1-p_s)\tau_{\rm resupply}\right]/T_y.
\label{eq:reserveavailabilityfull}
\end{equation}
This expression explicitly couples spare stock and availability.  Increasing robotic speed has little value when $1-p_s$ is dominated by stockout.

\subsection{Correlated-event capacity distribution}

Let independent node losses be represented by $X(t)$ and a domain event remove capacity fraction $F_d$ for duration $T_d$.  To first order in rare overlap,
\begin{equation}
1-A_{\rm cap}\simeq(1-A_{\rm ind})+
\frac{\lambda_d}{T_y}\mathbb E[F_d T_d].
\label{eq:correlatedsum}
\end{equation}
For multiple domain classes $d$, sum the second term.  If events overlap or one event changes subsequent hazard, a semi-Markov or discrete-event model is required.  The common engineering mistake is to calculate excellent independent-node availability and then omit the domain term entirely.

\subsection{Planned maintenance and workload drain}

For planned task class $i$ at rate $n_i$ per year, drained power $p_i$, user-visible migration time $t_{m,i}$, and physical maintenance time $t_{s,i}$, reserve decouples service from physical work.  If reserve is available,
\begin{equation}
1-A_{\rm plan}\simeq
\frac{1}{P_\Sigma T_y}\sum_i n_i p_i t_{m,i},
\label{eq:plannedavailability}
\end{equation}
not $n_i p_i t_{s,i}/(P_\Sigma T_y)$.  The physical service duration instead drives robot capacity and reserve occupancy.  This separation is one of the principal benefits of modular design.

\section{Failure, refresh, and population-flow model}
\label{app:failuremodel}

\subsection{Nonhomogeneous hazard}

For component class $i$, individual survival is
\begin{equation}
S_i(a)=\exp\Big[-\int_0^a h_i(u)\,du\Big],
\label{eq:survivalhazard}
\end{equation}
where $a$ is age.  A useful bathtub representation is
\begin{equation}
h_i(a)=h_{0,i}+h_{I,i}e^{-a/\tau_{I,i}}
+h_{W,i}\left(\frac{a}{\eta_i}\right)^{\beta_i-1},
\label{eq:agepopulationhazard}
\end{equation}
with constant random hazard, infant-mortality term, and Weibull-like wearout.  The expected event rate of a fleet with age density $n_i(a,t)$ is \Eq{eq:agepopulation}.  After a synchronized installation, the age density is narrow and a wearout knee can create a maintenance surge.  Rolling refresh broadens $n_i$ and reduces peak demand.

For a pure Weibull lifetime with scale $\eta$ and shape $\beta$,
\begin{equation}
S(a)=\exp[-(a/\eta)^\beta],
\qquad
h(a)=\frac{\beta}{\eta}\left(\frac{a}{\eta}\right)^{\beta-1}.
\label{eq:weibull}
\end{equation}
Batteries, bearings, capacitors, pumps, and flexible mechanisms often require $\beta>1$ wearout models; random semiconductor catastrophic faults may be approximated by $\beta\simeq1$ only after infant mortality is removed.

\subsection{Planned replacement before failure}

If a unit is replaced at age $T_r$ or failure, whichever occurs first, its expected cycle length is
\begin{equation}
\mathbb E[T_c]=\int_0^{T_r}S(a)\,da,
\label{eq:replacementcycle}
\end{equation}
and failure probability within the cycle is $1-S(T_r)$.  The competing long-run event rates per installed position are
\begin{equation}
\nu_f(T_r)=\frac{1-S(T_r)}{\int_0^{T_r}S(a)\,da},\qquad
\nu_p(T_r)=\frac{S(T_r)}{\int_0^{T_r}S(a)\,da}.
\label{eq:competingrenewalrates}
\end{equation}
These rates, rather than $\lambda+1/T_r$, apply when corrective replacement resets age.  For cost $C_f$ after failure and $C_p$ for planned replacement, the long-run cost rate is
\begin{equation}
\bar C(T_r)=
\frac{C_f[1-S(T_r)]+C_pS(T_r)}{\int_0^{T_r}S(a)\,da}.
\label{eq:agepolicy}
\end{equation}
The optimal age policy minimizes \Eq{eq:agepolicy}, subject to technology obsolescence and synchronized-logistics constraints.  For compute cartridges, the refresh interval is often set by performance per watt rather than physical wearout, so $C_p$ includes the opportunity cost of operating obsolete hardware.

\subsection{Obsolescence-weighted delivered compute}

Let physical survival be $e^{-\lambda t}$ and frontier performance per watt grow at effective rate $g$.  A unit's frontier-relative delivered compute over horizon $T$ is
\begin{equation}
E^*_{\rm comp}=U_{\rm eff}P_{\rm IT}
\int_0^T e^{-(\lambda+g)t}\,dt
=U_{\rm eff}P_{\rm IT}
\frac{1-e^{-(\lambda+g)T}}{\lambda+g}.
\label{eq:frontiercompute}
\end{equation}
The corresponding penalty relative to a current, immortal device is
\begin{equation}
\Pi_{\lambda+g}=
\frac{(\lambda+g)T}{1-e^{-(\lambda+g)T}}.
\label{eq:obsolescencepenaltyapp}
\end{equation}
A modular architecture reduces the obsolescence penalty associated with $g$ by refreshing only the compute payload while retaining the bus, radiator, and power plant.

\subsection{Reference component calculation}

The direct random replacement mass in \Tab{tab:componentallocation} is
\begin{equation}
\Mdot_{\rm rnd}=\sum_i N_i\lambda_i m_i.
\label{eq:randommassappendix}
\end{equation}
Using the table values gives approximately \SI{1.14}{t/(MW.yr)} before system-level allowances.  Applying a 15\% commis-sioning/infant-mortality allowance and a 10--30\% allowance for harnesses, seals, fasteners, tools, and unmodeled items gives approximately \SIrange{1.3}{2.0}{t/(MW.yr)}.  The range should not be interpreted as statistical precision; it is a design reserve until flight data replace the assumed hazards.

\section{Inventory, launch delay, and spare-pooling derivations}
\label{app:inventory}

\subsection{Exact Poisson service level}

For Poisson lead-time demand $D\sim\mathrm{Poisson}(\mu_L)$, the smallest base stock meeting cycle service level $q$ is
\begin{equation}
S_q=\min\Big\{s:\sum_{k=0}^{s}e^{-\mu_L}\frac{\mu_L^k}{k!}\ge q\Big\}.
\label{eq:poissonquantile}
\end{equation}
The expected backorder quantity is
\begin{equation}
B(S)=\mathbb E[(D-S)_+]
=\sum_{k=S+1}^{\infty}(k-S)e^{-\mu_L}\frac{\mu_L^k}{k!}.
\label{eq:poissonbackorder}
\end{equation}
If carrying one spare costs $c_s$ per replenishment cycle and one backordered unit costs $c_b$, the optimum minimizes
\begin{equation}
C(S)=c_sS+c_bB(S).
\label{eq:inventorycost}
\end{equation}
The discrete marginal condition is equivalent to a critical fractile.  Rare survival-critical parts generally justify very high $q$ even when expected demand is much less than one.

\subsection{Uncertain and delayed launch}

If resupply lead time $L$ is random and independent of demand rate $\Lambda$, mixed-Poisson demand has
\begin{align}
\mathbb E[D]&=\Lambda\mathbb E[L],\\
\mathrm{Var}(D)&=\Lambda\mathbb E[L]+\Lambda^2\mathrm{Var}(L).
\label{eq:randomlead}
\end{align}
Launch slips therefore add a variance term that can dominate the ordinary Poisson variance.  A six-month mean lead time with a three-month standard deviation has coefficient of variation 0.5; for high-rate cartridges, the $\Lambda^2\mathrm{Var}(L)$ term is substantial.  Multiple launch providers reduce this term only if their delays are not strongly correlated by weather, range, regulation, or shared production.

\subsection{Pooling with transport delay}

Pooling reduces stock but adds orbital delivery time $T_{\rm move}$.  Let local stock cost per unit be $c_l$, central stock cost be $c_c$, and capacity-loss value from transport be $v_p p_i T_{\rm move}$.  Pooling class $i$ is favored when
\begin{equation}
(c_lS_{i,l}-c_cS_{i,c})
> \Lambda_i v_p p_i T_{\rm move}+C_{\rm transport}.
\label{eq:poolcriterion}
\end{equation}
High-frequency low-mass cartridges usually remain partly local; rare heavy terminals, robot assemblies, and full nodes are pooled.  A service depot that is several orbital planes away may require a tug or phasing delay and should not be treated as instantaneous common stock.

\subsection{Repairable-spare pipeline}

If failed units are refurbished with turnaround $L_r$ and yield $y_r$, the effective new-production demand is
\begin{equation}
\Lambda_{\rm new}=\Lambda(1-y_r),
\label{eq:refurbdemand}
\end{equation}
while the number of units tied up in the repair pipeline is approximately
\begin{equation}
N_{\rm pipe}=\Lambda y_r L_r
\label{eq:repairpipeline}
\end{equation}
by Little's law.  Refurbishment reduces Earth upmass only if the pipeline inventory, test equipment, kits, and lower reflight confidence do not exceed the avoided new hardware.  Fast diagnosis and modular board-level repair increase $y_r$ and reduce $L_r$.

\section{Robotic service capacity and reliability}
\label{app:robotics}

\subsection{Service-cycle time decomposition}

The main-text service durations are end-to-end productive times rather than pure manipulator motion.  Table~\ref{tab:servicephases} makes the assumed phases explicit.  ``Remove and stow'' includes transfer to a quarantine, return, or disposal location; ``install and mate'' includes spare retrieval and all mechanical, electrical, optical, or fluid mating.  The ranges are planning values for cooperative interfaces under nominal geometry and exclude queueing, resupply, structural reconstruction, and unresolved retries.

\begin{table}[htbp]
\caption{Representative productive robotic service-time decomposition.  L1, L2, and L3 denote cartridge, subsystem, and external replacement levels.  Values are hours per completed event and are intended for queueing and capacity sizing, not as user-visible outage time.}
\label{tab:servicephases}
\centering
\small
\renewcommand{\arraystretch}{1.08}
\setlength{\tabcolsep}{3.2pt}
\begin{tabularx}{\textwidth}{@{}Y C{0.105\textwidth} C{0.08\textwidth} C{0.10\textwidth} C{0.10\textwidth} C{0.10\textwidth} C{0.08\textwidth}@{}}
\toprule
Service class & Confirm / safe & Access & Remove / stow & Install / mate & Verify / return & Total \\
\midrule
L1 compute/storage/network cartridge & 0.10--0.25 & 0.05--0.15 & 0.15--0.30 & 0.20--0.40 & 0.25--0.60 & 0.75--1.70 \\
L2 converter/power-distribution/battery & 0.25--0.75 & 0.25--0.75 & 0.50--1.50 & 0.75--1.75 & 0.75--1.75 & 2.5--6.5 \\
L2 pump/valve/manifold & 0.50--1.00 & 0.50--1.00 & 1.0--2.5 & 1.0--3.0 & 1.0--3.0 & 4.0--10.5 \\
L3 radiator/photovoltaic/terminal & 0.50--1.50 & 1.0--3.0 & 2.0--6.0 & 2.0--5.0 & 2.0--5.0 & 7.5--20.5 \\
Major robot or joint package & 0.50--1.50 & 1.0--3.0 & 2.0--6.0 & 2.0--5.0 & 1.0--4.0 & 6.5--19.5 \\
\bottomrule
\end{tabularx}
\end{table}

\subsection{Task-class workload}

Let task classes $j$ occur at rate $\Lambda_j$ with mean service time $t_j$ and compatibility matrix $a_{jr}\in\{0,1\}$ indicating whether robot type $r$ can perform task $j$.  Robot-hour demand is
\begin{equation}
H_R=\sum_j\Lambda_j t_j,
\label{eq:robotworkappendix}
\end{equation}
but feasibility requires a schedule satisfying
\begin{equation}
\sum_j a_{jr}x_{jr}t_j\le H_{r,\rm avail},
\qquad
\sum_r x_{jr}=\Lambda_j.
\label{eq:robotassignment}
\end{equation}
A scalar robot-equivalent count can conceal incompatibility: abundant cartridge machines cannot replace a radiator panel, and a heavy arm may not have access to an internal bay.  Mission planning should solve a multi-class queueing or discrete-event assignment problem with robot downtime and travel time.

\subsection{Retries and post-recovery success}

Let first-attempt failure probability be $p_1$.  If an independent retry with the same robot fails with conditional probability $p_2$, an alternate tool fails with probability $p_3$, and an alternate robot fails with probability $p_4$, the post-internal-recovery exception probability is
\begin{equation}
\pex=p_1p_2p_3p_4
\label{eq:retryproduct}
\end{equation}
only if the failures are conditionally independent.  In practice, geometric damage or a bad interface makes them correlated.  A beta-factor approximation writes
\begin{equation}
\pex\simeq\beta p_1+(1-\beta)p_1p_2p_3p_4,
\label{eq:retrybeta}
\end{equation}
where $\beta$ is the fraction of first-attempt failures caused by an interface condition shared by all internal recovery attempts.  Reducing $\beta$ through secondary release features, alternate approach geometry, and accessible inspection is more valuable than adding repeated identical retries.  Neither \Eq{eq:retryproduct} nor \Eq{eq:retrybeta} is a model for $\pU$: terminal non-recovery must also include diagnosis, safing, visiting-servicer, external-tool, and any authorized human-contingency defenses, together with their common causes.

\subsection{Robot fleet with common mode}

The independent $k$-of-$n$ availability is \Eq{eq:kofn}.  If common-mode robot failure has probability $q_c$, fleet availability becomes approximately
\begin{equation}
A_{R,\rm fleet}\simeq(1-q_c)A_{k|n}.
\label{eq:robotcommonmode}
\end{equation}
For $A_{k|n}=0.9995$, even $q_c=10^{-3}$ reduces fleet availability below 0.999.  Independent power, recovery firmware, and passive mechanics are therefore necessary.  Numerical redundancy alone cannot meet the service-system allocation.

\subsection{Minimum service capacity under campaigns}

Scheduled refresh creates a known campaign load $H_c$ over window $T_c$ in addition to random workload rate $h_r$.  With capacity $n_R h_y/T_y$, the minimum robot count must satisfy
\begin{equation}
n_R\ge
\max\left[
\frac{h_r}{\rho_{\max}h_y/T_y},
\frac{H_c/T_c+h_r}{\rho_{\max}h_y/T_y}
\right],
\label{eq:campaigncapacity}
\end{equation}
where $\rho_{\max}\simeq0.6$--0.7 is a planning utilization.  Campaigns can be spread over time, but overly slow refresh increases obsolescence and mixes hardware generations.  Automated magazines and acceptance-test stations should therefore be sized by peak planned flow, not only annual mean.

\section{Human-presence economics and crew-support scaling}
\label{app:human}

\subsection{Capital-cost-equivalent comparison}

An annual crew-support cost $C_H$ can be converted to a capital equivalent through the capital-recovery factor
\begin{equation}
\CRF=\frac{r(1+r)^n}{(1+r)^n-1},
\label{eq:crfappendix}
\end{equation}
so that
\begin{equation}
C_{H,\rm cap}=\frac{C_H}{\CRF}.
\label{eq:crewcap}
\end{equation}
For $C_H=\$0.8$ billion/yr and $\CRF=0.10$, $C_{H,\rm cap}=\$8$ billion.  The allocation per delivered IT kilowatt is
\begin{equation}
c_{H,\rm cap,kW}=\frac{C_{H,\rm cap}}{1000\PMW}
\quad[\mathrm{\$/kW}],
\label{eq:crewcapperkW}
\end{equation}
where $1000\PMW$ is the number of delivered kilowatts.  At $\PMW=1$, 10, 100, 300, and 1000, this is 8000, 800, 80, 26.7, and 8 k\$/kW, respectively.  Comparing with terrestrial infrastructure benchmarks of order 10--40 k\$/kW shows independently why a standalone permanent crew is not economically commensurate below several hundred MW.

\subsection{Mass amortization}

For crew-facility mass $M_H$, the mass penalty is
\begin{equation}
m_{H,\rm kW}=1000\frac{M_H}{P_\Sigma}
\quad[\mathrm{kg/kW}],
\label{eq:crewmasspenalty}
\end{equation}
with $P_\Sigma$ in W and $M_H$ in kg.  This physical mass-per-power relation intentionally uses $P_\Sigma$ rather than the dimensionless $P_{\rm MW}$; it is distinct from the per-MW lifecycle intensities.  A \SI{100}{t} facility gives the scale reported in Sec.~\ref{sec:human}.  This mass becomes modest before the operating cost does; mass amortization alone is not a sufficient human-presence criterion.

\subsection{Crew consumables example}

For $N_H$ crew, packaged food rate $m_f$ per person-day, and mission duration $T$, food mass is
\begin{equation}
M_f=N_Hm_fT.
\label{eq:foodmass}
\end{equation}
For $N_H=4$, $m_f=1.82~\mathrm{kg/(person\,day)}$, and one year, $M_f\simeq\SI{2.66}{t}$ \cite{Perchonok2009FoodMass}.  Water recovery near 98\% greatly reduces water upmass but requires pumps, filters, membranes, reactors, sensors, maintenance, and make-up inventory \cite{NASA2023WaterRecovery}.  Food, ECLSS parts, clothing, hygiene, medical supplies, waste containers, and crew transportation remain an annual stream.

\subsection{Availability value for low-probability events}

Let unstructured events follow classes $j$ with rate $\Lambda_j$, affected power $P_j$, autonomous repair time $T_{A,j}$, and human repair time $T_{H,j}$.  Define $P_{j,\rm MW}=P_j/(\SI{1}{MW})$.  The annual value of human response is
\begin{equation}
V_H=\sum_j\Lambda_j v_jP_{j,\rm MW}(T_{A,j}-T_{H,j})
+V_{\rm asset}+C_{\rm robot,avoided}+C_{\rm logistics,avoided}.
\label{eq:humanvalueclasses}
\end{equation}
Permanent presence is justified if $V_H>C_H$.  Equation~\eqref{eq:humanbreakeven} is the intensive approximation obtained by aggregating the event classes into $b_H$ in \Eq{eq:humanbenefitintensity}.  The class form is preferred when one or two rare events dominate the value.

\section{Supplementary physical and numerical calculations}
\label{app:numerics}

\subsection{Reference power and radiator calculation}

For \SI{1}{MW} IT and $\alphaOH=1.25$,
\begin{equation}
P_{\rm tot}=\SI{1.25}{MW}.
\label{eq:ptotnumeric}
\end{equation}
At $\epsilon_{\rm rad}=0.90$, $\eta_{\rm view}=0.85$, $T_{\rm rad}=\SI{350}{K}$, and $q_{\rm env}=\SI{150}{W/m^2}$, net heat flux is
\begin{equation}
q_{\rm net}=\epsilon_{\rm rad}\eta_{\rm view}\sigma_{\rm SB}T_{\rm rad}^4-q_{\rm env}
\simeq\SI{500}{W/m^2},
\label{eq:qnetnumeric}
\end{equation}
which gives $A_{\rm rad}\simeq\SI{2.5e3}{m^2}$.  At effective areal density \SI{5}{kg/m^2}, radiator mass is approximately \SI{12.5}{t}.  Raising $T_{\rm rad}$ to \SI{400}{K} increases ideal emitted flux by $(400/350)^4\simeq1.71$, but only if package, cold-plate, and loop temperature drops preserve junction margin.

\subsection{Thermal response after local cooling loss}

A cartridge with effective heat capacity $m_c c_{p,c}$ and residual power $P_r$ warms at
\begin{equation}
\frac{dT}{dt}=\frac{P_r-Q_{\rm passive}}{m_cc_{p,c}}.
\label{eq:thermaltransientappendix}
\end{equation}
For $P_r=\SI{5}{kW}$, $m_c=\SI{25}{kg}$, $c_{p,c}=\SI{500}{J/(kg.K)}$, and negligible immediate passive rejection, $dT/dt\simeq\SI{0.4}{K/s}$.  A \SI{25}{K} margin lasts only about one minute.  Fast hardware current limiting and local thermal conduction are therefore required; ground commands are not a primary protection layer.

\subsection{Stored electrical energy}

A dc-link capacitance $C$ at voltage $V$ stores
\begin{equation}
E_C=\frac12CV^2.
\label{eq:capacitorenergy}
\end{equation}
For $C=\SI{10}{mF}$ and $V=\SI{400}{V}$, $E_C=\SI{800}{J}$.  Blind-mate interfaces need positive discharge confirmation and connector-first capture sequencing even after nominal isolation.

\subsection{Eclipse battery calculation}

With $t_{\rm ecl}=\SI{35}{min}$, $P_{\rm tot}=\SI{1.25}{MW}$, cell specific energy $e_b$, allowable depth of discharge $\rho_{\rm DoD}$, and end-of-life capacity fraction $f_{\rm EOL}$,
\begin{equation}
M_b=\frac{P_{\rm tot}t_{\rm ecl}}{e_b\rho_{\rm DoD}f_{\rm EOL}}.
\label{eq:batterynumeric}
\end{equation}
For $e_b=\SI{200}{Wh/kg}$, $\rho_{\rm DoD}=0.60$, and $f_{\rm EOL}=0.80$, $M_b\simeq\SI{7.6}{t/MW}$.  This excludes packaging, power electronics, redundancy, and thermal control.  High-sunlight orbit selection can therefore remove several tonnes per MW and a major refresh stream.

\subsection{Fission two-temperature rejection}

For the example in \Eq{eq:nuclearheat}, let \SI{3.75}{MW} of conversion waste be rejected at \SI{700}{K} with $\epsilon\eta=0.75$ and low environmental load.  The hot radiator area is approximately
\begin{equation}
A_h\simeq\frac{\SI{3.75}{MW}}
{0.75\sigma_{\rm SB}(\SI{700}{K})^4}
\simeq\SI{370}{m^2}.
\label{eq:hotradiator}
\end{equation}
The IT system still requires approximately \SI{2500}{m^2} near \SI{350}{K}.  The nuclear architecture may therefore have radiator area comparable to or somewhat larger than the PV case if high-temperature conversion is achieved, but it adds reactor and shielding mass.  Rejecting conversion heat at the IT temperature would be prohibitive.

\subsection{Reference parameter set}

Table~\ref{tab:parameterappendix} consolidates the numerical assumptions used in the worked examples and distinguishes physical scale anchors, nominal design points, stage-gate requirements, and stress cases.  It provides the reference parameter set for replacement by mission-specific test and flight data.

\begin{table}[htbp]
\caption{Reference parameters used in numerical examples.  Values are conditional architecture inputs.}
\label{tab:parameterappendix}
\centering
\small
\renewcommand{\arraystretch}{1.00}
\begin{tabularx}{\textwidth}{@{}L{0.23\textwidth} C{0.11\textwidth} C{0.23\textwidth} Y@{}}
\toprule
Parameter & Symbol & Reference & Comment \\
\midrule
Delivered IT power & $P_\Sigma$ & \SI{1}{MW} & physical power scale anchor \\
Power normalization & $P_{\rm MW}$ & 1 & dimensionless number of delivered MW \\
Platform overhead & $\alpha_{\rm OH}$ & 1.25 & total power \SI{1.25}{MW} \\
Node power & $p_n$ & \SI{100}{kW} & ten active domains/MW \\
Cartridge power & $p_c$ & \SI{5}{kW} & 200 active cartridges/MW \\
Active-node reserve & $r_n$ & 10\% first MW; 5--10\% pooled & one full node in pathfinder \\
Compute refresh interval & $T_{r,c}$ & 4 yr & parameter range 3--5 yr \\
Bus/structure life & $T_{\rm bus}$ & 10--15 yr & with subsystem- and external-unit replacement \\
Resupply lead time & $L$ & 0.5 yr & stochastic in mission analysis \\
Common-spare service level & $q$ & 0.99 & higher for survival-critical rare items \\
Robot operations & $N_{\rm op}$ & 393--419/(MW yr) & 70.2 random plus 323--349 planned \\
Robot labor & $\ell_R$ & 560--700 h/(MW yr) & lower-bound-compatible planned-task realization; excludes large contingency reconstruction \\
Post-internal-recovery exception objective & $p_{\rm exc}$ & $10^{-4}$ & requirement $\le10^{-3}$; external recovery assumed \\
Terminal-unresolved probability & $p_U$ & mission allocated & 1-MW, 12-year homogeneous diagnostic $\le1.07\times10^{-5}$ for 5\% any-event risk \\
Node catastrophic hazard & $\lambda_{\rm cat}$ & 0.01--0.03/yr & stage-gate target \\
Large-domain event rate & $\lambda_{\rm dom}$ & $\le10^{-3}$/yr & only with event fraction and duration specified \\
First-generation logistics & $\ell_M$ & 5.3--9.0 t/(MW yr) & nominal 6.6; mature objective 3--5 \\
Onsite human labor & $\ell_H$ & 23--90 h/(MW yr) & nominal 35; stress case 140 \\
Dedicated crew annual cost & $C_H$ & 0.5--1.5 B\$/yr & illustrative sensitivity range \\
\bottomrule
\end{tabularx}
\end{table}


\begin{thebibliography}{45}%
\makeatletter
\providecommand \@ifxundefined [1]{%
 \@ifx{#1\undefined}
}%
\providecommand \@ifnum [1]{%
 \ifnum #1\expandafter \@firstoftwo
 \else \expandafter \@secondoftwo
 \fi
}%
\providecommand \@ifx [1]{%
 \ifx #1\expandafter \@firstoftwo
 \else \expandafter \@secondoftwo
 \fi
}%
\providecommand \natexlab [1]{#1}%
\providecommand \enquote  [1]{``#1''}%
\providecommand \bibnamefont  [1]{#1}%
\providecommand \bibfnamefont [1]{#1}%
\providecommand \citenamefont [1]{#1}%
\providecommand \href@noop [0]{\@secondoftwo}%
\providecommand \href [0]{\begingroup \@sanitize@url \@href}%
\providecommand \@href[1]{\@@startlink{#1}\@@href}%
\providecommand \@@href[1]{\endgroup#1\@@endlink}%
\providecommand \@sanitize@url [0]{\catcode `\\12\catcode `\$12\catcode
  `\&12\catcode `\#12\catcode `\^12\catcode `\_12\catcode `\%12\relax}%
\providecommand \@@startlink[1]{}%
\providecommand \@@endlink[0]{}%
\providecommand \url  [0]{\begingroup\@sanitize@url \@url }%
\providecommand \@url [1]{\endgroup\@href {#1}{\urlprefix }}%
\providecommand \urlprefix  [0]{URL }%
\providecommand \Eprint [0]{\href }%
\providecommand \doibase [0]{https://doi.org/}%
\providecommand \selectlanguage [0]{\@gobble}%
\providecommand \bibinfo  [0]{\@secondoftwo}%
\providecommand \bibfield  [0]{\@secondoftwo}%
\providecommand \translation [1]{[#1]}%
\providecommand \BibitemOpen [0]{}%
\providecommand \bibitemStop [0]{}%
\providecommand \bibitemNoStop [0]{.\EOS\space}%
\providecommand \EOS [0]{\spacefactor3000\relax}%
\providecommand \BibitemShut  [1]{\csname bibitem#1\endcsname}%
\let\auto@bib@innerbib\@empty
\bibitem [{\citenamefont {Turyshev}(2026)}]{Turyshev2026ODC}%
  \BibitemOpen
  \bibfield  {author} {\bibinfo {author} {\bibfnamefont {S.~G.}\ \bibnamefont
  {Turyshev}},\ }\href {https://doi.org/10.48550/arXiv.2604.27197} {\bibinfo
  {title} {Orbital data centers: Spacecraft constraints and economic
  viability}} (\bibinfo {year} {2026}),\ \Eprint
  {https://arxiv.org/abs/2604.27197} {arXiv:2604.27197 [physics.gen-ph]}
  \BibitemShut {NoStop}%
\bibitem [{\citenamefont {{Ag{\"u}era y Arcas}}\ \emph
  {et~al.}(2025)\citenamefont {{Ag{\"u}era y Arcas}}, \citenamefont {Beals},
  \citenamefont {Biggs}, \citenamefont {Bloom}, \citenamefont {Fischbacher},
  \citenamefont {Gromov}, \citenamefont {K{\"o}ster}, \citenamefont
  {Pravahan},\ and\ \citenamefont {Manyika}}]{Aguera2025Suncatcher}%
  \BibitemOpen
  \bibfield  {author} {\bibinfo {author} {\bibfnamefont {B.}~\bibnamefont
  {{Ag{\"u}era y Arcas}}}, \bibinfo {author} {\bibfnamefont {T.}~\bibnamefont
  {Beals}}, \bibinfo {author} {\bibfnamefont {M.}~\bibnamefont {Biggs}},
  \bibinfo {author} {\bibfnamefont {J.~V.}\ \bibnamefont {Bloom}}, \bibinfo
  {author} {\bibfnamefont {T.}~\bibnamefont {Fischbacher}}, \bibinfo {author}
  {\bibfnamefont {K.}~\bibnamefont {Gromov}}, \bibinfo {author} {\bibfnamefont
  {U.}~\bibnamefont {K{\"o}ster}}, \bibinfo {author} {\bibfnamefont
  {R.}~\bibnamefont {Pravahan}},\ and\ \bibinfo {author} {\bibfnamefont
  {J.}~\bibnamefont {Manyika}},\ }\href
  {https://doi.org/10.48550/arXiv.2511.19468} {\bibinfo {title} {Towards a
  future space-based, highly scalable {AI} infrastructure system design}}
  (\bibinfo {year} {2025}),\ \Eprint {https://arxiv.org/abs/2511.19468}
  {arXiv:2511.19468 [cs.DC]} \BibitemShut {NoStop}%
\bibitem [{\citenamefont {Bargatin}\ \emph {et~al.}(2025)\citenamefont
  {Bargatin}, \citenamefont {Jin}, \citenamefont {Alansari},\ and\
  \citenamefont {Raney}}]{Bargatin2025Tether}%
  \BibitemOpen
  \bibfield  {author} {\bibinfo {author} {\bibfnamefont {I.}~\bibnamefont
  {Bargatin}}, \bibinfo {author} {\bibfnamefont {D.}~\bibnamefont {Jin}},
  \bibinfo {author} {\bibfnamefont {Z.}~\bibnamefont {Alansari}},\ and\
  \bibinfo {author} {\bibfnamefont {J.~R.}\ \bibnamefont {Raney}},\ }\href
  {https://doi.org/10.48550/arXiv.2512.09044} {\bibinfo {title} {Tether-based
  architecture for solar-powered orbital {AI} data centers}} (\bibinfo {year}
  {2025}),\ \Eprint {https://arxiv.org/abs/2512.09044} {arXiv:2512.09044
  [physics.space-ph]} \BibitemShut {NoStop}%
\bibitem [{\citenamefont {{European Space Policy
  Institute}}(2025)}]{Gutierrez2025ESPI}%
  \BibitemOpen
  \bibfield  {author} {\bibinfo {author} {\bibnamefont {{European Space Policy
  Institute}}},\ }\href@noop {} {\emph {\bibinfo {title} {Data Centres in
  Space: Orbital Backbone of the Second Digital Era?}}},\ \bibinfo {type}
  {Technical Report}\ \bibinfo {number} {ESPI Report 98}\ (\bibinfo
  {institution} {European Space Policy Institute},\ \bibinfo {address} {Vienna,
  Austria},\ \bibinfo {year} {2025})\BibitemShut {NoStop}%
\bibitem [{\citenamefont {Chabra}\ \emph {et~al.}(2026)\citenamefont {Chabra},
  \citenamefont {Li}, \citenamefont {Hsieh}, \citenamefont {Segarra},
  \citenamefont {Arzani}, \citenamefont {Olsen},\ and\ \citenamefont
  {Chandra}}]{Chabra2026OrbitalBrain}%
  \BibitemOpen
  \bibfield  {author} {\bibinfo {author} {\bibfnamefont {O.}~\bibnamefont
  {Chabra}}, \bibinfo {author} {\bibfnamefont {C.}~\bibnamefont {Li}}, \bibinfo
  {author} {\bibfnamefont {K.}~\bibnamefont {Hsieh}}, \bibinfo {author}
  {\bibfnamefont {S.}~\bibnamefont {Segarra}}, \bibinfo {author} {\bibfnamefont
  {B.}~\bibnamefont {Arzani}}, \bibinfo {author} {\bibfnamefont
  {P.}~\bibnamefont {Olsen}},\ and\ \bibinfo {author} {\bibfnamefont
  {R.}~\bibnamefont {Chandra}},\ }\bibfield  {title} {\bibinfo {title}
  {Orbitalbrain: A distributed framework for training {ML} models in space},\
  }in\ \href {https://doi.org/10.4230/OASIcs.NINeS.2026.5} {\emph {\bibinfo
  {booktitle} {1st New Ideas in Networked Systems (NINeS 2026)}}},\ \bibinfo
  {series} {Open Access Series in Informatics (OASIcs)}, Vol.\ \bibinfo
  {volume} {139},\ \bibinfo {editor} {edited by\ \bibinfo {editor}
  {\bibfnamefont {K.}~\bibnamefont {Argyraki}}\ and\ \bibinfo {editor}
  {\bibfnamefont {A.}~\bibnamefont {Panda}}}\ (\bibinfo  {publisher} {Schloss
  Dagstuhl -- Leibniz-Zentrum f{\"u}r Informatik},\ \bibinfo {address}
  {Dagstuhl, Germany},\ \bibinfo {year} {2026})\ pp.\ \bibinfo {pages}
  {5:1--5:32}\BibitemShut {NoStop}%
\bibitem [{\citenamefont {Thummala}\ and\ \citenamefont
  {Falco}(2025)}]{Thummala2025Compute}%
  \BibitemOpen
  \bibfield  {author} {\bibinfo {author} {\bibfnamefont {R.}~\bibnamefont
  {Thummala}}\ and\ \bibinfo {author} {\bibfnamefont {G.}~\bibnamefont
  {Falco}},\ }\href {https://doi.org/10.48550/arXiv.2512.17054} {\bibinfo
  {title} {When to compute in space}} (\bibinfo {year} {2025}),\ \Eprint
  {https://arxiv.org/abs/2512.17054} {arXiv:2512.17054 [cs.CE]} \BibitemShut
  {NoStop}%
\bibitem [{\citenamefont {Aili}\ \emph {et~al.}(2025)\citenamefont {Aili},
  \citenamefont {Choi}, \citenamefont {Ong},\ and\ \citenamefont
  {Wen}}]{Aili2025CarbonNeutral}%
  \BibitemOpen
  \bibfield  {author} {\bibinfo {author} {\bibfnamefont {A.}~\bibnamefont
  {Aili}}, \bibinfo {author} {\bibfnamefont {J.}~\bibnamefont {Choi}}, \bibinfo
  {author} {\bibfnamefont {Y.~S.}\ \bibnamefont {Ong}},\ and\ \bibinfo {author}
  {\bibfnamefont {Y.}~\bibnamefont {Wen}},\ }\bibfield  {title} {\bibinfo
  {title} {The development of carbon-neutral data centres in space},\ }\href
  {https://doi.org/10.1038/s41928-025-01476-1} {\bibfield  {journal} {\bibinfo
  {journal} {Nature Electronics}\ }\textbf {\bibinfo {volume} {8}},\ \bibinfo
  {pages} {1016} (\bibinfo {year} {2025})}\BibitemShut {NoStop}%
\bibitem [{\citenamefont {Weston}\ \emph {et~al.}(2025)\citenamefont {Weston},
  \citenamefont {Burkhard}, \citenamefont {Stupl}, \citenamefont {Ticknor},
  \citenamefont {Yost}, \citenamefont {Austin}, \citenamefont {Galchenko},
  \citenamefont {Newman},\ and\ \citenamefont
  {Soto}}]{NASA2025SmallSpacecraft}%
  \BibitemOpen
  \bibfield  {author} {\bibinfo {author} {\bibfnamefont {S.~V.}\ \bibnamefont
  {Weston}}, \bibinfo {author} {\bibfnamefont {C.~D.}\ \bibnamefont
  {Burkhard}}, \bibinfo {author} {\bibfnamefont {J.~M.}\ \bibnamefont {Stupl}},
  \bibinfo {author} {\bibfnamefont {R.~L.}\ \bibnamefont {Ticknor}}, \bibinfo
  {author} {\bibfnamefont {B.~D.}\ \bibnamefont {Yost}}, \bibinfo {author}
  {\bibfnamefont {R.~A.}\ \bibnamefont {Austin}}, \bibinfo {author}
  {\bibfnamefont {P.}~\bibnamefont {Galchenko}}, \bibinfo {author}
  {\bibfnamefont {L.~K.}\ \bibnamefont {Newman}},\ and\ \bibinfo {author}
  {\bibfnamefont {L.~S.}\ \bibnamefont {Soto}},\ }\href
  {https://ntrs.nasa.gov/citations/20250000142} {\emph {\bibinfo {title}
  {State-of-the-Art Small Spacecraft Technology}}},\ \bibinfo {type} {Technical
  Report}\ \bibinfo {number} {NASA NTRS 20250000142}\ (\bibinfo  {institution}
  {National Aeronautics and Space Administration},\ \bibinfo {year}
  {2025})\BibitemShut {NoStop}%
\bibitem [{\citenamefont {Gilmore}(2002)}]{Gilmore2002}%
  \BibitemOpen
  \bibinfo {editor} {\bibfnamefont {D.~G.}\ \bibnamefont {Gilmore}},\ ed.,\
  \href@noop {} {\emph {\bibinfo {title} {Spacecraft Thermal Control Handbook,
  Volume I: Fundamental Technologies}}},\ \bibinfo {edition} {2nd}\ ed.\
  (\bibinfo  {publisher} {American Institute of Aeronautics and Astronautics
  and The Aerospace Press},\ \bibinfo {address} {Reston, Virginia},\ \bibinfo
  {year} {2002})\BibitemShut {NoStop}%
\bibitem [{\citenamefont {Juhasz}(2002)}]{Juhasz2002}%
  \BibitemOpen
  \bibfield  {author} {\bibinfo {author} {\bibfnamefont {A.~J.}\ \bibnamefont
  {Juhasz}},\ }\href {https://ntrs.nasa.gov/citations/20020082957} {\emph
  {\bibinfo {title} {Design Considerations for Lightweight Space Radiators
  Based on Fabrication and Test Experience with a Carbon--Carbon Composite
  Prototype Heat Pipe}}},\ \bibinfo {type} {Tech. Rep.}\ \bibinfo {number}
  {NASA/TP-1998-207427/REV1}\ (\bibinfo  {institution} {NASA Glenn Research
  Center},\ \bibinfo {year} {2002})\ \bibinfo {note} {{NASA NTRS}
  20020082957}\BibitemShut {NoStop}%
\bibitem [{\citenamefont {Ginet}\ \emph {et~al.}(2013)\citenamefont {Ginet},
  \citenamefont {O'Brien}, \citenamefont {Huston}, \citenamefont {Johnston},
  \citenamefont {Guild}, \citenamefont {Friedel}, \citenamefont {Lindstrom},
  \citenamefont {Roth}, \citenamefont {Whelan}, \citenamefont {Quinn},
  \citenamefont {Madden}, \citenamefont {Morley},\ and\ \citenamefont
  {Su}}]{Ginet2013AE9AP9}%
  \BibitemOpen
  \bibfield  {author} {\bibinfo {author} {\bibfnamefont {G.~P.}\ \bibnamefont
  {Ginet}}, \bibinfo {author} {\bibfnamefont {T.~P.}\ \bibnamefont {O'Brien}},
  \bibinfo {author} {\bibfnamefont {S.~L.}\ \bibnamefont {Huston}}, \bibinfo
  {author} {\bibfnamefont {W.~R.}\ \bibnamefont {Johnston}}, \bibinfo {author}
  {\bibfnamefont {T.~B.}\ \bibnamefont {Guild}}, \bibinfo {author}
  {\bibfnamefont {R.}~\bibnamefont {Friedel}}, \bibinfo {author} {\bibfnamefont
  {C.~D.}\ \bibnamefont {Lindstrom}}, \bibinfo {author} {\bibfnamefont {C.~J.}\
  \bibnamefont {Roth}}, \bibinfo {author} {\bibfnamefont {P.}~\bibnamefont
  {Whelan}}, \bibinfo {author} {\bibfnamefont {R.~A.}\ \bibnamefont {Quinn}},
  \bibinfo {author} {\bibfnamefont {D.}~\bibnamefont {Madden}}, \bibinfo
  {author} {\bibfnamefont {S.~K.}\ \bibnamefont {Morley}},\ and\ \bibinfo
  {author} {\bibfnamefont {Y.-J.}\ \bibnamefont {Su}},\ }\bibfield  {title}
  {\bibinfo {title} {{AE9}, {AP9} and {SPM}: New models for specifying the
  trapped energetic particle and space plasma environment},\ }\href
  {https://doi.org/10.1007/s11214-013-9964-y} {\bibfield  {journal} {\bibinfo
  {journal} {Space Science Reviews}\ }\textbf {\bibinfo {volume} {179}},\
  \bibinfo {pages} {579} (\bibinfo {year} {2013})}\BibitemShut {NoStop}%
\bibitem [{\citenamefont {Guild}\ \emph {et~al.}(2021)\citenamefont {Guild},
  \citenamefont {Davis}, \citenamefont {Boyd}, \citenamefont {O'Brien},
  \citenamefont {Fennell}, \citenamefont {Mazur}, \citenamefont {Brinkman},
  \citenamefont {Peterson},\ and\ \citenamefont
  {Jenkin}}]{Guild2021SpaceEnvironment}%
  \BibitemOpen
  \bibfield  {author} {\bibinfo {author} {\bibfnamefont {T.~B.}\ \bibnamefont
  {Guild}}, \bibinfo {author} {\bibfnamefont {S.~C.}\ \bibnamefont {Davis}},
  \bibinfo {author} {\bibfnamefont {A.~J.}\ \bibnamefont {Boyd}}, \bibinfo
  {author} {\bibfnamefont {T.~P.}\ \bibnamefont {O'Brien}}, \bibinfo {author}
  {\bibfnamefont {J.~F.}\ \bibnamefont {Fennell}}, \bibinfo {author}
  {\bibfnamefont {J.~E.}\ \bibnamefont {Mazur}}, \bibinfo {author}
  {\bibfnamefont {D.~G.}\ \bibnamefont {Brinkman}}, \bibinfo {author}
  {\bibfnamefont {G.~E.}\ \bibnamefont {Peterson}},\ and\ \bibinfo {author}
  {\bibfnamefont {A.~B.}\ \bibnamefont {Jenkin}},\ }\href@noop {} {\emph
  {\bibinfo {title} {Best Practices for Generating Space Environment
  Specifications with Modern Tools}}},\ \bibinfo {type} {Technical Report}\
  \bibinfo {number} {TOR-2022-00016}\ (\bibinfo  {institution} {The Aerospace
  Corporation},\ \bibinfo {year} {2021})\BibitemShut {NoStop}%
\bibitem [{\citenamefont {Felix}\ \emph {et~al.}(2024)\citenamefont {Felix},
  \citenamefont {Slater}, \citenamefont {Landauer}, \citenamefont {Pinson},\
  and\ \citenamefont {Rutherford}}]{Felix2024Jetson}%
  \BibitemOpen
  \bibfield  {author} {\bibinfo {author} {\bibfnamefont {M.~A.}\ \bibnamefont
  {Felix}}, \bibinfo {author} {\bibfnamefont {W.~S.}\ \bibnamefont {Slater}},
  \bibinfo {author} {\bibfnamefont {D.~C.}\ \bibnamefont {Landauer}}, \bibinfo
  {author} {\bibfnamefont {R.~E.}\ \bibnamefont {Pinson}},\ and\ \bibinfo
  {author} {\bibfnamefont {B.~B.~W.}\ \bibnamefont {Rutherford}},\ }\bibfield
  {title} {\bibinfo {title} {Total ionizing dose radiation testing of {NVIDIA
  Jetson Orin NX} system on module},\ }in\ \href@noop {} {\emph {\bibinfo
  {booktitle} {2024 IEEE Space Computing Conference (SCC)}}}\ (\bibinfo
  {publisher} {IEEE},\ \bibinfo {year} {2024})\ pp.\ \bibinfo {pages}
  {116--121}\BibitemShut {NoStop}%
\bibitem [{\citenamefont {Rodriguez-Ferrandez}\ \emph
  {et~al.}(2025)\citenamefont {Rodriguez-Ferrandez}, \citenamefont {Blasco},
  \citenamefont {Kosmidis}, \citenamefont {Tali},\ and\ \citenamefont
  {Steenari}}]{Rodriguez2025Jetson}%
  \BibitemOpen
  \bibfield  {author} {\bibinfo {author} {\bibfnamefont {I.}~\bibnamefont
  {Rodriguez-Ferrandez}}, \bibinfo {author} {\bibfnamefont {E.~R.}\
  \bibnamefont {Blasco}}, \bibinfo {author} {\bibfnamefont {L.}~\bibnamefont
  {Kosmidis}}, \bibinfo {author} {\bibfnamefont {M.}~\bibnamefont {Tali}},\
  and\ \bibinfo {author} {\bibfnamefont {D.}~\bibnamefont {Steenari}},\
  }\bibfield  {title} {\bibinfo {title} {Radiation effects on {NVIDIA Jetson
  SoCs}: Insights from heavy-ion irradiation},\ }in\ \href@noop {} {\emph
  {\bibinfo {booktitle} {2025 IEEE International Symposium on Defect and Fault
  Tolerance in VLSI and Nanotechnology Systems (DFT)}}}\ (\bibinfo  {publisher}
  {IEEE},\ \bibinfo {year} {2025})\ pp.\ \bibinfo {pages} {1--4}\BibitemShut
  {NoStop}%
\bibitem [{\citenamefont {Vavrin}\ \emph {et~al.}(2023)\citenamefont {Vavrin},
  \citenamefont {Manis}, \citenamefont {Seago}, \citenamefont {Gates},
  \citenamefont {Anz-Meador}, \citenamefont {Matney},\ and\ \citenamefont
  {Liou}}]{Vavrin2023ORDEM32}%
  \BibitemOpen
  \bibfield  {author} {\bibinfo {author} {\bibfnamefont {A.}~\bibnamefont
  {Vavrin}}, \bibinfo {author} {\bibfnamefont {A.}~\bibnamefont {Manis}},
  \bibinfo {author} {\bibfnamefont {J.}~\bibnamefont {Seago}}, \bibinfo
  {author} {\bibfnamefont {D.}~\bibnamefont {Gates}}, \bibinfo {author}
  {\bibfnamefont {P.}~\bibnamefont {Anz-Meador}}, \bibinfo {author}
  {\bibfnamefont {M.}~\bibnamefont {Matney}},\ and\ \bibinfo {author}
  {\bibfnamefont {J.-C.}\ \bibnamefont {Liou}},\ }\href
  {https://ntrs.nasa.gov/citations/20230014989} {\emph {\bibinfo {title} {NASA
  Orbital Debris Engineering Model {ORDEM} 3.2: Software User Guide}}},\
  \bibinfo {type} {Tech. Rep.}\ \bibinfo {number} {NASA/TP-20230014989}\
  (\bibinfo  {institution} {NASA Johnson Space Center},\ \bibinfo {year}
  {2023})\BibitemShut {NoStop}%
\bibitem [{\citenamefont {Banks}\ \emph {et~al.}(2003)\citenamefont {Banks},
  \citenamefont {Miller}, \citenamefont {de~Groh},\ and\ \citenamefont
  {Demko}}]{Banks2003AO}%
  \BibitemOpen
  \bibfield  {author} {\bibinfo {author} {\bibfnamefont {B.~A.}\ \bibnamefont
  {Banks}}, \bibinfo {author} {\bibfnamefont {S.~K.~R.}\ \bibnamefont
  {Miller}}, \bibinfo {author} {\bibfnamefont {K.~K.}\ \bibnamefont
  {de~Groh}},\ and\ \bibinfo {author} {\bibfnamefont {R.}~\bibnamefont
  {Demko}},\ }\href {https://ntrs.nasa.gov/citations/20030062195} {\emph
  {\bibinfo {title} {Atomic Oxygen Effects on Spacecraft Materials}}},\
  \bibinfo {type} {Tech. Rep.}\ \bibinfo {number} {NASA/TM-2003-212484}\
  (\bibinfo  {institution} {NASA Glenn Research Center},\ \bibinfo {year}
  {2003})\ \bibinfo {note} {{NASA NTRS} 20030062195}\BibitemShut {NoStop}%
\bibitem [{\citenamefont {de~Groh}\ and\ \citenamefont
  {Banks}(2025)}]{deGroh2025MISSE}%
  \BibitemOpen
  \bibfield  {author} {\bibinfo {author} {\bibfnamefont {K.~K.}\ \bibnamefont
  {de~Groh}}\ and\ \bibinfo {author} {\bibfnamefont {B.~A.}\ \bibnamefont
  {Banks}},\ }\href {https://ntrs.nasa.gov/citations/20240000755} {\emph
  {\bibinfo {title} {Space Environmental Exposure of the {MISSE} 9--15 Polymers
  and Composites Experiment 1--4 ({PCE} 1--4)}}},\ \bibinfo {type} {Tech.
  Rep.}\ \bibinfo {number} {NASA/TM-20240000755/REV1}\ (\bibinfo  {institution}
  {National Aeronautics and Space Administration},\ \bibinfo {year}
  {2025})\BibitemShut {NoStop}%
\bibitem [{\citenamefont {Coffin}(1954)}]{Coffin1954}%
  \BibitemOpen
  \bibfield  {author} {\bibinfo {author} {\bibfnamefont {J.}~\bibnamefont
  {Coffin}, \bibfnamefont {L.~F.}},\ }\bibfield  {title} {\bibinfo {title} {A
  study of the effects of cyclic thermal stresses on a ductile metal},\
  }\href@noop {} {\bibfield  {journal} {\bibinfo  {journal} {Journal of Fluids
  Engineering}\ }\textbf {\bibinfo {volume} {76}},\ \bibinfo {pages} {931}
  (\bibinfo {year} {1954})}\BibitemShut {NoStop}%
\bibitem [{\citenamefont {Manson}(1953)}]{Manson1953}%
  \BibitemOpen
  \bibfield  {author} {\bibinfo {author} {\bibfnamefont {S.~S.}\ \bibnamefont
  {Manson}},\ }\href {https://ntrs.nasa.gov/citations/19930083626} {\emph
  {\bibinfo {title} {Behavior of Materials Under Conditions of Thermal
  Stress}}},\ \bibinfo {type} {Tech. Rep.}\ \bibinfo {number} {NACA TN 2933}\
  (\bibinfo  {institution} {National Advisory Committee for Aeronautics},\
  \bibinfo {year} {1953})\ \bibinfo {note} {{NASA NTRS}
  19930083626}\BibitemShut {NoStop}%
\bibitem [{\citenamefont {Miner}(1945)}]{Miner1945}%
  \BibitemOpen
  \bibfield  {author} {\bibinfo {author} {\bibfnamefont {M.~A.}\ \bibnamefont
  {Miner}},\ }\bibfield  {title} {\bibinfo {title} {Cumulative damage in
  fatigue},\ }\href@noop {} {\bibfield  {journal} {\bibinfo  {journal} {Journal
  of Applied Mechanics}\ }\textbf {\bibinfo {volume} {12}},\ \bibinfo {pages}
  {A159} (\bibinfo {year} {1945})}\BibitemShut {NoStop}%
\bibitem [{\citenamefont {Fusaro}(1999)}]{Fusaro1999Mechanisms}%
  \BibitemOpen
  \bibfield  {author} {\bibinfo {author} {\bibfnamefont {R.~L.}\ \bibnamefont
  {Fusaro}},\ }\href {https://ntrs.nasa.gov/citations/20050192114} {\emph
  {\bibinfo {title} {NASA Space Mechanisms Handbook: Lessons Learned
  Documented}}},\ \bibinfo {type} {Tech. Rep.}\ (\bibinfo  {institution} {NASA
  Lewis Research Center},\ \bibinfo {year} {1999})\ \bibinfo {note} {{NASA
  NTRS} 20050192114}\BibitemShut {NoStop}%
\bibitem [{\citenamefont {Morgan}(2005)}]{Morgan2005FaultProtection}%
  \BibitemOpen
  \bibfield  {author} {\bibinfo {author} {\bibfnamefont {P.~S.}\ \bibnamefont
  {Morgan}},\ }\bibfield  {title} {\bibinfo {title} {Fault protection
  techniques in {JPL} spacecraft},\ }in\ \href
  {https://ntrs.nasa.gov/citations/20060044316} {\emph {\bibinfo {booktitle}
  {First International Forum on Integrated System Health Engineering and
  Management in Aerospace}}}\ (\bibinfo {address} {Napa, California},\ \bibinfo
  {year} {2005})\ \bibinfo {note} {{NASA NTRS} 20060044316}\BibitemShut
  {NoStop}%
\bibitem [{\citenamefont {Rasmussen}(2008)}]{Rasmussen2008FaultProtection}%
  \BibitemOpen
  \bibfield  {author} {\bibinfo {author} {\bibfnamefont {R.~D.}\ \bibnamefont
  {Rasmussen}},\ }\bibfield  {title} {\bibinfo {title} {{GN\&C} fault
  protection fundamentals},\ }in\ \href
  {https://ntrs.nasa.gov/citations/20100020171} {\emph {\bibinfo {booktitle}
  {31st Annual AAS Guidance and Control Conference}}}\ (\bibinfo {address}
  {Breckenridge, Colorado},\ \bibinfo {year} {2008})\ \bibinfo {note} {{NASA
  NTRS} 20100020171}\BibitemShut {NoStop}%
\bibitem [{\citenamefont {Aaseng}\ \emph {et~al.}(2023)\citenamefont {Aaseng},
  \citenamefont {Do}, \citenamefont {Frank}, \citenamefont {Fry},\ and\
  \citenamefont {Sweet}}]{Aaseng2023VSM}%
  \BibitemOpen
  \bibfield  {author} {\bibinfo {author} {\bibfnamefont {G.}~\bibnamefont
  {Aaseng}}, \bibinfo {author} {\bibfnamefont {M.}~\bibnamefont {Do}}, \bibinfo
  {author} {\bibfnamefont {J.}~\bibnamefont {Frank}}, \bibinfo {author}
  {\bibfnamefont {C.}~\bibnamefont {Fry}},\ and\ \bibinfo {author}
  {\bibfnamefont {A.}~\bibnamefont {Sweet}},\ }\bibfield  {title} {\bibinfo
  {title} {Integrating planning, diagnosis and execution for vehicle systems
  management},\ }in\ \href {https://ntrs.nasa.gov/citations/20230004265} {\emph
  {\bibinfo {booktitle} {ICAPS 2023 Workshop on Integrated Planning, Acting,
  and Execution}}}\ (\bibinfo {address} {Prague, Czech Republic},\ \bibinfo
  {year} {2023})\ \bibinfo {note} {{NASA NTRS} 20230004265}\BibitemShut
  {NoStop}%
\bibitem [{\citenamefont {Scholl}\ and\ \citenamefont
  {Suloway}(2023)}]{Scholl2023NIST8270}%
  \BibitemOpen
  \bibfield  {author} {\bibinfo {author} {\bibfnamefont {M.}~\bibnamefont
  {Scholl}}\ and\ \bibinfo {author} {\bibfnamefont {T.}~\bibnamefont
  {Suloway}},\ }\href {https://doi.org/10.6028/NIST.IR.8270} {\emph {\bibinfo
  {title} {Introduction to Cybersecurity for Commercial Satellite
  Operations}}},\ \bibinfo {type} {Tech. Rep.}\ \bibinfo {number} {NIST IR
  8270}\ (\bibinfo  {institution} {National Institute of Standards and
  Technology},\ \bibinfo {year} {2023})\BibitemShut {NoStop}%
\bibitem [{\citenamefont {McCarthy}\ \emph {et~al.}(2023)\citenamefont
  {McCarthy}, \citenamefont {Mamula}, \citenamefont {Brule}, \citenamefont
  {Meldorf}, \citenamefont {Jennings}, \citenamefont {Wiltberger},
  \citenamefont {Thorpe}, \citenamefont {Dombrowski}, \citenamefont {Lattin},\
  and\ \citenamefont {Sepassi}}]{NIST2023HSN}%
  \BibitemOpen
  \bibfield  {author} {\bibinfo {author} {\bibfnamefont {J.}~\bibnamefont
  {McCarthy}}, \bibinfo {author} {\bibfnamefont {D.}~\bibnamefont {Mamula}},
  \bibinfo {author} {\bibfnamefont {J.}~\bibnamefont {Brule}}, \bibinfo
  {author} {\bibfnamefont {K.}~\bibnamefont {Meldorf}}, \bibinfo {author}
  {\bibfnamefont {R.}~\bibnamefont {Jennings}}, \bibinfo {author}
  {\bibfnamefont {J.}~\bibnamefont {Wiltberger}}, \bibinfo {author}
  {\bibfnamefont {C.}~\bibnamefont {Thorpe}}, \bibinfo {author} {\bibfnamefont
  {J.}~\bibnamefont {Dombrowski}}, \bibinfo {author} {\bibfnamefont
  {O.}~\bibnamefont {Lattin}},\ and\ \bibinfo {author} {\bibfnamefont
  {S.}~\bibnamefont {Sepassi}},\ }\href {https://doi.org/10.6028/NIST.IR.8441}
  {\emph {\bibinfo {title} {Cybersecurity Framework Profile for Hybrid
  Satellite Networks ({HSN})}}},\ \bibinfo {type} {Tech. Rep.}\ \bibinfo
  {number} {NIST IR 8441}\ (\bibinfo  {institution} {National Institute of
  Standards and Technology},\ \bibinfo {year} {2023})\BibitemShut {NoStop}%
\bibitem [{\citenamefont {{Consultative Committee for Space Data
  Systems}}(2022)}]{CCSDS2022SDLS}%
  \BibitemOpen
  \bibfield  {author} {\bibinfo {author} {\bibnamefont {{Consultative Committee
  for Space Data Systems}}},\ }\href@noop {} {\emph {\bibinfo {title} {Space
  Data Link Security Protocol}}},\ \bibinfo {type} {Blue Book, Issue 2}\
  \bibinfo {number} {CCSDS 355.0-B-2}\ (\bibinfo  {institution} {Consultative
  Committee for Space Data Systems},\ \bibinfo {year} {2022})\BibitemShut
  {NoStop}%
\bibitem [{\citenamefont {Donahoe}(2022)}]{IDSS2022}%
  \BibitemOpen
  \bibfield  {author} {\bibinfo {author} {\bibfnamefont {S.~R.}\ \bibnamefont
  {Donahoe}},\ }\href {https://ntrs.nasa.gov/citations/20220011643} {\emph
  {\bibinfo {title} {International Docking System Standard ({IDSS}) Interface
  Definition Document ({IDD}), Revision F}}},\ \bibinfo {type} {Tech. Rep.}\
  \bibinfo {number} {IDSS IDD Revision F}\ (\bibinfo  {institution} {National
  Aeronautics and Space Administration},\ \bibinfo {year} {2022})\ \bibinfo
  {note} {{NASA NTRS} 20220011643}\BibitemShut {NoStop}%
\bibitem [{\citenamefont {{International Organization for
  Standardization}}(2022)}]{ISO24330}%
  \BibitemOpen
  \bibfield  {author} {\bibinfo {author} {\bibnamefont {{International
  Organization for Standardization}}},\ }\href@noop {} {\emph {\bibinfo {title}
  {Space Systems---Rendezvous and Proximity Operations ({RPO}) and On-Orbit
  Servicing ({OOS})---Programmatic Principles and Practices}}},\ \bibinfo
  {type} {Tech. Rep.}\ \bibinfo {number} {ISO 24330:2022}\ (\bibinfo
  {institution} {International Organization for Standardization},\ \bibinfo
  {year} {2022})\BibitemShut {NoStop}%
\bibitem [{\citenamefont {{National Aeronautics and Space
  Administration}}(2026{\natexlab{a}})}]{NASA2026RRM}%
  \BibitemOpen
  \bibfield  {author} {\bibinfo {author} {\bibnamefont {{National Aeronautics
  and Space Administration}}},\ }\href {https://www.nasa.gov/isam/rrm-1-2/}
  {\bibinfo {title} {{Robotic Refueling Mission 1 and 2}}},\ \bibinfo
  {howpublished} {NASA In-Space Servicing, Assembly, and Manufacturing program}
  (\bibinfo {year} {2026}{\natexlab{a}}),\ \bibinfo {note} {official mission
  page, updated March 23, 2026}\BibitemShut {NoStop}%
\bibitem [{\citenamefont {{National Aeronautics and Space
  Administration}}(2023{\natexlab{a}})}]{NASA2023Dextre}%
  \BibitemOpen
  \bibfield  {author} {\bibinfo {author} {\bibnamefont {{National Aeronautics
  and Space Administration}}},\ }\href
  {https://www.nasa.gov/international-space-station/mobile-servicing-system/}
  {\bibinfo {title} {Mobile servicing system: {Canadarm2}, {Dextre}, and the
  {Mobile Base System}}},\ \bibinfo {howpublished} {NASA International Space
  Station reference} (\bibinfo {year} {2023}{\natexlab{a}})\BibitemShut
  {NoStop}%
\bibitem [{\citenamefont {{National Aeronautics and Space
  Administration}}(2026{\natexlab{b}})}]{NASA2026HubbleSM4}%
  \BibitemOpen
  \bibfield  {author} {\bibinfo {author} {\bibnamefont {{National Aeronautics
  and Space Administration}}},\ }\href
  {https://science.nasa.gov/mission/hubble/observatory/missions-to-hubble/servicing-mission-4/}
  {\bibinfo {title} {{Servicing Mission 4} ({SM4})}},\ \bibinfo {howpublished}
  {NASA Hubble Space Telescope mission reference} (\bibinfo {year}
  {2026}{\natexlab{b}}),\ \bibinfo {note} {mission conducted May 11--24,
  2009}\BibitemShut {NoStop}%
\bibitem [{\citenamefont {{Northrop Grumman}}(2021)}]{Northrop2021MEV2}%
  \BibitemOpen
  \bibfield  {author} {\bibinfo {author} {\bibnamefont {{Northrop Grumman}}},\
  }\href@noop {} {\bibinfo {title} {Northrop grumman and intelsat make history
  with docking of second mission extension vehicle to extend life of
  satellite}},\ \bibinfo {howpublished} {Company mission release} (\bibinfo
  {year} {2021})\BibitemShut {NoStop}%
\bibitem [{\citenamefont {{Astroscale Japan
  Inc.}}(2026)}]{Astroscale2026ADRASJ}%
  \BibitemOpen
  \bibfield  {author} {\bibinfo {author} {\bibnamefont {{Astroscale Japan
  Inc.}}},\ }\href
  {https://www.astroscale.com/news/astroscales-adras-j-mission-completes-operations-begins-deorbit}
  {\bibinfo {title} {{ADRAS-J} mission completes operations, begins deorbit}},\
  \bibinfo {howpublished} {Official mission update} (\bibinfo {year}
  {2026})\BibitemShut {NoStop}%
\bibitem [{\citenamefont {Shortle}\ \emph {et~al.}(2018)\citenamefont
  {Shortle}, \citenamefont {Thompson}, \citenamefont {Gross},\ and\
  \citenamefont {Harris}}]{Gross2018Queueing}%
  \BibitemOpen
  \bibfield  {author} {\bibinfo {author} {\bibfnamefont {J.~F.}\ \bibnamefont
  {Shortle}}, \bibinfo {author} {\bibfnamefont {J.~M.}\ \bibnamefont
  {Thompson}}, \bibinfo {author} {\bibfnamefont {D.}~\bibnamefont {Gross}},\
  and\ \bibinfo {author} {\bibfnamefont {C.~M.}\ \bibnamefont {Harris}},\
  }\href@noop {} {\emph {\bibinfo {title} {Fundamentals of Queueing Theory}}},\
  \bibinfo {edition} {5th}\ ed.\ (\bibinfo  {publisher} {Wiley},\ \bibinfo
  {address} {Hoboken, New Jersey},\ \bibinfo {year} {2018})\BibitemShut
  {NoStop}%
\bibitem [{\citenamefont {{National Aeronautics and Space
  Administration}}(2017)}]{NASA2017RM}%
  \BibitemOpen
  \bibfield  {author} {\bibinfo {author} {\bibnamefont {{National Aeronautics
  and Space Administration}}},\ }\href@noop {} {\emph {\bibinfo {title} {NASA
  Reliability and Maintainability Standard for Spaceflight and Support
  Systems}}},\ \bibinfo {type} {Tech. Rep.}\ \bibinfo {number}
  {NASA-STD-8729.1A}\ (\bibinfo  {institution} {National Aeronautics and Space
  Administration},\ \bibinfo {year} {2017})\BibitemShut {NoStop}%
\bibitem [{\citenamefont {Modarres}\ \emph {et~al.}(2017)\citenamefont
  {Modarres}, \citenamefont {Kaminskiy},\ and\ \citenamefont
  {Krivtsov}}]{Modarres2017}%
  \BibitemOpen
  \bibfield  {author} {\bibinfo {author} {\bibfnamefont {M.}~\bibnamefont
  {Modarres}}, \bibinfo {author} {\bibfnamefont {M.}~\bibnamefont
  {Kaminskiy}},\ and\ \bibinfo {author} {\bibfnamefont {V.}~\bibnamefont
  {Krivtsov}},\ }\href@noop {} {\emph {\bibinfo {title} {Reliability
  Engineering and Risk Analysis: A Practical Guide}}},\ \bibinfo {edition}
  {3rd}\ ed.\ (\bibinfo  {publisher} {CRC Press},\ \bibinfo {address} {Boca
  Raton, Florida},\ \bibinfo {year} {2017})\BibitemShut {NoStop}%
\bibitem [{\citenamefont {Rausand}\ and\ \citenamefont
  {H{\o}yland}(2004)}]{Rausand2004}%
  \BibitemOpen
  \bibfield  {author} {\bibinfo {author} {\bibfnamefont {M.}~\bibnamefont
  {Rausand}}\ and\ \bibinfo {author} {\bibfnamefont {A.}~\bibnamefont
  {H{\o}yland}},\ }\href@noop {} {\emph {\bibinfo {title} {System Reliability
  Theory: Models, Statistical Methods, and Applications}}},\ \bibinfo {edition}
  {2nd}\ ed.\ (\bibinfo  {publisher} {Wiley},\ \bibinfo {address} {Hoboken, New
  Jersey},\ \bibinfo {year} {2004})\BibitemShut {NoStop}%
\bibitem [{\citenamefont {{National Aeronautics and Space
  Administration}}(2023{\natexlab{b}})}]{NASA3001V1}%
  \BibitemOpen
  \bibfield  {author} {\bibinfo {author} {\bibnamefont {{National Aeronautics
  and Space Administration}}},\ }\href@noop {} {\emph {\bibinfo {title} {NASA
  Spaceflight Human-System Standard, Volume 1: Crew Health}}},\ \bibinfo {type}
  {Tech. Rep.}\ \bibinfo {number} {NASA-STD-3001, Volume 1, Revision C}\
  (\bibinfo  {institution} {National Aeronautics and Space Administration},\
  \bibinfo {year} {2023})\BibitemShut {NoStop}%
\bibitem [{\citenamefont {{National Aeronautics and Space
  Administration}}(2025)}]{NASA3001V2}%
  \BibitemOpen
  \bibfield  {author} {\bibinfo {author} {\bibnamefont {{National Aeronautics
  and Space Administration}}},\ }\href@noop {} {\emph {\bibinfo {title} {NASA
  Spaceflight Human-System Standard, Volume 2: Human Factors, Habitability, and
  Environmental Health}}},\ \bibinfo {type} {Tech. Rep.}\ \bibinfo {number}
  {NASA-STD-3001, Volume 2, Revision E}\ (\bibinfo  {institution} {National
  Aeronautics and Space Administration},\ \bibinfo {year} {2025})\BibitemShut
  {NoStop}%
\bibitem [{\citenamefont {Gaskill}(2023)}]{NASA2023WaterRecovery}%
  \BibitemOpen
  \bibfield  {author} {\bibinfo {author} {\bibfnamefont {M.~L.}\ \bibnamefont
  {Gaskill}},\ }\href
  {https://www.nasa.gov/missions/station/iss-research/nasa-achieves-water-recovery-milestone-on-international-space-station/}
  {\bibinfo {title} {Nasa achieves water recovery milestone on international
  space station}},\ \bibinfo {howpublished} {NASA International Space Station
  Research Communications} (\bibinfo {year} {2023})\BibitemShut {NoStop}%
\bibitem [{\citenamefont {Leach}\ and\ \citenamefont
  {Ewert}(2021)}]{Leach2021ISSLogistics}%
  \BibitemOpen
  \bibfield  {author} {\bibinfo {author} {\bibfnamefont {H.}~\bibnamefont
  {Leach}}\ and\ \bibinfo {author} {\bibfnamefont {M.}~\bibnamefont {Ewert}},\
  }\bibfield  {title} {\bibinfo {title} {Analysis of historical international
  space station logistical mass delivery},\ }in\ \href
  {https://ntrs.nasa.gov/citations/20210000437} {\emph {\bibinfo {booktitle}
  {2021 IEEE Aerospace Conference}}}\ (\bibinfo  {publisher} {IEEE},\ \bibinfo
  {year} {2021})\ \bibinfo {note} {{NASA NTRS} 20210000437}\BibitemShut
  {NoStop}%
\bibitem [{\citenamefont {Wieland}(2005)}]{Wieland2005ECLSS}%
  \BibitemOpen
  \bibfield  {author} {\bibinfo {author} {\bibfnamefont {P.~O.}\ \bibnamefont
  {Wieland}},\ }\href {https://ntrs.nasa.gov/citations/20060005209} {\emph
  {\bibinfo {title} {Designing for Human Presence in Space: An Introduction to
  Environmental Control and Life Support Systems}}},\ \bibinfo {type} {Tech.
  Rep.}\ \bibinfo {number} {NASA/TM-2005-214007}\ (\bibinfo  {institution}
  {NASA Marshall Space Flight Center},\ \bibinfo {year} {2005})\ \bibinfo
  {note} {{NASA NTRS} 20060005209}\BibitemShut {NoStop}%
\bibitem [{\citenamefont {Perchonok}\ and\ \citenamefont
  {Stoklosa}(2009)}]{Perchonok2009FoodMass}%
  \BibitemOpen
  \bibfield  {author} {\bibinfo {author} {\bibfnamefont {M.~H.}\ \bibnamefont
  {Perchonok}}\ and\ \bibinfo {author} {\bibfnamefont {A.~M.}\ \bibnamefont
  {Stoklosa}},\ }\href {https://ntrs.nasa.gov/citations/20090006804} {\emph
  {\bibinfo {title} {Food Mass Reduction Trade Study}}},\ \bibinfo {type}
  {Tech. Rep.}\ \bibinfo {number} {JSC-17653}\ (\bibinfo  {institution} {NASA
  Johnson Space Center},\ \bibinfo {year} {2009})\ \bibinfo {note} {{NASA NTRS}
  20090006804}\BibitemShut {NoStop}%
\bibitem [{\citenamefont {Sheridan}(1992)}]{Sheridan1992Teleoperation}%
  \BibitemOpen
  \bibfield  {author} {\bibinfo {author} {\bibfnamefont {T.~B.}\ \bibnamefont
  {Sheridan}},\ }\href@noop {} {\emph {\bibinfo {title} {Telerobotics,
  Automation, and Human Supervisory Control}}}\ (\bibinfo  {publisher} {MIT
  Press},\ \bibinfo {address} {Cambridge, Massachusetts},\ \bibinfo {year}
  {1992})\BibitemShut {NoStop}%
\end{thebibliography}

%

\end{document}